\documentclass[twoside,a4paper,12pt,openright]{scrbook}

\usepackage[english]{babel}
\usepackage[utf8]{inputenc}
\usepackage{xspace}
\usepackage{import}

\usepackage{classicthesis}
\usepackage{graphicx}
\usepackage{geometry}

\newcommand{\myTitle}{Generalizing inference systems by coaxioms}
\newcommand{\myName}{Francesco Dagnino}
\newcommand{\myUni}{Università degli Studi di Genova}
\newcommand{\mySchool}{Scuola di Scienze Matematiche, Fisiche e Naturali}

\newcommand{\IntroName}{Introduction}

\PassOptionsToPackage{hyperfootnotes=false,pdfpagelabels}{hyperref}
    \usepackage{hyperref}  

\hypersetup{
    colorlinks=true, linktocpage=true, pdfstartview=FitV,%
    breaklinks=true, pdfpagemode=UseNone, pageanchor=true, pdfpagemode=UseOutlines,%
    plainpages=false, bookmarksnumbered, bookmarksopen=true, bookmarksopenlevel=1,%
    hypertexnames=true, pdfhighlight=/O,
    urlcolor=webbrown, linkcolor=RoyalBlue, citecolor=webgreen, 
    pdftitle={\myTitle},%
    pdfauthor={\textcopyright\ \myName, \myUni},%
    pdfsubject={},%
    pdfkeywords={},%
    pdfcreator={pdfLaTeX},%
    pdfproducer={LaTeX with hyperref and classicthesis}%
}

\usepackage{amsmath}
\usepackage{stmaryrd}
\usepackage{tikz-cd}
\usepackage{amsthm}

\DeclareMathOperator*{\argmin}{arg\,min}

\newtheoremstyle{smallcaps}
{}
{}
{}
{}
{}%
{.}
{.5em}
{\textsc{\thmname{#1}\thmnumber{ #2}\thmnote{ (#3)}}}%
\theoremstyle{smallcaps}

\newtheorem{definition}{Definition}[chapter]
\newtheorem{theorem}{Theorem}[chapter]
\newtheorem{lemma}{Lemma}[chapter]
\newtheorem{corollary}{Corollary}[chapter]
\newtheorem{proposition}{Proposition}[chapter]

\usepackage{csquotes}
\PassOptionsToPackage{
	backend=bibtex8,
	language=auto,%
    style=authoryear-comp, 
    bibstyle=authoryear,dashed=false, 
    maxbibnames=10, 
    maxcitenames=2,
    natbib=true 
}{biblatex}
    \usepackage{biblatex}

\renewcommand{\cite}[1]{\parencite{#1}}    

\usepackage{listings} 
\usepackage{etoolbox}

\newcommand{\chapSpace}{5.5pt}
\newcommand{\sectSpace}{1.5pt}

\pretocmd{\chapter}{\addtocontents{toc}{\protect\vspace{\chapSpace}}}{}{}
\pretocmd{\section}{\addtocontents{toc}{\protect\vspace{\sectSpace}}}{}{}

\newcommand{\Space}{\hskip 0.7em}
\newcommand{\BigSpace}{\hskip 1.5em}

\newcommand{\refToFigure}[1]{Figure~\ref{fig:#1}}
\newcommand{\refToSection}[1]{Section~\ref{sect:#1}}
\newcommand{\refToChapter}[1]{Chapter~\ref{chapter:#1}}
\newcommand{\refToTheorem}[1]{Theorem~\ref{theo:#1}}
\newcommand{\refToCorollary}[1]{Corollary~\ref{cor:#1}}
\newcommand{\refToLemma}[1]{Lemma~\ref{lem:#1}}
\newcommand{\refToDefinition}[1]{Definition~\ref{def:#1}}
\newcommand{\refToProposition}[1]{Proposition~\ref{prop:#1}}

\newcommand{\fun}[3]{{#1}:{#2}\rightarrow{#3}}
\newcommand{\Id}[1]{\textsf{id}_{#1}}
\newcommand{\dom}{\mathit{dom}}

\newcommand{\Tuple}[1]{\left({#1}\right)}
\newcommand{\Pair}[2]{\Tuple{{#1},\,{#2}}}
\newcommand{\Triple}[3]{\Tuple{{#1},\,{#2},\,{#3}}}

\newcommand{\N}{\mathbb{N}}
\newcommand{\Z}{\mathbb{Z}}

\newcommand{\judg}{\mathit{j}}
\newcommand{\prem}{\textit{Pr}}
\newcommand{\cons}{\textit{c}}
\newcommand{\is}{{\cal I}}
\newcommand{\universe}{{\cal U}}
\newcommand{\myrule}{\Rule{\prem}{\cons}}
\newcommand{\Rule}[2]{
	\displaystyle                  
	\frac{#1}{#2}     
}
\newcommand{\CoAxiom}[1]{
	\Rule{\bullet}{#1}     
}
\newcommand{\Op}[1]{{\textit{F}_{#1}}}
\newcommand{\IterOp}[2]{\textit{F}^{#2}_{#1}}
\newcommand{\Ind}[1]{\textit{Ind}(#1)}
\newcommand{\CoInd}[1]{\textit{CoInd}(#1)}
\newcommand{\Generated}[2]{\textit{Gen}(#1,#2)}
\newcommand{\coaxioms}{\gamma}

\newcommand{\DefSet}{\mathcal{D}}
\newcommand{\Spec}{\mathcal{S}}

\newcommand{\CoIndPrinciple}{\textsc{\small (CoInd)}\xspace}
\newcommand{\IndPrinciple}{\textsc{\small (Ind)}\xspace}
\newcommand{\SpecAllPos}{\Spec^\textit{allPos}}

\newcommand{\Type}[2]{{#1}\,{:}\,{#2}}
\newcommand{\Nat}{\textsf{Nat}}
\newcommand{\Succ}[1]{\textsf{s}(#1)}

\newcommand{\List}[1]{\textsf{List}(#1)}
\newcommand{\EList}{\Lambda}
\newcommand{\Cons}[2]{{#1}{::}{#2}}
\newcommand{\LSet}{\mathbb{L}}
\newcommand{\LInfSet}{\LSet^\infty}
\newcommand{\Bool}{\mathbb{B}}
\newcommand{\True}{\textsf{T}}
\newcommand{\False}{\textsf{F}}

\newcommand{\xs}{\mathit{xs}}
\newcommand{\elemsName}{\textit{elems}}
\newcommand{\memberName}{\textit{member}}
\newcommand{\allPosName}{\textit{allPos}}
\newcommand{\maxElemName}{\textit{maxElem}}

\newcommand{\elems}[2]{\elemsName{\Pair{#1}{#2}}}
\newcommand{\member}[3]{\memberName{\Triple{#1}{#2}{#3}}}
\newcommand{\memberPred}[2]{\memberName{\Pair{#1}{#2}}}
\newcommand{\allPos}[2]{\allPosName{\Pair{#1}{#2}}}
\newcommand{\allPosPred}[1]{\allPosName{\Tuple{#1}}}
\newcommand{\posPred}[1]{\textit{pos}{\Tuple{#1}}}
\newcommand{\maxElem}[2]{\maxElemName{\Pair{#1}{#2}}}

\newcommand{\Stream}[1]{\textsf{Stream}(#1)}
\newcommand{\Head}[1]{\textsf{head}(#1)}
\newcommand{\Tail}[1]{\textsf{tail}(#1)}

\newcommand{\Sub}[1]{\textsf{Sub}(#1)}
\newcommand{\GroupGen}[1]{\left\langle {#1} \right\rangle}

\newcommand{\node}{\textit{v}}
\newcommand{\anode}{\textit{u}}
\newcommand{\nodeset}{{\cal N}}
\newcommand{\Visit}[2]{#1{\stackrel{\star}{\rightarrow}}#2}
\newcommand{\adj}{\textit{adj}}
\newcommand{\Nodes}{\textit{V}}
\newcommand{\Labels}{{\cal L}}
\newcommand{\T}{{\cal T}}
\newcommand{\CTree}[1]{\T^{\textsf{ci}}(#1)}
\newcommand{\Tree}[2][{}]{\T_{#1}{(#2)}}
\newcommand{\GraphFun}[1]{\mathit{G}_{#1}}
\newcommand{\Path}{\textsf{P}}
\newcommand{\Paths}[1]{\textsf{path}{(#1)}}
\newcommand{\dsub}{\textit{dsub}}
\newcommand{\chl}{\textit{chl}}
\newcommand{\dist}[3]{\textit{dist}{\Triple{#1}{#2}{#3}}}
\newcommand{\minPath}[4]{\textit{spath}{\Tuple{{#1},\,{#2},\,{#3},\,{#4}}}}
\newcommand{\Edges}{\textit{E}}
\newcommand{\TOrder}{\triangleleft}
\newcommand{\ApproxTOrder}[1]{\TOrder_{#1}}
\newcommand{\TLub}{\bigvee}
\newcommand{\ApproxEq}[1]{\bowtie_{#1}}

\newcommand{\String}[1]{{#1}^\star}
\newcommand{\EString}{\varepsilon}
\newcommand{\Len}[1]{\left | {#1} \right |}

\newcommand{\nonterminal}{\textit{A}}
\newcommand{\firstset}{{\cal F}}
\newcommand{\first}[1]{\textit{first}(#1)}
\newcommand{\First}[2]{\textit{first}(#1,#2)}
\newcommand{\produzioneinline}[2]{#1::=#2}
\newcommand{\simb}{\sigma} 

\newcommand{\StarOne}{$(\star)$\xspace}
\newcommand{\StarTwo}{$(\star\star)$\xspace}

\newcommand{\isin}{\ensuremath{\mathit{is\_in}}}
\newcommand{\isinZero}{\ensuremath{\mathit{is\_in0}}}
\newcommand{\pathZero}{\ensuremath{\mathit{path0}}}
\newcommand{\tree}{\ensuremath{\mathit{tree}}}
\newcommand{\add}{\ensuremath{\mathit{add}}}
\newcommand{\sem}[1]{\ensuremath{\left\llbracket{#1}\right\rrbracket}}

\newcommand{\infv}{\ensuremath{v_\infty}}
\newcommand{\eval}[2]{\ensuremath{{#1}\Rightarrow{#2}}}
\newcommand{\subs}[3]{\ensuremath{{#1}[{#2}\leftarrow{#3}]}}
\newcommand{\RuleName}[3]{
\displaystyle                  
\mbox{({#1})}\,\frac{#2}{#3}     
}
\newcommand{\CoAxiomName}[2]{
\displaystyle                  
\mbox{({#1})}\,\frac{\bullet}{#2}     
}

\newcommand{\order}{\sqsubseteq}
\newcommand{\dualOrder}{\sqsupseteq}
\newcommand{\poset}{\mathit{P}}
\newcommand{\aposet}{\mathit{Q}}
\newcommand{\LowSet}[1]{{\downarrow}\,{#1}}
\newcommand{\UpSet}[1]{{\uparrow}\,{#1}}
\newcommand{\ub}[1]{\textsf{ub}(#1)}
\newcommand{\lb}[1]{\textsf{lb}(#1)}
\newcommand{\lub}{\bigsqcup}
\newcommand{\glb}{\bigsqcap}
\newcommand{\function}{\mathit{F}}

\newcommand{\lattice}{\mathit{L}}
\newcommand{\op}{\star}
\newcommand{\meet}{\sqcap}
\newcommand{\join}{\sqcup}

\newcommand{\Pre}[1]{\textsf{pre}(#1)}
\newcommand{\Post}[1]{\textsf{post}(#1)}
\newcommand{\Fix}[1]{\textsf{fix}(#1)}
\newcommand{\lfp}{\mu}
\newcommand{\gfp}{\nu}

\newcommand{\Iterate}[2]{\mathit{I}_{{#1}, {#2}} }

\newcommand{\CSys}{\mathit{C}}
\newcommand{\KSys}{\mathit{K}}
\newcommand{\closure}{\nabla}
\renewcommand{\ker}{\Delta}
\newcommand{\bound}{\beta}
\newcommand{\Restricted}[2]{{#1_{{\sqcap}#2}}}
\newcommand{\Extended}[2]{{#1_{{\sqcup}#2}}}

\newcommand{\Clause}[2]{{#1} \leftarrow {#2}}
\newcommand{\FJ}{\textsc{FJ}\xspace}
\newcommand{\coFJ}{\textsc{coFJ}\xspace}
\newcommand{\LP}{\textsc{LP}\xspace}
\newcommand{\coLP}{\textsc{coLP}\xspace}
\newcommand{\InAr}{\textsf{in}}
\newcommand{\FinAr}{\textsf{out}}

\newcommand{\LambdaCalculus}{$\lambda$-calculus\xspace}

\newif\ifsubmit
\submittrue
\ifsubmit
\newcommand{\EZ}[1]{{#1}} 
\newcommand{\FD}[1]{{#1}} 
 
\newcommand{\EZComm}[1]{} 
\newcommand{\DAComm}[1]{} 
\newcommand{\FDComm}[1]{} 
\else
\newcommand{\FD}[1]{\textcolor{red}{#1}} 
\newcommand{\EZ}[1]{\textcolor{blue}{#1}} 
 
\newcommand{\EZComm}[1]{{\scriptsize\textcolor{blue}{[\textbf{Elena: }#1}]}}
\newcommand{\DAComm}[1]{{\scriptsize\textcolor{magenta}{[\textbf{Davide: }#1}]}}
\newcommand{\FDComm}[1]{{\scriptsize\textcolor{red}{[\textbf{Francesco: }#1}]}}
\fi

\begin{document}

\frontmatter
\newgeometry{margin=2.5cm}

\begin{titlepage}

\newcommand{\HRule}{\rule{\linewidth}{0.3mm}} 

\center 

\textsc{\LARGE \myUni}\\[0.5cm] 
\textsc{\large \mySchool}\\[1.0cm] 

\textsc{\large Master Thesis}\\[0.5cm] 


\HRule \\[0.4cm]
{ \huge   \myTitle }\\[0.4cm]  
\HRule \\[1.5cm]


\textsc{\Large \myName} \\[1cm] 

\begin{minipage}{0.4\textwidth}
\begin{flushleft} \large
\emph{Examiner}\\
Prof. Eugenio \textsc{Moggi} 
\end{flushleft}
\end{minipage}
~
\begin{minipage}{0.45\textwidth}
\begin{flushright} \large
\emph{Supervisors} \\
Prof. Davide \textsc{Ancona}  \\
Prof. Elena \textsc{Zucca} 
\end{flushright}
\end{minipage}\\[2cm]



\includegraphics[height=7.5cm]{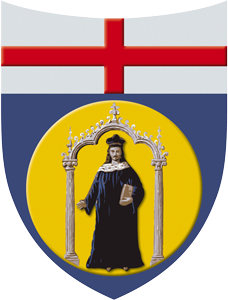}\\[1cm] 

\vspace{1ex}

{\large Academic year 2016/2017} 


\vfill 

\end{titlepage}

\restoregeometry

\cleardoublepage
\pdfbookmark[1]{Abstract}{Abstract}
\begingroup
\let\clearpage\relax
\let\cleardoublepage\relax
\let\cleardoublepage\relax

\chapter*{Abstract}

After surveying classical results, we introduce a generalized notion of inference system to support 
structural recursion on non-well-founded data types.
Besides  axioms and inference rules with the usual meaning, a generalized inference system
allows \emph{coaxioms}, which are, intuitively, axioms which can only be applied ``at infinite depth'' in a proof tree. 
This notion nicely subsumes standard inference systems and their inductive and coinductive interpretation, while providing more flexibility.
Indeed, the classical results can be extended to our generalized framework, interpreting recursive definitions as fixed points which are not necessarily the least, nor the greatest one. 
This allows formal reasoning in cases where the inductive and coinductive interpretation do not provide the intended meaning, or  are mixed together.

\vfill


\endgroup			


\cleardoublepage\pdfbookmark[1]{\contentsname}{tableofcontents}
\setcounter{tocdepth}{2} 
\setcounter{secnumdepth}{3} 
\manualmark
\markboth{\spacedlowsmallcaps{\contentsname}}{\spacedlowsmallcaps{\contentsname}}
\tableofcontents 
\automark[section]{chapter}
\renewcommand{\chaptermark}[1]{\markboth{\spacedlowsmallcaps{#1}}{\spacedlowsmallcaps{#1}}}
\renewcommand{\sectionmark}[1]{\markright{\thesection\enspace\spacedlowsmallcaps{#1}}}

\mainmatter
\cleardoublepage \chapter{\IntroName}

\emph{Induction} is the fundamental building block of a large part of both mathematics and computer science. 
We can mention a number of examples.
Peano's arithmetic has among its axioms an induction principle. G\"odel's recursive functions can be constructed in an inductive way and this feature is crucial to  increase the expressiveness of the theory. \FD{Orders} satisfying the descending chain condition enjoy the {well-founded} induction principle  that is widely used in commutative and computer algebra.
In formal language theory grammars are inductive structures and {operational semantics of programming languages} are usually defined in an inductive way.
In addition, almost every programming language supports in some form a notion of  inductive definitions of types and/or functions, {and} especially in declarative paradigms {this} is essential to write {non-}trivial programs. 
These are only few examples of the presence of induction in mathematics and computer science;
it is so important, widely used and well-established that it is taught since the early years of  any degree in such fields.

Induction allows us to define structures, i.e., data types, and provides a powerful and natural (in the sense that it is {driven} by the definition)  reasoning style to deal with such structures.
Usually inductive definitions are formulated through \emph{rules} that state under which {hypotheses} the judgement we are defining is valid.
{Let us consider} an example:
probably the most well-known one is the inductive definition of natural numbers reported below{:}
\[\Rule{}{\Type{0}{\Nat}} \BigSpace \Rule{\Type{n}{\Nat}}{\Type{\Succ{n}}{\Nat}}\]
Here we are defining the judgement $\Type{n}{\Nat}$ that says "$n$ is a natural number".
The above definition can be read in this way: $0$ is a natural number, and, if $n$ is a natural number, then $\Succ{n}$  is also a natural number. 

When the defining judgement{, like the previous one, describes a data type}, we can {read} the definition from another point of view: we can say that it shows how we can build objects of that type using some \emph{constructors}, in the above example $0$ and $\Succ{-}$.
This means that all natural numbers are built starting from $0$ and repeatedly (zero or more finitely many times)  applying the successor constructor $\Succ{-}$. 

As mentioned before, one of the strengths of inductive definitions is that {they} induce a natural reasoning style.
Indeed, following the rule structure we derive a powerful proof principle: 
if we want to prove {that} a predicate $P$ is satisfied by all {the} judgements valid in the inductive definition, {we can} just  consider each rule separately, assume that $P$ holds for every premise and prove that it holds also for the consequence of that rule.
For instance, in the case of natural numbers, we get the following proof principle.

\begin{quote}
Given a predicate $P$, it holds for any natural number $n$ if we prove $P(0)$,  and, for each $n$,  $P(\Succ{n})$ assuming $P(n)$, as formalized below{:}
\[\Rule
{P(0) \Space \forall \Type{n}{\Nat}.\,P(n) \Rightarrow P(\Succ{n}) }
{\forall \Type{n}{\Nat}.\, P(n)}  
\]
\end{quote}

Another typical example of inductively defined data type are finite lists.
We start from a given type $A$  and denote {by} $\List{A}$ the type of finite lists with elements of type $A$, defined as follows{:}
\[
\Rule{}{ \Type{\EList}{\List{A}} } 
\BigSpace  
\Rule{ \Type{x}{A} \Space \Type{l}{\List{A}} }{ \Type{\Cons{x}{l}}{\List{A}} }
\]
This definition can be read as follows: $\EList$ (the empty list) is a list, and, if $x$ is an element of type $A$ and $l$ is a list, then $\Cons{x}{l}$ is a list. 
Here {the} constructors are $\EList$ and $\Cons{-}{-}$ and again each list is built starting from $\EList$ and applying finitely many times the constructor $\Cons{-}{-}$, that is, a typical list has shape $\Cons{x_1}{\Cons{x_2}{\Cons{\ldots}{\Cons{x_n}{\EList}}}}$. 
The induction principle associated to this data type is the following{.}

\begin{quote}
Given a predicate $P$, it holds for any list $l$if we prove $P(\EList)$,  and, for each list $l$ and element $x$,  $P(\Cons{x}{l})$ assuming $P(l)$, as formalized below{:}
 
\[\Rule
{ 
  P(\EList) 
  \Space 
  \forall \Type{l}{\List{A}}.\, \forall \Type{x}{A}.\, P(l) \Rightarrow P(\Cons{x}{l}) 
}
{\forall \Type{l}{\List{A}}.\, P(l)}  
\]
\end{quote}

Again the structure of this proof principle is guided by  the structure of the data type.
Following this structure we can also define other judgements involving \FD{inductively defined data types}.
For instance, \FD{for lists,} we can define the predicate $\memberPred{x}{l}$ that states that the element $x$  appears in the list $l$.
Here we abstract from the type of the elements since it is not relevant.
The definition is given by the following rules
\[
\Rule{}{ \memberPred{x}{\Cons{x}{l}} }
\BigSpace
\Rule{ \memberPred{y}{l} }
{ \memberPred{y}{\Cons{x}{l}} }
\]
{and} states that the head of {a} list belongs to the list itself, and,  if $x$ belongs to the tail of {a} list, it also belongs to the whole list.
The definition is {correct} since intuitively $x$ belongs to $l$ if and only if removing iteratively the head of the list, in finitely many times we find $x$, that is, we find a list $\Cons{x}{l'}$.

In these simple examples we find a common feature: everything is in some sense finite. 
More precisely, in an inductive definition a judgement is valid if we reach a base case (rules with no premises) in finitely many times. In this situation we say that the definition or the data type is {well-founded}.
We will make {this} more precise in \refToChapter{is} where we will {present} inductive definitions in a rigorous way.\\

\FD{Beside} induction there is another {more} mysterious pattern: \emph{coinduction}, that is in some sense dual to induction.
Coinduction is {less well-known} than induction, however in many cases coinductive reasoning is very useful, notably when dealing with {non-well-founded} or circular structures such as graphs, infinite lists, infinite trees, etc. \\
A coinductive definition  shows how objects can be observed through \emph{destructors}, so the focus is more on the behaviour than on the structure.
For this reason, in literature coinduction has been firstly used to reason about {the behaviour} of concurrent and non-deterministic systems, in particular on bisimilarity \cite{Milner80, MilnerPW92}.
The focus on dynamics and behaviour of systems is present also when the perspective on coinduction is more abstract and general \cite{Aczel88, Rutten00}.

Let us consider an example of coinductive definition of a data type.
We define the type of streams, which are infinite sequences of elements of a given type.
The definition is again expressed using rules but the interpretation is very different{.}
\[
\Rule{\Type{x}{A} \Space \Type{s}{\Stream{A}}}
{\Type{\Cons{x}{s}}{\Stream{A}}}
\]
This definition is very similar to that of lists, \FD{indeed streams are also called infinite lists,} but there is no base case; thus interpreting it inductively makes no sense, because an inductive definition requires to reach a base case in finitely many steps, that is clearly not possible in this case.
Coinduction allows infinitely many steps, so a base case is not necessary and we can build infinite sequences of elements of type $A$.

{As already mentioned,} in coinductive definition{s} the {focus} is more on the behaviour, that is, on how we can decompose the structure we are defining.
{Destructors are not explicit in the rules,} but can be {derived by reading rules} bottom-up: {in the example,} starting from a stream $s$, there always exist an element $x$ (the head) and a stream $s'$ (the tail) such that $s = \Cons{x}{s'}$. 
So we can define destructors $\Head{s}$ and $\Tail{s}$ that given a stream $s$ observe its head and its tail{,} respectively.
Note that {such} destructors {could not} be defined {on finite lists,} since we are not guaranteed that every list has a head and a tail{. In} particular the empty list, that is the base case, has neither a head nor a tail.

Like for inductive definitions, associated to coinductive one{s} there is a proof principle. 
{I}t is not easy to express {such principle} in an informal way, probably due to the less spread and popularity of this definition and reasoning style. 
An attempt to formulate {the} coinduction principle in an informal way is done in \cite{KozenSilva16}, but we are still far from a natural informal explanation.
It is common practice to state coinduction principles referring to the particular framework in which one {works} (category theory, type theory, set theory, {etc.}), so we postpone this aspect to \refToChapter{is} where we will introduce the formal framework we have chosen.

Again {like} for inductive definitions, we can exploit the {structure} of a \FD{coinductively defined data type} in order to define more complex judgements.
For instance, {the} rules for the predicate $\memberPred{x}{l}$  given for the data type of lists work also for streams.
Note that in this case the data type is coinductive while the predicate definition is inductive.
Indeed, induction is enough, since, in order to establish whether this predicate holds, we need to inspect only finitely many elements of the stream.
Instead, if we need to inspect the whole stream, that is, we need infinitely many steps, a coinductive definition {is necessary}.
For instance, considering streams over integers, the predicate $\allPosPred{s}$, that holds if {all the} elements in $s$ are  positive, must be defined coinductively as follows{:}
\[
\Rule{ \allPosPred{s} }{ \allPosPred{\Cons{x}{s}} } x>0
\]
{where} the side condition specifies that the rule applies only when $x$ is a strictly positive integer.\\

We have briefly described the two dual {ways} in which recursive definitions can be interpreted.
However, there are cases in which neither induction nor coinduction precisely capture the intended semantics of the definition, \FD{especially when we are defining more complex judgement\EZ{s} on coinductive data types exploiting their structure.}
Roughly speaking, the problem is that inductive definitions are too strict for dealing with {non-well-founded} data types, but coinductive definitions are too weak: they consider too many judgements as valid.
So there is the need to make the interpretation of definitions more flexible, overcoming {the} dichotomy between induction and coinduction.

In the present {thesis} we address this problem: we choose a particular formal framework in which {to express} (co)inductive definitions and extend it in order to support more flexible interpretations.
This extension has been inspired by some works on operational semantics {of} \FD{language constructs for corecursive definitions of predicates and functions}  \cite{AnconaZucca12, AnconaZucca13, Ancona13}.
The first motivating idea {has been to develop} an abstract framework {for} better {understanding} such operational models and {providing} a more abstract semantics \FD{in order to allow formal reasoning on them.}

This thesis \FD{is an extended version of the work we have done in \cite{AnconaDZ17esop} and} is organized as follows.
In \refToChapter{lattice} we briefly \FD{report} some \FD{standard and well-known notions} from \emph{lattice theory}, a general mathematical theory, studying structures induced by partial orders, that will be useful to define the semantics of our framework.
In \refToChapter{is} we introduce the notion of \emph{inference system} and explain its syntax and semantics. This framework is based on the notion of \emph{inference rule} that allows us to talk about (co)inductive definitions in a very intuitive way, having at the same time a precise and quite simple formal semantics.
\FD{Results presented in this chapter are standard, however {some parts of the presentation are,} at the {best} of our knowledge, original.
In particular, since we {did} not find in literature a rigorous \FD{enough} treatment of the proof-theoretic semantics of inference systems, we have {developed} it autonomously.
The key notions are those of tree and graph presented in \refToSection{trees-graphs}, where we prove \refToTheorem{graph-tree}, that {allows} us to give a new proof of the equivalence between proof-theoretic and fixed point semantics in the coinductive case (see \refToSection{sem-eq}). }

In \refToChapter{coaxioms} we develop an extension of inference systems, both in syntax and semantics, built on the notion of \emph{coaxioms}, that are particular rules used to control the semantics of the whole definition.
\FD{We define both a model-theoretic and a proof-theoretic semantics for coaxioms{.}
The former is based on the new notion of bounded fixed point and the latter on standard and newly defined approximated proof trees.
We also present proof techniques to reason with coaxioms and several examples of applications.}
This chapter presents in a more detailed and complete way the notions and results in \cite{AnconaDZ17esop}. 
\FD{Notably, we have discussed in more \EZ{detail} concepts related to closures and kernels,  in order to better frame the bounded fixed point in lattice theory.
Furthermore, we have provided a better proof-theoretic characterization of the interpretation generated by coaxioms (\refToTheorem{approx-sequence}), thanks to a more formal treatment of proof trees.
We have also consider\EZ{ed} further examples to show the effectiveness of coaxioms.  }

\FD{Finally, in \refToChapter{related} we summarize related work and in \refToChapter{conclu} we conclude the work, {also discussing}  further developments. }

\cleardoublepage \chapter{Topics in lattice theory} \label{chapter:lattice}

\emph{Lattice theory} \cite{Nation98} is a \EZ{well-}established mathematical framework widely used both in mathematics and in computer science, especially for studying semantics.
The subject of this theory is a particular class of \emph{partially ordered sets} with a rich algebraic structure.
Indeed lattices arise very frequently in various branches of both mathematics (algebra, topology, logic, \EZ{etc.}) and computer science (semantics \EZ{of programming languages}, formal verification, abstract interpretation, \EZ{etc.}).

In this framework semantics is usually \EZ{expressed} as a fixed point of a function \FD{that depends on} the language\footnote{This term is used here in a very broad sense, it does not \EZ{necessarily} indicate \FD{a language defined by a formal syntax}.} we are studying.
From this point of view, an important quality of lattice theory is that it \EZ{provides} a good expressiveness, \EZ{yet} keeping the theory quite simple: it allows to assign a semantics to a broad class of recursive definitions using elementary mathematical tools.
However usually, in order to reason about the semantics of programs, more powerful and complex frameworks are adopted, for instance, in the \EZ{order-theoretic} setting, \emph{domain theory} \cite{AbramskiJung94} is often used. 
It has been shown in \cite{Scott80} that objects studied by domain theory are able to capture the semantics of the \LambdaCalculus, thus of every computable \EZ{function. Hence}, in order to deal with complete languages, domain theory is the right choice.
However, for our aims, lattices are \EZ{powerful} enough, since we will deal with inference systems, that allow us to develop a simpler theory.

In this chapter we will summarize some standard and \EZ{well-}known notions and results about lattices, in order to be \EZ{self-contained}, providing all necessary concepts for understanding our work.
We will mainly discuss \emph{complete lattices} that are the structure we need in the rest of the thesis, and in particular we will focus on fixed point theorems, that is, on theorems that ensure the existence and/or give a characterization of fixed points of particular classes of functions.

The chapter is organized as follows.
In \refToSection{poset} we introduce the notion of partial order and monotone function that are the elementary bricks of lattice theory.
In \refToSection{lattice} we introduce lattices as an algebraic structure, we provide a canonical way to turn a lattice into  a partially ordered set and introduce the notion of complete lattice.
\refToSection{KT}  is devoted to the first and most important fixed point theorem: the Knaster-Tarski fixed point theorem \cite{Tarski55}.
In \refToSection{continuity} we discuss the notion of continuity from the \EZ{order-theoretic} point of view, and, relying on it, we give an alternative characterization of fixed points of continuous functions. 

\section{Partial orders and monotone functions} \label{sect:poset}

In this section we will introduce some basic concepts from order theory, fixing a uniform notation for the the rest of the thesis.
We will assume basic notions from set theory (set, relation, function, elementary set constructions etc.) and we will \EZ{generally} use an infix notation for binary relations and operations.

\begin{definition}\label{def:poset}
Given a set $\poset$, a binary relation $\order$ on $\poset$ is a \emph{partial order} if it satisfies the following properties\EZ{:}
\begin{description}
\item[Reflexivity] for all $x \in \poset$, $x\order x$ 
\item[Antisymmetry] for all $x, y \in \poset$, if $x\order y$ and $y \order x$ then $x=y$ 
\item[Transitivity] for all $x, y, z \in \poset$, if $x\order y$ and $y \order z$, then $x \order z$
\end{description}
A pair $\Pair{\poset}{\order}$ where $\poset$ is a set and $\order$ is a partial order on $\poset$ is called \emph{partially ordered set} or simply \emph{poset}\EZ{.}
\end{definition}

\EZ{The most well-known} example of partial order is standard \EZ{set} inclusion. 
Indeed for any set $X$ the pair $\Pair{\wp(X)}{\subseteq}$ is a poset: 
\begin{description}
\item[Reflexivity] for all $A \in \wp(X)$, $A\subseteq A$ trivially
\item[Antisymmetry] for all $A, B \in \wp(X)$, $A\subseteq B$ and $B\subseteq A$ implies $A=B$ by the extensionality axiom
\item[Transitivity] for all $A, B, C \in \wp(X)$, $A\subseteq B$ and $B\subseteq C$ implies $A\subseteq C$ since for each $x\in A$, $x \in B$ \EZ{by} $A\subseteq B$ and then $x \in C$ by $B\subseteq C$
\end{description}

Starting from this partial order we can build other orders, for instance considering algebraic structures on sets. 
We consider as an example the group structure, but analogous \EZ{definitions} can be \EZ{given} for every algebraic structure.

A \emph{group} is a triple  $\Triple{G}{\cdot}{1}$, where $G$ is a set with a binary operation $\cdot$ that is associative\EZ{, that is}
\[\forall x, y, z \in G.\ (x\cdot y) \cdot z = x \cdot (y \cdot z)\]
and  $1 \in $ is an identity for $\cdot$\EZ{, that is}
\[\forall x\in G.\ x\cdot 1 = 1 \cdot x = x\]
\EZ{In} addition every element $x \in G$ is required to have an inverse with respect to $\cdot$, that is
\[\forall x \in G.\ \exists x^{-1} \in G.\ x\cdot x^{-1} = x^{-1} \cdot x = 1\]
A subset $H \subseteq G$ is a \emph{subgroup} of $G$ if $\Triple{H}{\cdot}{1}$ is a group.
We denote \EZ{by} $\Sub{G}$ the set of all subgroups of $G$.
Clearly $\Sub{G} \subseteq \wp(G)$, thus $\Pair{\Sub{G}}{\subseteq}$ is a poset. \\
This last statement is actually a particular instance of a general property of posets, indeed if $\Pair{\poset}{\order}$ is a poset and $\aposet \subseteq \poset$, 
then the restriction $\order_\aposet$ of $\order$ to $\aposet$ is a partial order on $\aposet$.

Let us note that if $\Pair{\poset}{\order}$ is a partially ordered set, then we can consider the dual ordering $\dualOrder$, that makes $\Pair{\poset}{\dualOrder}$  a poset.
We will rely on this observation in the following, avoiding \EZ{to give} the dual of several definitions, \EZ{assuming} that these definitions are the same yet given in the dual ordering.

Let us now fix \EZ{some} terminology.

\begin{definition}\label{def:order-lang}
Let $\Pair{\poset}{\order}$ be a partially ordered set\EZ{. Then:}
\begin{enumerate}
\item A subset $A \subseteq \poset$ is a \emph{lower set} if for all $x \in A$ and $y \in \poset$,  $y\order x$ implies $y \in A$. The dual notion is named \emph{upper set}\EZ{.}
\item Given a subset $A \subseteq \poset$, the \emph{lower set generated by $A$} is the intersection of all lower sets $X\subseteq \poset$ such that $A\subseteq X$, and is denoted by $\LowSet{A}$. 
Dually the \emph{upper set generated by $A$} is denoted by $\UpSet{A}$.
If $A=\{x\}$ then we will write $\LowSet{x}$ and $\UpSet{x}$\EZ{.}
\item Given a subset $A\subseteq \poset$, an element $x \in \poset$ is an \emph{upper bound} of $A$ if, for all $y \in A$, $y \order x$\EZ{.
We} denote \EZ{by} $\ub{A}$ the set of upper bounds of $A$.
The dual notion is named \emph{lower bound} and $\lb{A}$ \EZ{denotes} the set of all lower bounds of $A$\EZ{.}
\item Given a subset $A \subseteq \poset$, $x \in A$ is a \emph{bottom (least) element} of $A$ if\EZ{,} for all $y \in A$, $x \order y$. 
The dual notion is named \emph{top (greatest) element}.
If $A = \poset$ usually bottom and top elements are denoted by $\bot$ and $\top$\EZ{, respectively.}
\item Given a subset $A\subseteq \poset$, the \emph{least upper bound} or \emph{join} of $A$ is the bottom element of \EZ{$\ub{A}$} and is denoted by $\lub A$\EZ{.}
The \emph{greatest lower bound} or \emph{meet} is defined dually and denoted by $\glb A$\EZ{.}
\end{enumerate}
\end{definition}

When $A = \{x_1, \ldots, x_n\}$\EZ{,} we will often write $\lb{x_1, \ldots, x_n}$ and $\ub{x_1, \ldots, x_n}$\EZ{,} omitting curly braces. 
Below we report some trivial \EZ{facts} that point out the behaviour of concepts introduced in \refToDefinition{order-lang} and \EZ{the} relationships \EZ{among} them.
\begin{itemize}
\item \EZ{Given} $A\subseteq \poset$ and $x \in \poset$\EZ{,} we have
\begin{align*}
\LowSet{x} &= \{y \in \poset \mid y \order x\} & \UpSet{x} &= \{y \in \poset \mid x \order y\}\\
\LowSet{A} &= \bigcup_{a \in A} \LowSet{a} & \UpSet{A} &= \bigcup_{a \in A} \UpSet{a} \\
\lb{A} &= \bigcap_{a \in A} \LowSet{a} & \ub{A} &= \bigcap_{a\in A} \UpSet{a}
\end{align*}
\item Given $A\subseteq \poset$\EZ{,} the least and the greatest elements of $A$, if they exist, are unique, and thus \EZ{the} least upper bound and greatest lower bound of $A$ are unique too\EZ{.}
\item Given $A\subseteq B \subseteq \poset$\EZ{, we have} $\lub A \order \lub B$ and $\glb B \order \glb A$
\item Given $A\subseteq \poset$, if $A = \bigcup A_i$ for $i \in I$,  then 
\[\glb A = \glb_{i \in I} \glb A_i \qquad \lub A = \lub_{i \in I} \lub A_i\]
\end{itemize}

Let us now introduce another fundamental  notion.
\EZ{Every time} we study a mathematical structure\EZ{,} at some point we have to talk about maps preserving such structure: \emph{homomorphisms}.
In our case\EZ{,} the structure is given by a partial order, thus \EZ{structure} preserving maps are those that preserve the order.

\begin{definition} \label{def:monotone-fun}
Let $\Pair{\poset}{\order_\poset}$ and $\Pair{\aposet}{\order_\aposet}$ be partially ordered sets. 
A function $\fun{\function}{\poset}{\aposet}$ is \emph{monotone} if\EZ{,} for all $x, y \in \poset$, $x \order_\poset y$ implies $\function(x) \order_\aposet \function(y)$\EZ{.}
\end{definition}

We will always omit subscripts when they are clear from the context.
We can find several examples of monotone functions\EZ{.}
\begin{enumerate}
\item Considering the poset $\Pair{\wp(A)}{\subseteq}$ for a given set $A$, the function $\fun{f}{\wp(A)}{\wp(A)}$ given by $f(X) = X \cup C$ with $C \in \wp(A)$
 is clearly monotone since if $X\subseteq Y$ then $X\cup C \subseteq Y \cup C$.
\item Considering the poset $\Pair{\N}{\le}$ of natural numbers with  the usual total ordering, the successor function ($n\mapsto n+1$), \EZ{the} multiplication by a constant $k\in \N$ ($n \mapsto kn$) and the factorial ($n\mapsto n!$) are monotone\EZ{.}
\item Considering two groups $G_1$ and $G_2$\EZ{,} a group homomorphism is a function $\fun{f}{G_1}{G_2}$ such that\EZ{,} for all $x, y \in G_1$, $f(x \cdot y) = f(x) \cdot f(y)$. 
With an abuse of notation we denote \EZ{by} $f(X)$ the image of $X$, that is, the set $\{f(x) \mid x \in X\}$ where $X \subseteq G_1$.
It is easy to see that if $H\in \Sub{G_1}$ then $f(H) \in \Sub{G_2}$, indeed $f(1) = 1$, so $1 \in f(H)$, and if $f(x), f(y) \in f(H)$ then $f(x) \cdot f(y) = f(x \cdot y) \in f(H)$ and finally if $f(x) \in f(H)$, $f(x)^{-1} = f(x^{-1}) \in f(H)$.
Therefore the map $\fun{f}{\Sub{G_1}}{\Sub{G_2}}$ is \EZ{well-}defined and is clearly monotone.
\end{enumerate}

As usual we define \emph{isomorphisms} between posets $\Pair{\poset}{\order}$ and $\Pair{\aposet}{\order}$ as monotone functions $\fun{\function}{\poset}{\aposet}$ which have an inverse, that is, a monotone function $\fun{\function^{-1}}{\aposet}{\poset}$ such that $\function \circ \function^{-1} = \Id{\aposet}$ and $\function^{-1} \circ \function = \Id{\poset}$. 
We can equivalently characterize isomorphisms as follows\EZ{.}

\begin{proposition}
A monotone function $\fun{\function}{\poset}{\aposet}$ is an isomorphism iff it is surjective and\EZ{,} for all $x, y \in \poset$, $x\order y$ iff $\function(x) \order \function(y)$\EZ{.}
\end{proposition}
\begin{proof}
$(\Rightarrow)$. Since $\function$ is an isomorphism it is bijective and so surjective. 
Since $\function$ is monotone\EZ{,} for all $x, y \in \poset$, if $x\order y$ then $\function(x) \order \function(y)$, so we have only to show the reverse implication.
But if $\function(x) \order \function(y)$, $\function^{-1}(\function(x)) \order \function^{-1}(\function(y))$ since $\function^{-1} $ is monotone; hence $x \order y$.\\
$(\Leftarrow)$. We have only to show that $\function$ is injective and its inverse is monotone.
If $\function(x) = \function(y)$, then $\function(x) \order \function(y)$ and $\function(y) \order \function(x)$, this implies that $x\order y$ and $y\order x$ and so by \EZ{antisymmetry} we get $x=y$, that shows $\function$ to be injective. 
Since $\function $ is also \EZ{surjective}, it has an inverse $\fun{\function^{-1}}{\aposet}{\poset}$.
To show that $\function^{-1}$ is monotone we consider $x, y \in \aposet$ such that $x\order y$. 
By \EZ{surjectivity} of $\function$ there are $x', y' \in \poset$ such that $x=\function(x')$ and $y=\function(y')$, hence by hypothesis we get $x' \order y'$, that shows the thesis.
\end{proof}

 \section{Semilattices, lattices and complete lattices} \label{sect:lattice}

In this section we will {introduce} richer structures involving partial orders.
In particular we will consider algebraic structures having a natural ordering with good properties.
These {structures} are \emph{semilattices}, \emph{lattices} and \emph{complete lattices}.

\begin{definition} \label{def:semilattice}
A pair $\Pair{\lattice}{\op}$, where $\lattice$ is a set and $\op$ is a binary operation on it, is a \emph{semilattice} if the following properties hold{:}
\begin{description}
\item[Idempotence] for all $x \in \lattice$, $x\op x = x$
\item[Commutativity] for all $x, y \in \lattice$, $x\op y = y\op x$
\item[Associativity] for all $x, y, z \in \lattice$, $(x\op y) \op z = x \op (y \op z)$
\end{description}
\end{definition}

In other words{,} a semilattice is an idempotent commutative semigroup.
A natural example of semilattice is the pair $\Pair{\wp(X)}{\cap}$ where the binary operation is set intersection.
Analogously the pair $\Pair{\wp(X)}{\cup}$ is also a semilattice with set union as binary operation. 

As mentioned {above,} we are interested in algebraic structures with a natural ordering, so let us introduce such ordering  for semilattices. Given a semilattice $\Pair{\lattice}{\op}${,} for each $x, y \in \lattice$ we define 
\[ x \order y \Longleftrightarrow x \op y = x\]
{Then} the following theorem holds{.}

\begin{theorem} \label{theo:semilattice-order}
Let $\Pair{\lattice}{\op}$ be a semilattice, then $\Pair{\lattice}{\order}$ is a partially ordered set where each pair of elements $x, y \in \lattice$ has a greatest lower bound{.}
\end{theorem}
\begin{proof}
We have to show that $\order$ verifies axioms of partial orders{.}
\begin{description}
\item[Reflexivity] for all $x \in \lattice$, $x\order x$ since $x\op x = x$ by idempotence
\item[Antisymmetry] for all $x, y \in \lattice$, if $x\order y$ and $y\order x$, we have $x\op y=x$ and $y\op x = y$, then by commutativity we get $x = x\op y = y \op x = y$ as needed
\item[Transitivity] for all $x, y, z \in \lattice$, if $x\order y $ and $y\order z $, we have $x\op y = x$ and $y \op z = y$, hence $x\op z = (x\op y)\op z = x\op (y\op z) = x\op y = x$ by associativity, thus $x\order z$ 
\end{description}
Let us now consider two elements $x, y \in \lattice$, we will show that $x\op y$ is the greatest lower bound of $\{x, y\}$. 
First we show that $x\op y \in \lb{x, y}$, i.e.{,} $x\op y \order x$ and $x\op y \order y$. 
Indeed $(x\op y)\op y = x\op (y\op y) = x\op y$ that shows $x\op y \order y$. The other inequality {can be shown analogously}.\\
Now we have to show that $x\op y$ is the greatest element of $\lb{x, y}$.
So let $z \in \lb{x, y}$, then $z\order x$ and $z\order y$, that is, $z\op x=z$ and $z\op y = z$.
Hence $z\op (x\op y) = (z\op x) \op y = z\op y = z$ as needed.
\end{proof}

Semilattices with this ordering are usually called \emph{meet semilattices} and the binary operation is denoted by $\meet$.

We can show also the converse of \refToTheorem{semilattice-order}, that is, we can build a semilattice starting from a partial order where each  pair of elements has a meet.

\begin{theorem} \label{theo:order-semilattice}
Let $\Pair{\lattice}{\order}$ be a partially ordered set where each pair of elements $x, y \in \lattice$ has a greatest lower bound, say $x\meet y$. 
{Then,} $\Pair{\lattice}{\meet}$ is a semilattice{.}
\end{theorem}
\begin{proof}
We have to show that the operation $\meet$ satisfies axioms of semilattice{s.}
\begin{description}
\item[Idempotence] for all $x \in \lattice$, $x\meet x = x$ since $\lb{x} = \LowSet{x}$ that has as largest element $x$ itself
\item[Commutativity] for all $x, y \in \lattice$, $x\meet y = y \meet x$ since for {the} lower bound the order of the elements {does not} matter
\item[Associativity] for all $x, y, z \in \lattice$, $(x\meet y) \meet z = x \meet(y\meet z)$ since 
\[\glb \{x, y\} \meet z = \glb \{x, y, z\} = x \meet \glb \{y, z\}\]
\end{description}
\end{proof}

Analogous {definitions} can be {given} considering the dual ordering 
\[x\order y \Longleftrightarrow x\op y = y\]
In this case $\Pair{\lattice}{\order}$ is a poset where each pair of elements has a least upper bound.
These semilattices are usually called \emph{join semilattices} and the binary operation is {called} $join$. 

Relying on \refToTheorem{order-semilattice}{,} from now on we will denote the greatest lower bound and the least upper bound of $\{x, y\}$ {by} respectively $x\meet y$ and $x\join y$. 

Combining a meet semilattice and a join semilattice in a proper way we get a \emph{lattice}.

\begin{definition} \label{def:lattice}
A triple $\Triple{\lattice}{\meet}{\join}$ is a \emph{lattice} if 
\begin{enumerate}
\item $\Pair{\lattice}{\meet}$ is a meet semilattice
\item $\Pair{\lattice}{\join}$ is a join semilattice
\item for all $x, y \in \lattice$, $x\meet (x\join y) = x$ and $x\join (x\meet y) = x$
\end{enumerate}
\end{definition}

The last conditions, called \emph{abstraction laws},  are needed in order to ensure that orderings induced by the join and the meet semilattice structures agree with each other.
More precisely{,} abstraction laws imply the following proposition{:}

\begin{proposition} \label{prop:ordering-agree}
If $\Triple{\lattice}{\meet}{\join}$ is a lattice, then $x\meet y = y$ iff $x\join y = y${.}
\end{proposition}
\begin{proof}
$(\Rightarrow)$ If $x\meet y = x$, then $x\join y = (x\meet y)\join y = y$ by the second abstraction law.
$(\Leftarrow)$ If $x\join y = y$, then $x\meet y = x \meet (x\join y) = x$ by the first abstraction law{.}
\end{proof}

Thus a lattice $\Triple{\lattice}{\meet}{\join}$ has a unique natural ordering.
We can state a theorem analogous to \refToTheorem{order-semilattice} that allows us to build a lattice from a {partially} ordered set where each pair of elements has both a meet and a join. 

The most natural example of lattice is given by the triple $\Triple{\wp(X)}{\cap}{\cup}$ of all subsets of a given set $X$ with set intersection and union. 

Let us now introduce the notion of completeness. 
We start defining a \emph{complete lattice}{.}

\begin{definition} \label{def:complete-lattice}
A \FD{non-empty} partially ordered set $\Pair{\lattice}{\order}$ is a \emph{complete lattice} if every \FD{non-empty} subset $A\subseteq \lattice$ has both least upper bound $\lub A$ and greatest lower bound $\glb A$.
\end{definition}

Clearly{,} a complete lattice is a lattice where for all $x, y \in \lattice$, $x\meet y = \glb\{x, y\}$ and $x\join y = \lub \{x, y\}$.
In addition{,} note that a complete lattice {has} both a top and a bottom element, namely, $\top = \lub \lattice$ and $\bot = \glb \lattice$.
\FD{Therefore we can define least upper bound and greatest lower bound also for the empty set as follows: $\lub \emptyset = \bot$ and $\glb \emptyset = \top$. }
Again an example of complete lattice is  the power set of a given set $X$ with set inclusion $\Pair{\wp(X)}{\subseteq}$.

Another important and immediate {property} of a complete lattice $\Pair{\lattice}{\order}$ is that{,} for each $x \in \lattice${,} both $\UpSet{x}$ and $\LowSet{x}$ are complete lattices with the same meet and join operations as $\lattice$. 

Actually requiring the existence of either the least upper bound or the greatest lower bound  of every subset is enough to turn a partially ordered set into a complete lattice, since {these notions} are closely related{.}

\begin{theorem} \label{theo:complete-lattice}
A partially ordered set $\Pair{\lattice}{\order}$ such that every subset $A\subseteq \lattice$ has a least upper bound $\lub A$ is a complete lattice{.}
\end{theorem}
\begin{proof}
\FD{Note that $\lattice$ is not empty, since $\bot = \lub \emptyset \in \lattice$. }
We define $\glb A = \lub \lb{A}$. 
Clearly by definition $\glb A$ is an upper bound of $\lb{A}$, so it is enough to show that $\glb A \in \lb{A}$ to prove that it is  the greatest lower bound of $A$.\\
By definition{,} for every set $X\subseteq \lattice $, $\lub X$ is the least element of $\ub{X}$, hence $\lub \lb{A}$ is the least element of $\ub{\lb{A}}$.
Therefore{,} for all $x \in \ub{\lb{A}}$, $\lub \lb{A} \order x$.
By definition of $\lb{A}$ every $x \in A$ is an upper bound of $\lb{A}$, that is, $A\subseteq \ub{\lb{A}}$.
Thus $\glb A = \lub \lb{A} \order x$ for all $x \in A$, namely, $\glb A \in \lb{A}$.
\end{proof}

Obviously a dual construction can be done if greatest lower bounds exist for every subset $A\subseteq \lattice$.
\FD{Note that \refToTheorem{complete-lattice} requires the existence of the least upper bound for every subset, namely, also for the empty set.
This is equivalent to require that the partially order\EZ{ed} set has a bottom element, and every non-empty subset has least upper bound. }

Consider for instance a group $G$, we already know that $\Pair{\Sub{G}}{\subseteq}$ is a partial order, we will show that it is a complete lattice.

First of all we show that{,} given a  family $(H_i)_{i \in I} \subseteq \Sub{G}$ of subgroups of $G$, the intersection $H = \bigcap_{i \in I} H_i$ is a subgroup of $G$. {Indeed:}
\begin{itemize}
\item $1 \in H$ since $1 \in H_i$ for all $i \in I$, because they are subgroups
\item if $x, y \in H$, then for all $i \in I$, $x, y \in H_i$, and since they are subgroups, $x\cdot y \in H_i$, hence $x\cdot y \in H$
\item if $x \in H$, then $x \in H_i$ for all $i \in I$, and since they are subgroups there is an inverse $x^{-1} \in H_i$, hence $x^{-1} \in H$
\end{itemize}
{This} shows that $H\in \Sub{G}$.
Clearly this operation computes the greatest lower bound of the family $(H_i)$, thus from \refToTheorem{complete-lattice} $\Pair{\Sub{G}}{\subseteq}$ is a complete lattice.

Finally we note that meet and join operations in $\Sub{G}$ are different from those in $\wp(G)$, indeed given a set $X \subseteq \Sub{G}$ we have
\[\glb X = \bigcap X \qquad  \lub X = \bigcap \{ H \in \Sub{G} \mid \bigcup X \subseteq H\}\]
{In} other words $\lub X$ is the \emph{subgroup generated by $\bigcup X$}.

We conclude this section presenting a result that shows how a monotone function \FD{acts on  meets and joins of subsets of a complete lattice}.
We fix the following notation: given a function $\fun{f}{A}{B}$, {and} $X \subseteq A$, we denote {by} $f(X)$ the set $\{f(x) \in B \mid x \in X\}$.

\begin{proposition} \label{prop:monotone-lub-glb}
Let $\Pair{\lattice}{\order}$ and $\Pair{\lattice'}{\order'}$  be complete lattices, $\fun{\function}{\lattice}{\lattice'}$ be a monotone function and $X \subseteq \lattice$.
The following facts hold{:}
\begin{enumerate}
\item $\lub \function(X) \order' \function(\lub X)$
\item $\function(\glb X) \order' \glb \function(X)$
\end{enumerate} 
\end{proposition}
\begin{proof}
We prove only 1, the proof for 2 is symmetric.
First note that, for all $x \in X$, we have $x \order \lub X$.
Hence by monotonicity we get that $\function(x) \order' \function(\lub X)$, so $\function(\lub X)$ is an upper bound of $\function(X)$.
Therefore by definition of least upper bound we get $\lub \function(X) \order' \function(\lub X)$.
\end{proof}

We present {the above} result for complete lattices in order not to care about the existence of meet and join.
Actually this result holds in any partially ordered set provided that the needed least upper bounds and greatest lower bounds exist.

\section{A fixed point theorem} \label{sect:KT}

In this section we will analyse properties of monotone functions defined over a complete lattice.
In particular we will state and prove the Knaster-Tarski theorem \cite{Tarski55}, that is the fundamental mathematical foundation of the whole thesis.

We start introducing the notions we deal with in this section\EZ{.}

\begin{definition}\label{def:fixed-point}
Let $\Pair{\poset}{\order}$ be a partially ordered set and $\fun{\function}{\poset}{\poset}$ a monotone function.
Consider an element $x \in \poset$, then 
\begin{itemize}
\item $x$ is a \emph{pre-fixed point} if $\function(x) \order x$
\item $x$ is a \emph{post-fixed point} if $x\order \function(x)$
\item $x$ is a \emph{fixed point} if $\function(x) = x$
\end{itemize}
\end{definition}

We will denote \EZ{by} $\Pre{\function}$, $\Post{\function}$ and $\Fix{\function}$ the subsets of $\poset$  \EZ{of respectively} pre-fixed, post-fixed and fixed points of $\function$. 

Note that, thanks to the antisymmetry of the order relation,   $\function(x) = x$ is equivalent to $\function(x)\order x$ and $x \order \function(x)$, 
thus a fixed point is a point that is both pre-fixed and post-fixed.
This implies that properties valid for either post-fixed or pre-fixed points also hold for fixed points.

The first observation we do concerns the restriction of a monotone function. 

\begin{proposition} \label{prop:monotone-restrict}
Let $\Pair{\poset}{\order}$ be a partially ordered set and $\fun{\function}{\poset}{\poset}$ a monotone function on $\poset$.
Consider $x\in \poset$\EZ{,} then 
\begin{enumerate}
\item if $x$ is pre-fixed, then  $\fun{\function}{\LowSet{x}}{\LowSet{x}}$ is \EZ{well-}defined and monotone
\item if $x$ is post-fixed, then  $\fun{\function}{\UpSet{x}}{\UpSet{x}}$ is \EZ{well-}defined and monotone
\end{enumerate}
\end{proposition}
\begin{proof}
We prove only point 1, the other is symmetric.
We have to show that $\function$ \EZ{maps} $\LowSet{x}$ into itself, that is, for each $z \in \LowSet{x}$, $\function(z) \in \LowSet{x}$. 
This is straightforward noting that $z \in \LowSet{x}$ means $z\order x$, thus $\function(z) \order \function(x)$ and, since $x$ is pre-fixed, $\function(x) \order x$.
Therefore by transitivity of $\order$ we get $\function(z) \order x$, that is, $\function(z) \in \LowSet{x}$.\\
The monotonicity follows immediately by hypothesis.
\end{proof}

We now study the structure of $\Pre{\function}$ and $\Post{\function}$ when $\function$ is defined over a complete lattice $\Pair{\lattice}{\order}$. 
Surely \EZ{they} are partially ordered with the same ordering as $\lattice$, indeed this fact holds also when $\lattice$ is simply a poset, but what about the lattice structure and the completeness?
The following proposition shows that if $\lattice$ is complete, then we have a least upper bound or greatest lower bound operation on these sets.

\begin{proposition} \label{prop:glb-lub-pre-post}
Let $\Pair{\lattice}{\order}$ be a complete lattice and $\fun{\function}{\lattice}{\lattice}$ a monotone function.
The following facts hold\EZ{:}
\begin{enumerate}
\item if $X\subseteq \Pre{\function}$ then $\glb X \in \Pre{\function}$, that is, $\glb X$ is a pre-fixed point
\item if $X\subseteq \Post{\function}$ then $\lub X \in \Post{\function}$, that is, $\lub X$ is a post-fixed point
\end{enumerate}
\end{proposition}
\begin{proof}
We only prove the first statement, the proof for the second is  symmetric.
Consider a subset $X\subseteq \Pre{\function}$, so every $x \in X$ is pre-fixed, that is, $\function(x)\order x$.
Since $\lattice$ is complete $\glb X$ exists, so we have only to prove that it is pre-fixed.\\
Let us denote \EZ{by} $\function(X)$ the set $\{ \function(x) \mid x \in X\}$. 
By definition of greatest lower bound (\refToDefinition{order-lang}) we have that\EZ{,} for each $x \in X$, $\glb X \order x$.
By the monotonicity of $\function$ we get that $\function(\glb X) \order \function(x)$ for each $x \in X$.
In other words $\function(\glb X)$ is a lower bound of $\function(X)$, so it is below the greatest lower bound of $\function(X)$, namely, $\function(\glb X) \order \glb \function(X)$.\\
Now, since $\glb \function(X)$ is a lower bound of $\function(X)$, we get that\EZ{,} for each $x \in X$, $\glb\function(X) \order \function(x)$. 
Thus, as $x$ is pre-fixed, by transitivity we get $\glb \function(X) \order x$ for each $x \in X$, namely, \EZ{$\glb \function(X)$} is a lower bound of $X$.
Therefore we get $\glb \function(X) \order \glb X$ and so, by transitivity we get $\function(\glb X) \order \glb X$ as needed. 
\end{proof}

\begin{corollary} \label{cor:complete-pre-post}
If $\Pair{\lattice}{\order}$ is a complete lattice and $\fun{\function}{\lattice}{\lattice}$ is monotone, then both $\Pair{\Pre{\function}}{\order}$ and $\Pair{\Post{\function}}{\order}$ are complete lattices\EZ{.}
\end{corollary}
\begin{proof}
It follows immediately by \refToProposition{glb-lub-pre-post} and \refToTheorem{complete-lattice}.
\end{proof}

Note that even if $\Pre{\function}$ and $\Post{\function}$ are complete lattices and they are contained in $\lattice$, they are not \EZ{``}sublattices'', since meet and join operations are not exactly as those of $\lattice$.
Consider for instance $\Pre{\function}$, here the meet operation is the same, and the join is defined as usual as $\lub_{\Pre{\function}} X = \glb \lb{X}_{\Pre{\function}}$ for $X \subseteq \Pre{\function}$.
But note that $\lb{X}_{\Pre{\function}} \subseteq \lb{X}$ and $\lb{x}_{\Pre{\function}} \ne \lb{X}$ in general, thus we have that $\lub X \order \lub_{\Pre{\function}} X$ but they are not necessarily equal.  
Let us clarify this \EZ{issue} with an example.

Consider a group $G$  and the poset $\Pair{\wp(G)}{\subseteq}$.
Given a subset $X\subseteq G$ we denote \EZ{by} $\GroupGen{X}$ the \EZ{subgroup} of $G$ generated by $X$, that is, the intersection of all subgroups of $G$ containing $X$. 
Clearly the function $\fun{\GroupGen{-}}{\wp(G)}{\wp(G)}$ is monotone.
It can be \EZ{easily} seen that $\Sub{G} = \Pre{\GroupGen{-}}$, namely, pre-fixed points are subgroups of $G$. 

Let us now set $G = \Z$ with the group structure given by the sum operation  with identity $0 \in \Z$, and consider subgroups $\GroupGen{3}$ and $\GroupGen{5}$. 
In $\Sub{\Z}$ we have that $\GroupGen{3} \join \GroupGen{5} = \GroupGen{3, 5} = \Z$.
However $\Z \ne \GroupGen{3} \cup \GroupGen{5}$ since \EZ{for} instance $2 \notin \GroupGen{3} \cup \GroupGen{5}$.
Therefore the join operation \EZ{in} $\Pre{\GroupGen{-}}$ is different from the join operation in $\wp(\Z)$. 

We conclude the section stating and proving the Knaster-Tarski theorem \cite{Tarski55}\EZ{,} that is the fundamental   result on which the whole thesis relies.

\begin{theorem} [Knaster-Tarski] \label{theo:KT}
Let $\Pair{\lattice}{\order}$ be a complete lattice and $\fun{\function}{\lattice}{\lattice}$ a monotone function.
Then $\function$ admits both a least and a greatest fixed point, $\lfp \function$ and $\gfp \function$ respectively, with 
\[\lfp\function = \glb \Pre{\function} \qquad   \gfp \function = \lub \Post{\function}\]
\end{theorem}
\begin{proof}
We will prove the statement for the least fixed point, for the greatest one the proof is symmetric. 
Set $z = \glb \Pre{\function}$.
From \refToProposition{glb-lub-pre-post} we know that $z$ is pre-fixed, so we have only to show that it is post-fixed to conclude the proof. \\
Since $z$ is pre-fixed we know that $\function(z)\order z$, and since $\function$ is monotone we get $\function(\function(z)) \order \function(z)$, that is, $\function(z)$ is pre-fixed, so $\function(z) \in \Pre{\function}$.
By definition of greatest lower bound (\refToDefinition{order-lang}) we get that $z \order \function(z)$, namely, $z$ is post-fixed. \\
To show that $z$ is the least fixed point, it is enough to note that an arbitrary fixed point $x$ is by definition a pre-fixed point, so $x \in \Pre{\function}$, and since $z$ is the greatest lower bound of $\Pre{\function}$ we get $z \order x$.
\end{proof}

\EZ{The above} theorem is of fundamental importance since it not only ensures the existence of the least and the greatest fixed point of every monotone function, but it \EZ{provides} an explicit characterization of them.
Moreover from this characterization immediately follow two techniques for proving inequalities involving $\lfp \function$ and $\gfp \function$.
Indeed given $x \in \lattice $ the following principles hold\EZ{:}
\begin{description}
\item[Induction]  if $x$ if pre-fixed, that is, $\function(x) \order x$, then $\lfp \function \order x$
\item[Coinduction] if $x$ is post-fixed, that is, $x\order \function(x)$, then $x \order \gfp \function$
\end{description} 
These two principles are widely used in practice, since they \EZ{make} much easier proving inequalities involving least and greatest fixed points: we can completely forget of these fixed \EZ{points}, because inequalities only depend on the point $x$ and its \EZ{properties} with respect to the monotone function $\function$.

\section{Chains and continuity} \label{sect:continuity}

In this section we will introduce the notion of continuity for functions over partially ordered sets and we will prove a theorem giving an alternative characterization of the least and the greatest fixed points when the function is continuous. 

In order to discuss continuity we have to introduce the notion of \emph{chain} in a partially ordered set.

\begin{definition} \label{def:chain}
Let $\Pair{\poset}{\order}$ be a partially ordered set.
A subset $C \subseteq \poset$ is a \emph{chain} if it is totally ordered, that is, for each $x, y \in C$, either $x\order y$ or $y\order x$.
An $\omega$-chain is a countable chain.
\end{definition}

Since an $\omega$-chain is countable, it can be indexed over natural numbers, so we will often use the notation $(x_i)$  for $\omega$-chains, assuming $i$ to range over $\N$.
We will also omit the prefix $\omega$  when it is clear \EZ{from} the context.

A chain $(x_i)$ is called \emph{ascending} if for all $i \in \N$, $x_i \order x_{i+1}$, and it is called \emph{descending} if for all $i\in \N$, $x_{i+1} \order x_i$.
Clearly an $\omega$-chain can be  either ascending or descending depending on the choice of indexes.

\begin{definition} \label{def:continuous}
Let $\Pair{\lattice}{\order}$ be a complete lattice. 
A \EZ{function} $\fun{\function}{\lattice}{\lattice}$ is called \emph{upward continuous}  if for every chain $C = (x_i)$, $\function(\lub C) = \lub \function(C)$, 
\emph{downward continuous} if for every chain $C=(x_i)$, $\function(\glb C) = \glb \function(C)$.
\end{definition}

Conditions for continuity will be \EZ{also} expressed as  follows 
\[
\function \left( \lub_{i\in \N} x_i \right) = \lub_{i \in \N} \function(x_i) 
\qquad
\function \left( \glb_{i \in \N} x_i \right) = \glb_{i \in \N} \function(x_i)
\]

We have given the definition of continuous function for functions defined on a complete lattice, actually the definition works in a more general settings:
it suffices to have a function $\fun{\function}{\poset}{\aposet}$ where $\poset$ and $\aposet$ are both partially ordered  and have least upper bounds (respectively greatest lower bounds) for ascending (respectively. descending) chains.

\begin{proposition} \label{prop:continuous-monotone}
Let $\Pair{\lattice}{\order}$ be a complete lattice and $\fun{\function}{\lattice}{\lattice}$  a continuous function, then $\function$ is monotone\EZ{.}
\end{proposition} 
\begin{proof}
Assume that $\function$ is upward continuous, the proof for the downward case is symmetric.
Consider $x, y \in \lattice$ with $x\order y$. The set $C = \{x, y\}$ is clearly a chain, with $\lub C  = x \join y = y$. 
So we have that 
\[\function(y) = \function(x \join y) = \function(x) \join \function(y)\]
\EZ{and,} since by definition of join $\function(x) \order \function(x) \join \function (y)$\EZ{,} we get $\function(x) \order \function(y)$ as needed.
\end{proof}

We conclude the section presenting \EZ{and discussing} another fixed point theorem: the Kleene theorem.
Actually this result has several names and \EZ{variations}, a summary can be found in \cite{LassezKS82}, we will discuss it in the context of complete lattices, since that is the framework in which we work.

First of all let us introduce a useful notation.
Given a complete lattice $\Pair{\lattice}{\order}$\EZ{,} a monotone function $\fun{\function}{\lattice}{\lattice}$\EZ{,} and an element $x \in \lattice$, we denote \EZ{by} $\Iterate{\function}{x}$ the set $\{ \function^n(x) \mid n \in \N\}$ of iterative \EZ{applications} of $\function$ on $x$. 
The following lemma holds\EZ{.}

\begin{lemma} \label{lem:iterate-chain}
Let $\Pair{\lattice}{\order}$ be a complete lattice and $\fun{\function}{\lattice}{\lattice}$ a monotone function. 
If $x \in \lattice$ is either pre-fixed or post-fixed, then $\Iterate{\function}{x}$ is a chain\EZ{.}
\end{lemma}
\begin{proof}
Assume $x$ to be pre-fixed, the other case is symmetric.
We show by induction on $n$ that\EZ{, for all $n \in \N$,} $\function^n(x) \order \function^{n+1}(x)$\EZ{. This} implies that $\Iterate{\function}{x}$ is an ascending chain. \\
If $n=0$, we have $\function^0(x) = x \order \function(x)$ since $x$ is pre-fixed.\\
Assume the thesis for $n$, thus $\function^n(x) \order \function^{n+1}(x)$; since $\function$ is monotone, we have that $\function(\function^n(x)) \order \function(\function^{n+1}(x))$ from \EZ{which} we get $\function^{n+1}(x) \order \function^{n+2}(x)$ as needed.
\end{proof}

In particular the lemma shows that $\Iterate{\function}{\bot}$ and $\Iterate{\function}{\top}$ are both chains, since $\bot$ is post-fixed and $\top$ is pre-fixed.
Also the following lemma holds.

\begin{lemma} \label{lem:iterate-fp}
Let $\Pair{\lattice}{\order}$ be a complete lattice and $\fun{\function}{\lattice}{\lattice}$ a monotone function. Then the following conditions hold\EZ{:}
\begin{enumerate}
\item if $\lub \Iterate{\function}{\bot}$ is pre-fixed, then it is the least fixed point of $\function$
\item if $\glb \Iterate{\function}{\top}$ is post-fixed, then it is the greatest fixed point of $\function$
\end{enumerate}
\end{lemma}
\begin{proof}
We prove 1, the proof for 2 is symmetric.
To show that $\lub \Iterate{\function}{\bot}$ is the least fixed point, we first prove that every pre-fixed point is an upper bound of $\Iterate{\function}{\bot}$.
Hence, given $x \in \Pre{\function}$, we show by induction on $n$ that\EZ{, for all $n \in \N$,} $\function^n(\bot) \order x$.
If $n=0$, we have $\function^0(\bot) = \bot \order x$ by definition of bottom element.
Assume the thesis for $n$, so $\function^n(\bot) \order x$.
Then by the monotonicity of $\function$ we get $\function^{n+1}(\bot) \order \function (x)$, but since $x$ is pre-fixed, $\function(x) \order x$, so by transitivity we get $\function^{n+1}(\bot) \order x$ as needed.\\
In other words we have proved that $\Pre{\function} \subseteq \ub{\Iterate{\function}{\bot}}$, that implies $\lub \Iterate{\function}{\bot} = \glb \ub{\Iterate{\function}{\bot}} \order \glb \Pre{\function}$.
However by hypothesis $\lub \Iterate{\function}{\bot} \in \Pre{\function}$, hence we get also  the opposite inequality, that implies $\lub \Iterate{\function}{\bot} = \glb \Pre{\function}$.
Recall that by \refToTheorem{KT} we have $\lfp \function = \glb \Pre{\function}$, hence $\lub \Iterate{\function}{\bot} = \lfp \function$ as needed.
\end{proof}

We can now prove the Kleene theorem.

\begin{theorem}[Kleene] \label{theo:Kleene}
Let $\Pair{\lattice}{\order}$ be a complete lattice and $\fun{\function}{\lattice}{\lattice}$ a function. \EZ{Then the following conditions hold:}
\begin{enumerate}
\item if $\function$ is upward continuous, then $\lub \Iterate{\function}{\bot}$ is the least fixed point of $\function$
\item if $\function$ is downward continuous, then $\glb \Iterate{\function}{\top}$ is the greatest fixed point of $\function$
\end{enumerate}
\end{theorem}
\begin{proof}
We prove 1, the proof for 2 is symmetric.
Since $\function$ is upward continuous and $\Iterate{\function}{\bot}$ is a chain by \refToLemma{iterate-chain},  we get that 
\[\function \left( \lub \Iterate{\function}{\bot} \right )
	= \lub \function (\Iterate{\function}{\bot}) 
	= \lub_{n \in \N} \function^{n+1} (\bot)
\]
Since $\bot \order \function^n(\bot)$ for all $n \in \N$, $\bot \join \lub \function(\Iterate{\function}{\bot}) = \lub \function(\Iterate{\function}{\bot})$;
moreover, since $\bot = \function^0(\bot)$, $\bot \join \lub \function^{n+1}(\bot) = \lub \function^n(\bot) = \lub \Iterate{\function}{\bot}$ for $n \in \N$. 
Therefore we get $\function(\lub \Iterate{\function}{\bot} ) = \lub \Iterate{\function}{\bot}$, namely, it is a fixed point.\\
In particular $\lub \Iterate{\function}{\bot}$ is pre-fixed, hence by \refToLemma{iterate-fp} we get that it is the least fixed point.
\end{proof}

An important consequence of this theorem is that it \EZ{provides} another way to prove the same inequalities that we can prove with induction and coinduction principles presented in \refToSection{KT}.
Indeed, assuming that $\function$ is continuous in the appropriate sense,  we have the following proof principles\EZ{:}
\begin{itemize}
\item for all $x \in \lattice$, if\EZ{,} for all $n \in \N$, $\function^n(\bot) \order x$ implies $\function^{n+1}(\bot) \order x$, then $\lfp \function \order x$
\item for all $x \in \lattice$, if\EZ{,}s for all $n \in \N$, $x\order \function^n(\bot)$ implies $x \order \function^{n+1}(\bot)$, then $x\order \gfp \function $
\end{itemize}
These principles only require a proof by arithmetic induction and do not impose \EZ{any constraint} on the element $x$, while induction and coinduction principles require it to be pre-fixed or post-fixed.
The correctness of these principles is trivial:  they say \EZ{in other words} that if $x$ is an upper bound (respectively. a lower bound) of the chain $\Iterate{\function}{\bot}$ (respectively. $\Iterate{\function}{\top}$) its least upper bound $\lub \Iterate{\function}{\bot}$ (respectively. its greatest lower bound $\glb \Iterate{\function}{\top}$) is below (respectively. above) $x$. 

This looks like a great simplification with respect to induction and coinduction, however  we have an additional hypothesis: $\function$ must be continuous and this is not easy to prove in practice.

\cleardoublepage \chapter{A framework for recursive definitions} \label{chapter:is}

In the introduction we have discussed inductive and coinductive definitions in an informal way.
We have represented them by means of rules, that relate premises to a consequence in a recursive way, that is, the {shape of the judgement} in the consequence {is the same of those in} premises.
We have used rules since they {provide} a very intuitive understanding of (recursive) definitions, even if we {have not} described {yet} their precise semantics.

In this chapter we discuss in a rigorous way {rule-based} definitions, following the framework introduced in \cite{Aczel77} and become very popular: the theory of \emph{inference systems}.
This framework is widely used in several fields: in programming languages, operational semantics and type systems are usually expressed through an inference system, in mathematical logic the logical consequence relation is also usually defined using rules and  in type theory  rules to define and manipulate types are again often expressed in this way.

The notions and results we will introduce in this chapter are all standard  and are aimed to assign a formal meaning to concepts, such as definition, rule, inductive and coinductive interpretation, and several others{,}  that form the framework we will use for the rest of this work.

\FD{Although, \EZ{as} we said, this chapter \EZ{presents well-known notions and results}, some parts of the presentation are, at the best of our knowledge, original.
In particular, since we did not find in literature a rigorous enough treatment of the proof-theoretic semantics of inference systems, we have developed it autonomously, starting from a precise notion   of tree and graph presented in \refToSection{trees-graphs}.
There, we also prove \refToTheorem{graph-tree}, that allows us to give a new  proof of the equivalence between proof-theoretic and fixed point semantics in the coinductive case (see \refToSection{sem-eq}). }

The rest of this chapter is organized as follows.
In \refToSection{is} {we} introduce the notion of inference system{,} describing {its} syntax and semantics. 
The semantics is provided in a {proof-theoretic} fashion, by means of proof trees, that are just trees representing a proof for the validity of a judgement.
To this aim we will briefly discuss some concepts and results about trees and graphs.
\refToSection{fp-sem} describes a {model-theoretic} semantics of inference systems as fixed {points} of a monotone function, on a {particular} complete lattice, induced by each inference system.
Exploiting this semantics we will also introduce induction and coinduction principles to reason about {inference systems}.
Finally{,} we will prove the equivalence between {proof-theoretic} and {model-theoretic} semantics.
In \refToSection{iterate-is} we will discuss sufficient and necessary conditions on inference systems that allows {one} to compute their semantics in an iterative way.

\section{Inference systems: syntax and semantics} \label{sect:is}

Let us assume a \emph{universe}  $\universe$  whose elements are called \emph{judgements}, ranged over by $\judg$.

\begin{definition} \label{def:is}
An \emph{inference rule}, or simply {\emph{rule}},  is a pair $\myrule$ with $\prem \subseteq \universe$ and $\cons \in \universe$.
A rule $\myrule$ is an \emph{axiom} if $\prem = \emptyset$. \\
An \emph{inference system} $\is$ is a set of rules, that is, $\is \subseteq \wp(\universe) \times \universe$.
\end{definition}

Let us {show} some examples to illustrate this concept.
We denote {by} $\Z$ the set of integers and {by} $\LSet$ the set of finite lists of integers, that can be constructed as in the introduction: $\EList$ is the empty list and every list has shape $\Cons{x_1}{ \Cons{\ldots}{ \Cons{x_n}{\EList} } }$ with $x_1, \ldots, x_n \in \Z$. 
We consider the definition of the predicate $\memberPred{x}{l}${, mentioned in the introduction,} that holds if the element $x$ {occurs} in $l$. 
In this case the universe can be the set $\{\memberPred{x}{l} \mid x \in \Z, l \in \LSet\}$, so for instance judgements like $\memberPred{1}{\EList}$, $\memberPred{3}{\Cons{1}{\Cons{3}{\EList}}}$ or $\memberPred{1}{\Cons{1}{\Cons{3}{\Cons{2}{\EList}}}}$ are in the universe. 

Rules that define the predicate $\memberPred{x}{l}$ are {necessarily} infinitely many, therefore it is not possible to write down all rules  in an extensional way; we have to represent them in some intensional way.
To this aim {it} is a standard practice {to} use \emph{\EZ{meta-rule}s} or \emph{rule schemes}, that is, we {show} all possible shapes that a rule can assume rather {than} all possible rules. 
So we need to use {(meta-)}variables to range over  base elements, in this case integers and finite lists of integers.
Therefore the definition of the predicate $\memberPred{x}{l}$ through an inference system looks like the following{:}
\[
\Rule{}{ \memberPred{x}{\Cons{x}{l}} } 
\BigSpace
\Rule{ \memberPred{x}{l} }{ \memberPred{x}{\Cons{y}{l}} } 
\]
where $x, y \in \Z$ and $l \in {\LSet}$.
Actual rules can be obtained from these schemes {by} instantiating variables with actual elements.

Another example, again mentioned in the introduction, is the predicate $\allPosPred{l}$, that holds if all elements in $l$   are strictly positive integers. 
The universe in this case can be $\{\allPosPred{l} \mid l \in \LSet\}$ and the definition as inference system is the following{:}
\[
\Rule{}{ \allPosPred{\EList} }
\BigSpace
\Rule{ \allPosPred{l} }{ \allPosPred{\Cons{x}{l}} } x>0
\]
{This example shows} another {important} feature of \EZ{meta-rule}s: \emph{side conditions}.
Aside the second rule we have specified a condition ($x>0$), called side condition, that $x$ must satisfy. 
{That is,} side conditions restrict the set of values on which variables range over, reducing the number of rules that can be obtained as instances of a \EZ{meta-rule}.
They are extremely useful in order to provide a finer control on instances of rule schemes, and without them many definition are very difficult to express as inference systems. 
For instance the definition of $\allPosPred{l}$ without  side {conditions} reported below requires an additional predicate $\posPred{x}$, that holds if $x$ is positive.
\[
\Rule{}{ \posPred{1} }
\BigSpace
\Rule{ \posPred{x} }{ \posPred{x+1} }
\BigSpace
\Rule{}{ \allPosPred{\EList} }
\BigSpace
\Rule{ \posPred{x} \Space \allPosPred{l} }{ \allPosPred{\Cons{x}{l}} }
\]
However for this definition we need to change the universe in order to include judgements of shape $\posPred{x}$, hence the universe is ${\{\allPosPred{l} \mid l \in \LSet\}}\cup \{\posPred{x} \mid x \in \Z\}$.

We report another example of a judgement regarding lists.
We define the judgement $\maxElem{l}{x}$ that holds if $x$ is the maximum element {occurring} in $l$. 
Note that if $l$ is not empty this element surely exists (every finite set of integers has a maximum) and is unique, thus this judgement actually represents a (total) function from $\LSet \setminus \{\EList\}$ to $\Z$. 
The inference system is the following{:}
\[
\Rule{}{ \maxElem{\Cons{x}{\EList}}{x} }
\BigSpace
\Rule{ \maxElem{l}{y} }{ \maxElem{\Cons{x}{l}}{z} } z = \max\{x, y\}
\]
Again the side condition is crucial to obtain a correct definition, but it can be removed changing a bit the inference system. \\

Until now {we have focused} on the syntax of inference {systems}, explaining how {definitions} given through them should be read.
{{That is,} we have relied on reader's intuition to convince {her/him} that a given inference system actually defines an intended predicate or function. }
In order to formally prove the correctness of such definitions we need to define in a rigorous way  how an inference system can be interpreted, that is, its semantics. 

We first address this {issue} in a {\emph{proof-theoretic}} setting, that allows us to define a very intuitive semantics  of inference systems.
This semantics is based on the notion of \emph{proof tree} or \emph{derivation}, hence to discuss it, we need to say something about trees.

\subsection{A digression on graphs and trees} \label{sect:trees-graphs}
In this section we report some results about trees and graphs.
Although {such results} presented here are well-known, we have not found them presented in this way. 
In \cite{AdamekEtAl15} some similar results can be found, that have inspired this presentation, but  with a substantial difference in some definitions, for instance the definition of tree:
\FD{they consider trees as special graphs and do not consider labels.
Furthermore, {the} conditions they impose on graphs in order to be trees are very restrictive, {hence} many trees that for us are different for them {are the same}.
This restriction is due to the fact that the result they aim to prove is stronger than our \refToTheorem{graph-tree}: they want trees to form a final coalgebra for a suitable power-set functor. }

Along this section we denote by $\String{A}$ the set of finite \emph{strings}  on the \emph{alphabet} $A$, which is an arbitrary set of \emph{symbols}.
We use Greek letters $\alpha, \beta, \ldots$ to range over strings and Roman letters $a, b, \ldots$  to range over symbols in $A$ and we implicitly identify strings of length one and symbols.
Moreover we denote by juxtaposition \emph{string concatenation}, and by $\Len{\alpha}$ the \emph{length} of the string $\alpha$.
Finally $\EString$ is the empty string.
\FD{We also extend string concatenation to sets of {strings}, denoting, for $X, Y \subseteq \String{A}$, by $XY$ the set $\{\alpha \beta \in \String{A} \mid \alpha \in X, \beta \in Y\}$; moreover if either $X$ or $Y$ are singleton{s} we will omit curly braces, namely $\alpha Y = \{\alpha\}Y$. }

On the set $\String{A}$ we can define the \emph{prefixing relation} $\prec$ as follows: for any $\alpha, \beta \in \String{A}$, $\alpha\prec\beta$ if and only if there exists $\gamma \in \String{A}$ such that $\alpha\gamma = \beta$.
It can be shown that $\prec$ is a partial order and thus, for any $X\subseteq \String{A}${,} the restriction of $\prec$ to $X$ is well-defined and still a partial order.
We say that a subset $X\subseteq \String{A}$ is \emph{well-founded} with respect to prefixing if any chain $C \subseteq X$ is finite.

A non-empty subset $L \subseteq \String{A}$ is a \emph{tree language} if it satisfies the \emph{prefix property}, that is, if $\alpha a \in L$ then $\alpha \in L$.
In particular $\EString \in L$ for any tree language $L\subseteq \String{A}$. 
Now we are able to define trees following \cite{Courcelle83}.

\begin{definition} \label{def:tree}
Let $A$ be an alphabet, $L \subseteq \String{A}$ a tree language and $\Labels$ a set.
A \emph{tree} labelled in $\Labels$ is a function $\fun{t}{L}{\Labels}$.
The element $t(\EString)$ is called the \emph{root} of $t$.
\end{definition}

\FD{{T}he notion of tree in \refToDefinition{tree} is slightly different from that introduced by \citet{Courcelle83}. {Indeed, there $\N$ is taken as fixed alphabet and, moreover, a canonical choice is imposed on strings in a tree language, that is,} if $\alpha n\in L$ and $m \le n$, then $\alpha m \in L$.
Since, as we will see, the branching of the tree is bounded by the cardinality of the alphabet, with this choice \Citet{Courcelle83}  only considers trees with countable branching.
Because we have to use trees in the  context of inference systems, this restrictions is too strong for us: it would compel us to manage boring conditions on the cardinality of the universe $\universe$ or of rule premises, hence we simply remove it.
In this way, we also loose the canonical choice, since the alphabet $A$ may be not ordered, but this is not a real issue. }  

If $\fun{t}{L}{\Labels}$ is a tree, {then,} for any $\alpha \in L$, the \emph{subtree} rooted at $\alpha$ is the function $\fun{t_\alpha}{L_\alpha}{\Labels}$, where $L_\alpha=\{\beta \in \String{A} \mid \alpha\beta \in L\}$ and $t_\alpha(\beta) = t(\alpha\beta)$.
This notion is well-defined since $L_\alpha$ is a tree language, hence $t_\alpha$ is a tree.
Note that $t$ is itself a subtree, rooted at $\EString$.
Subtrees rooted at $\alpha$ with $\Len{\alpha}=1$ are called \emph{direct subtrees} of $t$.
Finally a tree $t$ is \emph{well-founded} if $\dom(t)$ is well-founded with respect to $\prec$.

{The notion of tree introduced in \refToDefinition{tree} is mathematically precise, but  not very intuitive.
A usual, and perhaps more intuitive,} way to {introduce} trees {is} as particular graphs. 
\FD{Intuitively, using a graph-like terminology, that we will make precise {below}, we can see the elements in the tree language $\dom(t)$ as nodes.
Actually, thanks to the {prefix} property, a node $\alpha \in \dom(t)$  represents also all nodes (its prefixes) we have to traverse to reach $\alpha$ starting from the root $\EString$.
For instance, if $\alpha = abc$, we know that $\EString, a, ab, abc \in dom(t)$, hence they are nodes of $t$ and they form the path from the root to $\alpha$. 
Therefore, requiring $t$ to be well-founded is equivalent to require that any sequence of prefixes is finite, hence it is equivalent to require that all paths in $t$ are finite. }

{To formally show that} indeed trees can be {seen} as particular graphs, {we} start {by} giving a definition of graph{.}

\begin{definition} \label{def:graph}
A \emph{graph} is a pair $\Pair{\Nodes}{\adj}$ where $\Nodes$ is the set of \emph{nodes} and $\fun{\adj}{\Nodes}{\wp(\Nodes)}$ is the \emph{adjacency function}.\\
A \emph{labelled {graph}}, with labels in a set $\Labels$, is a triple $\Triple{\Nodes}{\adj}{\ell}$, where $\Pair{\Nodes}{\adj}$ is a graph and $\fun{\ell}{\Nodes}{\Labels}$ is the \emph{labelling function}. 
\end{definition}

Taking a more abstract perspective a graph is a coalgebra for the power set functor carried by the set of nodes and a labelled graph is a coalgebra for the functor $\GraphFun{\Labels}$ defined by $\Nodes \mapsto \Labels \times \wp(\Nodes)$, again carried by the set of nodes.
Therefore the notion of {graph} homomorphism is simply definable as coalgebra homomorphism for these functors.

With this definition it is easy to assign a graph {structure} to a tree.
Let $\fun{t}{L}{\Labels}$ be a tree, we can represent it as a labelled graph with {set of nodes $L$, adjacency function $\chl(\alpha) = \{\beta \in L \mid \beta = \alpha a\}$ returning the \emph{children} of a node $\alpha$, and} labelling function given by $t$ itself.
Thanks to this graph structure we justify terminology {like} node and {adjacent} for trees: a node is a string $\alpha \in \dom(t)$ and given a node $\alpha$, {the set of its adjacents is $\chl(\alpha)$}.

We now analyse the role of the alphabet $A$ in the definition of tree (\refToDefinition{tree}).
First note that its elements are essentially not relevant, because the important thing are labels.
What actually {matters} {is the cardinality of $A$}, that determines the maximum \emph{branching} of the tree, that is, the {maximum number of children (hence subtrees) for each node $\alpha$.
In other words we have $|\chl(\alpha)| \le |A|$ for all $\alpha \in L$.
For instance, {we can build essentially the same trees if $A$ is either $\{1, 2, 3\}$ or $\{a, b, c\}$.}
However{,} the fact that they have both cardinality {3 is relevant, since trees built on $A$ have for each node at most 3 children.} }
\EZComm{forse il pezzetto successivo si potrebbe omettere}
\FD{More precisely if $|A|\le |B|$, we know that there exists an injection from $A$ to $B$, hence we can identify elements of $A$ with elements of $B$ through this injection.
In this way, we can consider trees built on $A$ as trees built on $B$, and so, making implicit these identifications, we can always assume that all trees are built on the same alphabet. }
For these reasons we will often abstract away $A$ and simply specify a cardinal number in order to {make} explicit the branching of the tree.

\FD{It is convenient, especially to discuss proof trees, as we will do later, to introduce a special choice for the alphabet $A$: we identify $A$ with the set of labels $\Labels$.
This choice is suitable for modelling proof trees, since they are labelled on judgements, {notably} nodes are (labelled by) consequences of rules and their children correspond to  set{s} of premises. 
In this way{,} each node can be identified by the path (string) of judgements/labels from the root to it.
This choice can be applied to all situations where children of each node have distinct labels, hence it is a canonical choice in these cases, as formalized below. }

{We say that a tree $\fun{t}{L}{\Labels}$ is \emph{children injective} if for all $\alpha \in \dom(t)$, the restriction of $t$ to the set $\chl(\alpha)$ is injective;
more explicitly, for all $\alpha \in \dom(t)$, if $\alpha a, \alpha b \in \dom(t)$ and $t(\alpha a) = t(\alpha b)$, then $a = b$.
In other words this means that all children of any node must have different labels. 
Note that all subtrees of a children injective tree are themselves children injective. }
\EZComm{di fatto \`e iniettiva la funzione da stringhe/nodi a stringhe di etichette, che \`e quello che serve, quella sui figli \`e la condizione che lo garantisce} \FDComm{Esatto}
Therefore, if $\fun{t}{L}{\Labels}$ is children injective, we can replace $A$ with $\Labels$, using the following function
\[
\fun{f}{L}{\String{\Labels}} \Space \left\{\begin{array}{ll}
f(\EString) &= \EString\\
f(\alpha a) &= f(\alpha) t(\alpha a)
\end{array} \right.
\]
\FD{Intuitively, the function $f$ maps each node $\alpha \in L$ to the string of labels encountered in the path from the root to $\alpha$. }
{It} is easy to see that {$f$ is injective} and  $f(L)$ is a tree language.
Therefore we can define the tree $\fun{t'}{f(L)}{\Labels}$ as follows
\[
\left\{\begin{array}{ll}
t'(\EString) &= t(\EString)\\
t'(\alpha a) &= a
\end{array} \right.
\]
and it is easy to check that for any $\alpha \in L$, $t'(f(\alpha)) = t(\alpha)$, hence they are essentially the same tree.

We can simplify the representation even more: note that the definition of $t'$ depends only on its domain $f(L)$ and  on the choice for the root label, hence we can forget of functions and represent $t$ as the pair $\Pair{t(\EString)}{f(L)}$.
\FD{Essentially a children injective tree is completely  determined by the label of its root and by the set of all paths (of labels) in it. }
Moreover the subtree rooted at $\alpha$, $t_\alpha$, is represented by $\Pair{t(\alpha)}{\{\beta \in \String{\Labels} \mid \alpha\beta \in f(L)\}}$.

We denote {by} $\Tree[\lambda]{\Labels}$ the set of all $\lambda$-branching trees labelled in $\Labels$, that is, trees  built on an alphabet of cardinality $\lambda$, where $\lambda$ is an arbitrary cardinal number.
{We omit $\lambda$ when it is not relevant.
We denote by $\CTree{\Labels}$ the set of children injective trees labelled in $\Labels$. 
If $\lambda = |\Labels|$ we have that $\CTree{\Labels} \subseteq \Tree[\lambda]{\Labels}$.}\\

The main result of this section is \refToTheorem{graph-tree}.
Before {stating} it we need to briefly say {something} about \emph{paths} on a (labelled) graph.
Since paths are independent from the labelling, we {introduce} them for graphs and everything immediately extends to labelled graphs.
Let $G=\Pair{\Nodes}{\adj}$ be a graph, a path in $G$ is {a} {non-empty} string $\node_0 \cdots \node_n \in \String{\Nodes}$ such that{, for all} $i \in \{0, \ldots, n-1\}$, $\node_{i+1} \in \adj(\node_i)$, that is, {for all pairs of subsequent nodes the latter is adjacent to the former}.
{We say that $\node_0\cdots \node_n$ is a path \emph{from $\node_0$ to $\node_n$}.
Note that the string $\node_0$ of length 1 is also a path, {from $\node_0$ to $\node_0$.}}
We denote {by} $\Paths{G}$ the set of paths in $G$.

{Note that $\Paths{G}$ is closed under {non-empty} prefixes, that is, if $\alpha a $ is a path and $\alpha$ is not empty, then $\alpha$ is a path too, and more generally if $\alpha \beta \in \Paths{G}$ and $\alpha$ and $\beta$ are not empty, then $\alpha, \beta \in \Paths{G}$.
Therefore we can easily lift $\Paths{G}$ to a tree language, by adding to it the empty string.
From these observations immediately follows that{,  for each $\alpha \in \Paths{G}$,} the set $\{\beta \in \String{\Nodes} \mid \alpha \beta \in \Paths{G}\} \subseteq \Paths{G} \cup \{\EString\}$ is a tree language. }

Another important observation is that the sets $\Tree[\lambda]{\Labels}$ and $\CTree{\Labels}$  both carry a labelled graph structure with the following adjacency function{:}
\[\dsub(t) = \{ t_\alpha \mid \alpha \in \dom(t),\, \Len{\alpha} = 1\}\]
and labelling given by $r(t) = t(\EString)$.

\FD{Thanks to this observation, we can now prove the following theorem, that will be essential to show the equivalence between proof-theoretic and fixed point semantics of inference systems (see \refToSection{sem-eq}).
Intuitively, this result allows us to associate, with any node in a graph, in a canonical way,   a tree rooted in it, preserving the graph structure. }

\begin{theorem} \label{theo:graph-tree}
The following facts hold{.}
\begin{enumerate}
\item Let $G=\Pair{\Nodes}{\adj}$ be a graph, then there exists a graph homomorphism $\fun{\Path}{\Nodes}{\CTree{\Nodes}}$ such the following diagram commutes{:}
\begin{center}
\begin{tikzcd}[column sep=huge, row sep=huge]
\Nodes \ar[d, "\adj"] \ar[r, "\Path"] & \CTree{\Nodes} \ar[d, "\dsub"] \\
\wp(\Nodes) \ar[r, "\wp(\Path)"] & \wp(\CTree{\Nodes})
\end{tikzcd}
\end{center}
\item Let $G=\Triple{\Nodes}{\adj}{\ell}$ be a labelled graph with labels in $\Labels$, then there exists a labelled graph homomorphism $\fun{\Path_\Labels}{\Nodes}{\Tree[\lambda]{\Labels}}$, with $\lambda = |\Nodes|$, such that the following diagram commutes{:}
\begin{center}
\begin{tikzcd}[column sep=huge, row sep=huge]
\Nodes \ar[d, "\Pair{\ell}{\adj}"] \ar[r, "\Path_\Labels"] & \Tree[\lambda]{\Labels} \ar[d, "\Pair{r}{\dsub}"] \\
\Labels \times \wp(\Nodes) \ar[r, "\GraphFun{\Labels}(\Path_\Labels)"] & \Labels \times \wp(\Tree[\lambda]{\Labels})
\end{tikzcd}
\end{center}
\end{enumerate}
\end{theorem}
\begin{proof}
We will give a complete proof only for 1, for 2 we will only define the function $\Path_\Labels$, then the proof is analogous.
\begin{enumerate}
\item The function $\Path$ computes for each node the \emph{path expansion} starting from this node, that is, it maps each node $\node$ to the set of all paths starting with $\node$.
More precisely the set of paths we compute for each node $\node$ is the following {:}
\[L_\node = \{\alpha \in \String{\Nodes} \mid \node\alpha\in \Paths{G}\}\]
Hence, using the representation of children injective trees as pairs $\Pair{r}{L}$ where $r$ is a label and $L$ is a tree language using labels as alphabet, we have that 
\[\Path(\node) = \Pair{\node}{L_\node}\]
Now we have to show that the diagram commutes, that is, for each node $\node$, $\wp(\Path)(\adj(\node)) = \dsub(\Path(\node))$. 
First note that each $\Pair{\anode}{L} \in \dsub(\Path(\node))$ is such that $L = \{\alpha \in \String{\Nodes} \mid \anode\alpha \in L_\node\}$, in other words $L = L_\anode$;
hence $\Pair{\anode}{L} = \Path(u)$.
Moreover, since $\anode \alpha \in L_\node$ we have that $\node\anode\alpha$ is a path in $G$, and so $\anode \in \adj(\node)$ and this shows the equality.
\item We only define the function $\Path_\Labels$. For each node $\node$ we have that $\Path_\Labels(\node) = t$ where $t$ is a tree defined as follows
\[
\fun{t}{L_\node}{\Labels} \Space \left\{ \begin{array}{ll}
t(\EString) &= \ell(\node) \\
t(\alpha \anode) &= \ell(\anode)
\end{array} \right.
\]
\end{enumerate}
\end{proof}

\subsection{A {proof-theoretic} semantics}
In this section we discuss a first way to define the semantics of an inference system.
We call it \emph{{proof-theoretic}} since it is based on a notion of proof of the validity of a judgement, that is, the construction of an object that witnesses that a judgement is valid.
These objects are named \emph{proof trees} or \emph{derivations} {and are} defined below

\begin{definition} \label{def:proof-tree}
Let $\is$ be an inference system, a \emph{proof tree} {(or \emph{derivation})} in $\is$ is a tree $\fun{t}{L}{\universe}$, such that{,} for each node (labelled with) $\cons$ having children (labelled) in $\prem$, the rule $\Rule{\prem}{\cons}$ is in $\is${.}
\end{definition}

More precisely, using {the} previously introduced {notations},  a {proof} tree $t$ is such that{,} for each node $\alpha \in L$ with $t(\alpha) = \cons$, children of $\alpha$ {are in bijection with a set $\prem \subseteq \universe$ such that $\Rule{\prem}{\cons} \in \is$. 
Therefore clearly a proof tree $\fun{t}{L}{\universe}$  is children injective, hence we can assume {the canonical choice} $L \subseteq \String{\universe}$ and $t(\alpha \judg) = \judg$ for each $\alpha\judg \in L$.
This {allows us} to rewrite the condition {that} a proof tree has to satisfy as follows:
for each $\alpha \in \dom(t)$ with $t(\alpha) = \cons$, there is a set $\prem \subseteq \universe$ such that $\chl(\alpha) = \alpha \prem$ and $\Rule{\prem}{\cons} \in \is$. }

{In the following we will often represent proof trees using stacks of rules, that is, if $\myrule \in \is$ and $\T$ is a set of proof trees such that for all $t \in \T$, $t(\EString) \in \prem$ and vice versa, we denote by $\Rule{\T}{\cons}$ the proof tree $t_\cons$ given by}
\[
\dom(t_\cons) = \{\EString\} \cup \bigcup_{t\in \T} t(\EString) \dom(t)
\BigSpace  
\left\{ \begin{array}{ll}
t_\cons(\EString) &= \cons\\
t_\cons(\alpha \judg) &= \judg
\end{array} \right.
\]

We say that a tree $t$ is a proof tree for a judgement $\judg \in \universe$ if it is a proof tree {rooted} in $\judg$.
With this terminology we can define {two} interpretations of an inference system{.}

\begin{definition} \label{def:ind-coind}
Let $\is$ be an inference system. Then{:}
\begin{itemize}
\item the \emph{inductive interpretation} of $\is$, denoted {by} $\Ind{\is}$ is the {set} of judgements having a {well-}founded proof tree
\item the \emph{coinductive interpretation} of $\is$, denoted {by} $\CoInd{\is}$ is the {set} of judgements having an arbitrary ({well-}founded or not) proof tree
\end{itemize}
\end{definition}

Clearly by definition $\Ind{\is} \subseteq \CoInd{\is}$ but the converse is not {necessarily} true; indeed when the two interpretations are equal we are in a special case with many pleasant properties.

Let us now discuss some examples {on} lists.
Recall the definitions of predicates $\memberPred{x}{l}$ and $\allPosPred{l}$
\[
\Rule{}{\memberPred{x}{\Cons{x}{l}}} 
\Space
\Rule{ \memberPred{x}{l} }{\memberPred{x}{\Cons{y}{l}}} 
\BigSpace
\Rule{}{ \allPosPred{\EList} }
\Space
\Rule{ \allPosPred{l} }{ \allPosPred{\Cons{x}{l}} } x>0
\]
where $l$ ranges over finite lists and $x, y $ on integers. 
We interpret these inference systems both inductively. The following are valid proof trees for some judgements
\[
\Rule{}{\memberPred{1}{\Cons{1}{\Cons{2}{\Cons{1}{\EList}}}}}
\BigSpace
\Rule{
	\Rule{
		\Rule{}{ \memberPred{1}{\Cons{1}{\EList}}}
	}{ \memberPred{1}{\Cons{2}{\Cons{1}{\EList}}}}
}{ \memberPred{1}{\Cons{1}{\Cons{2}{\Cons{1}{\EList}}}}}
\BigSpace
\Rule{
	\Rule{
		\Rule{
			\Rule{}{ \allPosPred{\EList} }
		}{ \allPosPred{\Cons{1}{\EList}}}
	}{ \allPosPred{\Cons{2}{\Cons{1}{\EList}}}}
}{ \allPosPred{\Cons{1}{\Cons{2}{\Cons{1}{\EList}}}}}
\]
Note that the same judgement can be proved with different proof trees, as for $\memberPred{1}{\Cons{1}{\Cons{2}{\Cons{1}{\EList}}}}$.
This is due to the nature of \EZ{meta-rule}s that are in some sense redundant: the second rule can be applied also in cases when the first suffices.
In order to remove this redundancy we can add a side condition to the second rule, to make the two rule mutually exclusive: the needed side condition is $x \ne y$. 
In this way the second tree shown above is not a proof tree since the first step is not justified by any rule.

Writing down these trees it is clearer what we {meant} when we said that these recursive definitions on {lists} {``}inspect'' the list.
Indeed at each step in the tree we go deeper in the list looking at its tail.
We can also see that for $\allPosPred{l}$ we need to inspect the whole list to complete a proof, while for $\memberPred{x}{l}$ we stop as soon as we find the first $x$ in $l$.

We can also reason a little bit on what {happens} for judgements that should not {hold,} like $\memberPred{1}{\EList}$ or $\allPosPred{\Cons{-1}{\EList}}$.
For the predicate $\memberPred{x}{l}${,} if $x$ {does not occur} in $l$, {a tentative proof} will continue unfolding elements from the list $l$ until it reaches the empty list for which there is no applicable rule.
For $\allPosPred{l}${,}  if there is an $x$ in $l$ with $x\le 0$, {a tentative proof} will unfold elements from $l$ until it reaches $x$, and here it stops since there is no applicable rule.

Let us now assume that $l$ ranges over both finite and infinite lists of integers. 
We represent regular\footnote{\EZ{That is, with a finite number of distinct sublists.}} lists with syntactic equations as it is standard, so the list $L = \Cons{1}{\Cons{2}{L}}$ is the infinite regular list in which 1 and 2 are repeated infinitely many times.

Now, what happens if we interpret both inference systems coinductively?
For $\memberPred{x}{l}$ all valid judgements are still provable, since{,} as we said, it suffices  to inspect finitely many elements of the list to find  $x$ in $l$; however we have some trouble with judgements that should not hold, for instance $\memberPred{0}{L}$ can be proved as follows
\[
\Rule{
	\Rule{
		\Rule{
			\vdots
		}{ \memberPred{0}{L} }
	}{ \memberPred{0}{\Cons{2}{L}} }
}{ \memberPred{0}{ \Cons{1}{\Cons{2}{L}} } }
\]
where the dots indicate that the proof continues indefinitely in the same way. 
This proof {is} an infinite {non-well-}founded) proof tree since each step is correctly justified by a rule, but {it} proves a judgement that should not hold.
Therefore we can conclude that even with infinite lists the correct interpretation for the inference system defining $\memberPred{x}{l}$ is the inductive one.

For the predicate $\allPosPred{l}$ the situation is quite different, \FD{in some sense symmetric,} indeed for judgements that should not hold it is still true that we reach in finitely many steps a {non-}positive element and so we do not have any applicable rule.
Moreover if $l$ is infinite, since the predicate should check all elements in the list, we cannot prove this judgement with a finite derivation, so we need an infinite proof tree like the following
\[
\Rule{
	\Rule{
		\Rule{
			\vdots
		}{ \allPosPred{L} }
	}{ \allPosPred{\Cons{2}{L}} }
}{ \allPosPred{ \Cons{1}{\Cons{2}{L}} } }
\]
Therefore we see that inductive and coinductive interpretations are both necessary to define judgements in a proper way.

We now discuss a last example on lists that shows a very important situation.
Consider the following inference system
\[
\Rule{}{ \maxElem{\Cons{x}{\EList}}{x} }
\BigSpace
\Rule{ \maxElem{l}{y} }{ \maxElem{\Cons{x}{l}}{z} } z = \max\{x, y\}
\]
defining a judgement that computes the maximum of a list if it exists, where $l$ ranges over finite and infinite lists.
How we should interpret this inference system?
Clearly to compute a maximum we need to inspect the whole list, so we need possibly infinitely many steps, hence inductive interpretation seems to be not enough.
Then {let us} try with the coinductive interpretation, the following are {two} valid derivations{:}
\[
\Rule{
	\Rule{
		\Rule{
			\vdots
		}{ \maxElem{L}{2} }
	}{ \maxElem{\Cons{2}{L}}{2} }
}{ \maxElem{ \Cons{1}{\Cons{2}{L}} }{2} }
\BigSpace
\Rule{
	\Rule{
		\Rule{
			\vdots
		}{ \maxElem{L}{3} }
	}{ \maxElem{\Cons{2}{L}}{3} }
}{ \maxElem{ \Cons{1}{\Cons{2}{L}} }{3} }
\]
The first derivation proves a judgement that is expected to hold, while the second {proves an invalid judgement}, since 3 does not belongs to the list $L$.
So also the coinductive interpretation is not suitable for this definition since it considers too many judgements as valid.
Our extension of inference systems, presented in \refToChapter{coaxioms}{,} is just designed to overcome this rigid dichotomy and capture a broader range of definitions.

Let us conclude this section showing {an} example dealing with another important {non-}well-founded structure: graphs.
{This is another case in which coinduction is needed in order to correctly define predicates and functions. }
We represent a graph as a pair $\Pair{\Nodes}{\adj}$ like in \refToDefinition{graph}  and we define the judgement $\dist{\node}{\anode}{\delta}$ stating that the distance between node $\node$ and node $\anode$ is $\delta$, with $\delta \in \N \cup\{\infty\}$. 
Here we {mean as distance} the minimum number of edges  we have to traverse to reach node $\anode$ starting from $\node$, {infinite if} we cannot reach $\anode$ from $\node$. 
The definition is the following, where we assume that $\min \emptyset = \infty$ and $n + \infty = \infty $ for all $n \in \N$.
\begin{small}
\[
\Rule{}{ \dist{\node}{\node}{0} }
\BigSpace
\Rule{
	\dist{\node_1}{\anode}{\delta_1} 
	\Space \ldots \Space
	\dist{\node_k}{\anode}{\delta_k}
}{ \dist{\node}{\anode}{1+\delta} }
{\begin{array}{l}
\node \ne \anode \\
\adj(\node) = \{\node_1, \ldots, \node_k\}\\
\delta = \min\{\delta_1, \ldots, \delta_k\}
\end{array} }
\]
\end{small}
This definition should be interpreted coinductively, since the graph structure is not well-founded, hence we have to deal with possibly infinite paths (e.g.{,} \FD{a finite path followed by a cycle})  and the inductive interpretation is not able to deal with such situations. 
Consider for instance the following graph 
\begin{center}
\begin{tikzcd}[column sep=large, row sep=large]
e        & b \ar[d, bend right]         &   \\
d \ar[r] & a \ar[u, bend right] \ar[ul] & c \ar[l]  
\end{tikzcd}
\end{center}
We need infinite proofs in order to derive{, for instance,} judgements like $\dist{c}{e}{2}$ or $\dist{b}{e}{2}$, since in both cases, to reach $e$, we need to pass through $a$ that is part of a cycle, as shown in \refToFigure{tree-dist-1}.
\begin{figure}
\caption{Proof trees for $\dist{c}{e}{2}$ and $\dist{c}{e}{2}$} \label{fig:tree-dist-1}
\begin{small}
\[
\Rule{
	\Rule{
		\Rule{}{ \dist{e}{e}{0} }
		\Space
		\Rule{
			\Rule{
				\vdots
			}{ \dist{a}{e}{1} }
		}{ \dist{b}{e}{2} }
	}{ \dist{a}{e}{1} }
}{ \dist{c}{e}{2} }
\BigSpace
\Rule{
	\Rule{
		\Rule{}{\dist{e}{e}{0} }
		\Space
		\Rule{
			\vdots
		}{ \dist{b}{e}{2} }
	}{ \dist{a}{e}{1} }
}{ \dist{b}{e}{2} }
\]
\end{small}
\end{figure}
Note that $\dist{e}{\node}{\infty}$ is the only valid judgement for all $\node \in \{a, b, c, d\}$, since there are no outgoing edges from $e$, hence {we can only instantiate} the second rule  {with $\{\delta_1, \ldots, \delta_k\}=\emptyset$ (no premises)} and so $\delta = \infty$.
Finally let us consider {judgements of shape} $\dist{d}{c}{\delta}$.
A derivation schema is shown in \refToFigure{tree-dist-2}.
\begin{figure}[h]
\caption{Proof tree for $\dist{d}{c}{\delta}$} \label{fig:tree-dist-2}
\[
\Rule{
	\Rule{
		\Rule{}{\dist{e}{c}{\infty} }
		\Space
		\Rule{
			\Rule{
				\vdots
			}{ \dist{a}{e}{\delta - 3} }
		}{ \dist{b}{c}{\delta - 2} }
	}{ \dist{a}{c}{\delta - 1} }
}{ \dist{d}{c}{\delta} }
\]
\end{figure}
Now, which value of $\delta$ makes the proof correct? 
Surely for $\delta = \infty$ the proof is valid, because it becomes cyclic.
Actually there is no other {possible} value, because going up in the proof tree, $\delta$ should indefinitely decrease, and this is not possible since $\delta$ is a natural number and so it cannot go below zero.
Therefore as expected $\dist{d}{c}{\infty}$ is the only derivable judgement, meaning that  we cannot reach $c$ starting from $d$.

\section{Fixed point semantics} \label{sect:fp-sem}

In this section we will introduce another way to assign a semantics to an inference system, relying on order{-theoretic} notions discussed in \refToChapter{lattice}.
In particular we will characterize inductive and coinductive interpretation{s} as fixed points of a monotone function on the power set complete lattice $\Pair{\wp(\universe)}{\subseteq}$, we will derive from this characterization proof principles to reason about these semantics and finally we will prove the equivalence with the {proof-theoretic} semantics.

The starting point is the observation that an inference system $\is$ induces a function on $\wp(\universe)$ {defined} as follows{:}
\[
\Op{\is}(S) = \{ \cons \in \universe \mid \prem \subseteq S,\, \Rule{\prem}{\cons} \in \is\}
\]
This function is called \emph{(one step) inference operator}, since intuitively $\Op{\is}(S)$ contains all judgements that can be derived from those in $S$ using a single rule.

For instance, considering {the} inference systems on finite lists of integers from the previous section we get the following inference operators{:}

\[
\begin{split} 
\Op{\memberName}(S) =& \{\memberPred{x}{\Cons{x}{l}} \mid x\in \Z,\, l \in \LSet\} \cup \\
				& \{\memberPred{x}{\Cons{y}{l}} \mid \memberPred{x}{l} \in S\} \\
\Op{\allPosName}(S) =& \{\allPosPred{\EList}\} \cup
				  \{\allPosPred{\Cons{x}{l}} \mid x>0,\, \allPosPred{l} \in S \}\\
\Op{\maxElemName}(S) =& \{\maxElem{\Cons{x}{\EList}}{x} \mid x \in \Z \} \cup \\
				 & \{\maxElem{\Cons{x}{l}}{z} \mid \maxElem{l}{y} \in S,\, z = \max\{x, y\} \}
\end{split}
\]

The inference operator of an inference system has the following key property{.}

\begin{proposition}
Let $\is$ be an inference system, then $\Op{\is}$ is monotone{.}
\end{proposition}
\begin{proof}
Let $X\subseteq Y \subseteq \universe$, we have to show that $\Op{\is}(X) \subseteq \Op{\is}(Y)$.
Consider a judgement $\cons \in \Op{\is}(X)$, by definition of $\Op{\is}$ there exists $\prem \subseteq X \subseteq Y$ with $\Rule{\prem}{\cons} \in \is$.
By transitivity we get $\prem \subseteq Y$ and again by definition of $\Op{\is}$ we get $\cons \in \Op{\is}(Y)$ as needed.
\end{proof}

Now since $\wp(\universe)$ carries a complete lattice structure and $\Op{\is}$ is monotone, we can apply the Knaster-Tarski theorem (\refToTheorem{KT}), that ensures us the existence of the least and the greatest fixed points, $\lfp \Op{\is}$ and $\gfp \Op{\is}$.

In the rest of this section we will justify the two following sentences
\begin{itemize}
\item $\lfp\Op{\is}$ is the inductive interpretation of $\is$
\item $\gfp\Op{\is}$ is the coinductive interpretation of $\is$
\end{itemize}

\subsection{Induction and coinduction principles}
An important feature of the fixed point semantics is that it  provides immediately two powerful proof principles as  described in \refToSection{KT}.
These {principles} are induction and coinduction, that can be used to \FD{show how the least and the greatest fixed point\EZ{,} respectively\EZ{,} are related to a set representing a given specification. } 
Let us reformulate them in the context of inference systems.

Assume that $\DefSet \subseteq \universe$ is a set {(inductively or coinductively)} defined by an inference system $\is$.
Typically we have an \emph{expected semantics} described as a set $\Spec$ (for specification) and we would like to prove that $\DefSet$ and $\Spec$ agree with each other in some way.
More precisely we are often interested to prove $\Spec \subseteq \DefSet$, i.e.{,} the \emph{completeness} of the definition, and/or $\DefSet \subseteq \Spec$, i.e.{,} the \emph{soundness} of the definition. 

We say that a set $\Spec\subseteq \universe$ is \emph{closed} if $\Op{\is}(\Spec) \subseteq \Spec$, and we say that $\Spec$ is \emph{consistent} if $\Spec \subseteq \Op{\is}(\Spec)$.
Then we can reformulate induction and coinduction principles as follows{:}
\begin{description}
\item[Induction] if $\Spec$ is closed, then $\lfp\Op{\is} \subseteq \Spec$
\item[Coinduction] if $\Spec$ is consistent, then $\Spec \subseteq \gfp\Op{\is}$
\end{description}
In other words{,} the induction {principle} allows us to prove soundness of an inductive definition, while the coinduction principle allows us to {prove} completeness of a coinductive definition.
Note that if $\Spec$ is both closed and consistent, that is, it is a fixed point of $\Op{\is}$, then we have $\lfp\Op{\is} \subseteq \Spec \subseteq \gfp\Op{\is}$, as expected.

For a set $\Spec \subseteq \universe$, being closed or consistent can be {easily} expressed in terms of {the} rules in the inference system{.}

\begin{proposition}
Consider a subset $\Spec \subseteq \universe$, then 
\begin{enumerate}
\item $\Spec$ is closed iff for each rule $\myrule \in \is$, $\prem \subseteq \Spec$ implies $\cons \in \Spec$
\item $\Spec$ is consistent iff for all $\cons \in \Spec$ there exists a rule $\myrule \in \is$ such that $\prem \subseteq \Spec$
\end{enumerate}
\end{proposition}
\begin{proof}
\hspace*{\fill}
\begin{enumerate}
\item We prove the two implications.\\
$(\Rightarrow)$. Consider a rule $\myrule \in \is$ such that $\prem \subseteq \Spec$. Then $\cons \in \Op{\is}(\Spec) \subseteq \Spec$, so $\cons \in \Spec$.\\
$(\Leftarrow)$. Consider $\cons \in \Op{\is}(\Spec)$. 
By definition of the inference operator we know that there is a rule $\myrule\in \is$ such that $\prem\subseteq \Spec$, hence by hypothesis $\cons \in \Spec$.
\item  We show the two implications.\\
$(\Rightarrow)$. Consider $\cons \in \Spec$, hence $\cons \in \Op{\is}$ by hypothesis.
By definition of the inference operator we know that there is a rule $\myrule\in \is$ such that $\prem\subseteq \Spec$.\\
$(\Leftarrow)$. Consider $\cons \in \Spec$.
By hypothesis there is a rule $\myrule \in \is$ such that $\prem \subseteq \Spec$, so $\cons \in \Op{\is}(\Spec)$.
\end{enumerate}
\end{proof}

Note that these characterizations are expressed in terms of rules, however if the inference system is defined using \EZ{meta-rule}s, we can reason on them rather than on plain rules, opportunely quantifying on variables occurring in \EZ{meta-rule}s.

We now show two examples of application of the induction and {coinduction} principles.
In particular we will show that the definition of $\memberPred{x}{l}$ is sound with respect to the expected semantics and that the definition of $\allPosPred{l}$  is complete with respect to the expected semantics. \\

\spacedlowsmallcaps{Soundness of} $\memberPred{x}{l}$. \hspace{2ex}
The expected semantics is given by the following set{, where} write $x \in l$ to {mean} that $x$ {occurs} at least once in $l$.
\[
\Spec = \{\memberPred{x}{l} \mid x \in l\}
\]
Set $\DefSet = \lfp \Op{\memberName}$, so we have to show that $\DefSet \subseteq \Spec$.
We prove {this} by induction, hence we have only to prove that $\Spec$ is closed. 
\begin{itemize}
\item Consider the axiom $\Rule{}{\memberPred{x}{\Cons{x}{l}}}$. Since there are no premises we have only to show that $\memberPred{x}{\Cons{x}{l}} \in \Spec$, {that} is obviously true.
\item Consider the rule $\Rule{\memberPred{x}{l}}{\memberPred{x}{\Cons{y}{l}}}$ and assume that $\memberPred{x}{l} \in \Spec$, that is, $x \in l$. 
Hence surely $x \in \Cons{y}{l}$ since all elements in $l$ {occur} also in $\Cons{y}{l}${.}
\end{itemize}

\spacedlowsmallcaps{Completeness of} $\allPosPred{l}$. \hspace{2ex}
The expected semantics is given by the following set{, where} write $x \in l$ to {mean} that $x$ {occurs} at least once in $l$.
\[
\Spec = \{ \allPosPred{l} \mid \forall x \in l.\,x>0 \}
\]
Set $\DefSet = \gfp \Op{\allPosName}$, so we have to show that $\Spec\subseteq \DefSet$.
We prove {this} by coinduction, hence we have only to prove that $\Spec$ is consistent. 
The proof is very easy. Consider $l$ such that $\allPosPred{l} \in \Spec$.
If $l$ is empty, then $\allPosPred{l}$  is the consequence of the {axiom}.
Otherwise, if $l = \Cons{x}{l'}$, we have that $x>0$, and $\forall y \in l'. y>0$, hence $\allPosPred{l'} \in \Spec$. 
Therefore the needed rule is $\Rule{\allPosPred{l'}}{\allPosPred{\Cons{x}{l'}}}$.

\subsection{Equivalence with the proof{-theoretic} semantics} \label{sect:sem-eq}
In this section we will come up with two results stating that proof{-theoretic} and fixed point semantics agree with each other.
These results are of paramount importance since {they} allow us to equivalently use one of the two characterizations depending on {which} is {more} suitable in each situation.
A sketch of the same proofs can be found in \cite{LeroyGrall09}, even if our proof for the equivalence in the coinductive case is quite different.

Proofs for the inductive and the coinductive case are different, but share a common underlying scheme: one inclusion is proved using induction and coinduction principles{,} respectively, the other one \FD{showing, using properties of trees, that $\Ind{\is}$ and $\CoInd{\is}$\EZ{,} respectively\EZ{,} enjoy the induction and coinduction principles. }
In particular for the inductive case we rely on the {well-foundedness of} trees to build an inductive reasoning, while for the coinductive one we will use the property of trees described in point $1$ of \refToTheorem{graph-tree}.

{Let us} start with the inductive case.
The following lemma states that the inductive interpretation of an inference system $\is$  in terms of proof trees, that is, $\Ind{\is}$, enjoys the induction principle{.}

\begin{lemma} \label{lem:ind-trees}
Let $\is$ be an inference system and $\Spec$ a closed subset of $\universe$, {then} $\Ind{\is} \subseteq \Spec${.}
\end{lemma}
\begin{proof}
Consider a judgement $\cons \in \Ind{\is}$, so it has a {well-}founded proof tree $t$. 
Rephrasing the statement we have to show that the root of $t$ belongs to $\Spec$ for any {well-}founded  proof tree $t$.
Since $t$ is a {well-}founded tree we can reason by {well-}founded induction on it, namely, to prove the thesis, we assume it for all subtrees of $t$ and prove it for $t$. 

By definition of proof tree we that there is a rule $\myrule\in \is$ such that judgements in $\prem$ are the children of $\cons$ in $t$. 
In other words each $\judg \in \prem$ is the root of a subtree of $t$. 
Therefore by inductive hypothesis $\judg \in \Spec$, hence $\prem \subseteq \Spec$.
Now, by hypothesis $\Spec$ is closed and this implies that $\cons \in \Spec$ as needed.
\end{proof}

We can now state and prove the equivalence {for the inductive case.}

\begin{theorem} \label{theo:ind-eq}
Let $\is$ be an inference system, {then} $\Ind{\is} = \lfp\Op{\is}${.}
\end{theorem}
\begin{proof}
First note that, since $\lfp\Op{\is}$ is a fixed point of $\Op{\is}$, it is in particular a pre-fixed point, namely, it is closed.
Therefore by \refToLemma{ind-trees} we get the first inclusion $\Ind{\is} \subseteq \lfp\Op{\is}$.

We show the other one by induction, hence we have to show that $\Ind{\is}$ is closed with respect to $\Op{\is}$.
To this aim consider a rule $\myrule \in \is$ such that $\prem \subseteq \Ind{\is}$. 
By definition of $\Ind{\is}${,} for each judgement $\judg \in \prem$ there is a {well-}founded proof tree $\fun{t_\judg}{L_\judg}{\universe}$ rooted in $\judg$. 
Therefore we can define a {well-}founded tree $t$ with $\dom(t) = \{\EString\} \cup \bigcup_{\judg \in \prem} \judg L_\judg$ as follows{:}
\[
\left\{\begin{array}{ll}
t(\EString) &= \cons\\
t(\judg\alpha) &= t_\judg(\alpha)
\end{array}\right.
\]
As a consequence we get that $\cons \in \Ind{\is}$, hence $\Ind{\is}$ is closed and this implies by induction that $\lfp\Op{\is} \subseteq \Ind{\is}$.
\end{proof}

We now address the coinductive case.
As mention{ed} above the proof scheme is very similar, so {let us} start with a lemma stating that $\CoInd{\is}$ enjoys the coinduction principle.

\begin{lemma} \label{lem:coind-trees}
Let $\is$ be an inference system and $\Spec$ a consistent subset of $\universe$, {then}  $\Spec \subseteq \CoInd{\is}${.}
\end{lemma}
\begin{proof}
By hypothesis $\Spec$ is consistent, so for each judgement $\judg \in \Spec$ we can choose a rule $\Rule{\prem_\judg}{\judg} \in \is$ such that $\prem_\judg \subseteq \Spec$. 
In other words we can define the map $\fun{\adj}{\Spec}{\wp(\Spec)}$ given by $\adj(\judg) = \prem_\judg$, that turns $\Spec$ into a graph as in \refToDefinition{graph}.

By \refToTheorem{graph-tree} there exists a map $\fun{\Path}{\Spec}{\CTree{\Spec}}$ making the following diagram commute.
\begin{center}
\begin{tikzcd} 
\Spec \ar[d, "\adj"] \ar[r, "\Path"] & \CTree{\Spec} \ar[d, "\dsub"] \\
\wp(\Spec) \ar[r, "\wp(\Path)"] & \wp(\CTree{\Spec})
\end{tikzcd}
\end{center}
Therefore{,} for each $\judg \in \Spec$, $\Path(\judg)$ is a tree rooted in $\judg$ and labelled in $\Spec$, hence in $\universe$.
Moreover $\Path(\judg)$ is a proof tree, since the commutativity of the diagram ensures that children of a node labelled with $\judg'$ in $\Path(\judg)$ are exactly labelled with judgements in $\adj(\judg')$, namely, in $\prem_{\judg'}$.
Therefore $\judg \in \CoInd{\is}$ as needed.
\end{proof}

In the end we conclude the section stating and proving the equivalence for the coinductive case, that is proved using the coinduction principle, again in analogy with the inductive case.

\begin{theorem} \label{theo:coind-eq}
Let $\is$ be an inference system, {then} $\CoInd{\is} = \gfp\Op{\is}${.}
\end{theorem}
\begin{proof}
First note that, since $\gfp\Op{\is}$ is a fixed point of $\Op{\is}$, it is in particular a post-fixed point, namely, it is consistent.
Therefore by \refToLemma{coind-trees} we get the first inclusion $\gfp\Op{\is} \subseteq \CoInd{\is}$.

We show the other one by coinduction, hence we have to show that $\CoInd{\is}$ is consistent with respect to $\Op{\is}$.
To this aim consider a judgement $\cons \in \CoInd{\is}$.
By definition of coinductive interpretation there exists a proof tree $t$ rooted in $\cons$.
All directed subtrees of $t$, that is, all $t_\judg \in \dsub(t)$, are themselves proof trees for their roots, namely{,} $t_\judg$ is a proof tree for $\judg$, hence $\judg \in \CoInd{\is}$. 
Moreover, since $t$ is a proof tree, the set $\prem = \{ \judg \in \universe \mid t_\judg \in \dsub(t)\}$ is such that $\Rule{\prem}{\cons} \in \is$ and $\prem \subseteq \CoInd{\is}$.
Therefore $\CoInd{\is}$ is consistent and this implies by coinduction that $\CoInd{\is} \subseteq \gfp\Op{\is}$.
\end{proof}

\section{Continuity and iteration} \label{sect:iterate-is}

In \refToSection{continuity} we have introduced an iterative characterization of the least and the greatest fixed point, provided that the function is either upward or downward continuous.
In this section we give some sufficient and necessary conditions condition on inference systems to ensure that the induced inference operator is continuous.
This is particularly relevant because the iterative characterization gives \EZ{us} other proof principles in addition to induction and coinduction, and  provides a way to compute these fixed points just iterating the inference operator.

We start with the inductive case.

\begin{definition}\label{def:is-finitary}
An inference system $\is$ is \emph{finitary} if for every rule $\myrule \in \is$, $\prem$ is a finite set. 
\end{definition}

All the examples we have provided so far are finitary inference systems, but in general it is not guaranteed  that every rule has a finite set of premises.
Finitary inference systems have a particularly pleasant property: they induce an upward continuous inference operator\EZ{.}

\begin{theorem} \label{theo:finitary-is}
Let $\is$ be a finitary inference system, \EZ{then} $\Op{\is}$ is upward continuous\EZ{.}
\end{theorem}
\begin{proof}
Let $C \subseteq \wp(\universe)$ be a chain.
We have to show that $\Op{\is}(\bigcup C) = \bigcup \Op{\is}(C)$.
By \refToProposition{monotone-lub-glb} we know that $\bigcup \Op{\is}(C) \subseteq \Op{\is}(\bigcup C) $, thus we have to prove only the other inclusion.

Consider a judgement $\cons \in \Op{\is}(\bigcup C)$, by definition of $\Op{\is}$ there is a rule $\myrule \in \is$ such that $\prem \subseteq \bigcup C$.
Since $\prem$ is finite by hypothesis, there exists a finite subset $X \subseteq C$ such that $\prem \subseteq \bigcup X$.\\
Note that $X$ is itself a chain, because \EZ{it} is a subset of a chain, and being finite there exists   the top element of $X$, that is, an element $M \in X$ such that\EZ{,} for each $S\in X$, $S \subseteq M$.
This implies that $\bigcup X = M$ and so $\prem \subseteq M$.\\
Therefore $\cons \in \Op{\is}(M)$ and, since $\Op{\is}(M) \in \Op{\is}(C)$\EZ{,} we get that $\cons \in \bigcup \Op{\is}(C)$, from which follows $\Op{\is}(\bigcup C) \subseteq \bigcup \Op{\is}(C)$.
\end{proof}

\FD{Actually the above proof shows that the \EZ{finitariness} condition implies  that the inference operator \EZ{is} continuous, but in a stronger way with respect to the notion of continuity introduced in \EZ{\refToDefinition{continuous}}.
Indeed, the inference operator induced by a finitary inference system  preserves the least upper bound of any chain, rather than only of countable ones. }

Note that this is a sufficient but not necessary condition.
Indeed there are non-finitary inference systems that induce upward continuous inference operator, for instance the \EZ{following} one.
\[
\Rule{}{0}
\BigSpace
\Rule{n}{n+1}
\BigSpace
\Rule{\N}{0}
\]
with universe $\universe = \Z$. 
The operator is continuous since the non-finitary rule, that might break continuity, is redundant, in the sense that if it is removed the inductive interpretation remains the same.
Actually this is a particular instance of a more general fact.
\FD{We say that an inference system $\is$ is \emph{countable} if all rules have countable premises.}

\begin{theorem} \label{theo:continuity-finitary}
\FD{Let $\is$ \EZ{be} a countable inference system} such that $\Op{\is}$ is upward continuous, then there exists a finitary inference system $\is' \subseteq \is$ such that $\Op{\is'}=\Op{\is}$.
\end{theorem}
\begin{proof}
Set $\is' = \{ \myrule \in \is \mid \prem \mbox{ is finite}\} \subseteq \is$.
Surely for all $S \subseteq \universe$, $\Op{\is'}(X) \subseteq \Op{\is}(X)$, by construction.
To prove the other inclusion \EZ{it} is enough to show that for each rule $\myrule \in \is$ with $\prem$ \FD{an infinite and  countable set}, there exists a rule $\Rule{\prem'}{\cons} \in \is$ with $\prem'\subseteq \prem$ a finite set.

\FD{Consider a rule $\myrule \in \is$ with $\prem$ an infinite and countable  set. 
Hence, there exists a bijection between $\N$ and $\prem$\EZ{. L}et $(x_i)_{i \in \N}$ be the enumeration of all elements in $\prem$, induced by such bijection. }
We construct a chain $C = (X_i)_{i \in \N}$ as follows:
\[X_0 = \emptyset \BigSpace X_{n+1} = X_n \cup \{x_n\}\]

It is easy to check that $X_n$ is a finite set for each $n \in \N$ ($|X_n| = n$) and $\bigcup C = \prem$.
By the continuity of $\Op{\is}$ we know that $\bigcup \Op{\is}(C) = \Op{\is}(\bigcup C) = \Op{\is}(\prem)$, hence $\cons \in \bigcup \Op{\is}(C)$,  and this implies that there exist $n \in \N$ such that $\cons \in \Op{\is}(X_n)$.
From the definition of $\Op{\is}$ it follows that there is a rule $\Rule{\prem'}{\cons}\in \is$ with $\prem' \subseteq X_n \subseteq \prem$, and, since $X_n$ is finite, also $\prem'$ is finite  as needed.
\end{proof}

\FD{It is immediate that if $\universe$ is a countable set, then every inference system on $\universe$ is countable; hence, in this case, we can say that for all inference system\EZ{s} $\is$, $\Op{\is}$ is upward continuous if and only if there exists a finitary inference system $\is' \subseteq \is$ such that $\Op{\is} = \Op{\is'}$. 
Actually, the same result can be proved also without the countability hypothesis, however this requires a stronger notion of continuity and transfinite induction. }

With \refToTheorem{finitary-is} and \refToTheorem{continuity-finitary} we have provided an almost complete characterization of inference systems that induce an upward continuous inference operator, at least in the case of a countable universe.

This characterization gives us a way to establish if we can compute the least fixed point by iterative application of the inference operator to the empty set.
This iterative computation corresponds to the intuitive idea that we start assuming that every judgement is not valid, then, applying once the inference operator, we get what is surely valid (the consequences of axioms), then we compute what we can derive from axioms and so on.
In other words  in $\IterOp{\is}{n}(\emptyset)$ there are the judgements that we can prove in $n$ steps.

Another observation \FD{we can do under this additional hypothesis is that, since} $\Ind{\is} = \bigcup \Iterate{\Op{\is}}{\emptyset}$, each judgement $\judg \in \Ind{\is}$ belongs to some $\IterOp{\is}{n}(\emptyset)$, that is, it must be provable in a finite number of steps.
\FD{This implies, from the proof-theoretic perspective, that all judgements are derivable by a well-founded proof tree with finite depth. } \\

Let us consider the case of downward continuity.
This is much more complicated, indeed we will not give a characterization for inference systems that induce a downward continuous inference operator, but we will characterize a more restrictive class of them.

\begin{definition} \label{def:is-deterministic}
An inference system $\is$ is \emph{deterministic} if for any two rules $\Rule{\prem}{\cons}, \Rule{\prem'}{\cons'} \in \is$ if $\cons = \cons'$ then $\prem = \prem'$\EZ{.}
\end{definition}

We will show that deterministic inference systems induce an inference operator that preserves the meet of any set, that is, for $X \subseteq \wp(\universe)$, $\Op{\is}(\bigcap X) = \bigcap \Op{\is}(X)$. 
Clearly, since chains are particular subsets of $\wp(\universe)$, this condition implies downward continuity.

\begin{theorem} \label{theo:deterministic-is}
Let $\is$ be a deterministic inference system, then\EZ{,} for each $X\subseteq \wp(\universe)$, $\Op{\is}(\bigcap X) = \bigcap \Op{\is}(X)$\EZ{.}
\end{theorem}
\begin{proof}
From \refToProposition{monotone-lub-glb} we know that $\Op{\is}(\bigcap X) \subseteq \bigcap \Op{\is}(X)$, for $X\subseteq \wp(\universe)$, so we have to prove only the other inclusion.

Consider a judgement $\cons \in \bigcap \Op{\is}(X)$, hence for all $A \in X$ there is a rule $\Rule{\prem_A}{\cons} \in \is$ such that $\prem_A \subseteq A$.
Since the inference system is deterministic, for all $A, B \in X$, $\prem_A = \prem_B = \prem$, and so $\prem \subseteq \bigcap X$.
This implies that $\cons \in \Op{\is}(\bigcap X)$ as needed.
\end{proof}

\EZ{Examples} of deterministic inference system\EZ{s} are those for $\allPosPred{l}$ and $\memberPred{x}{l}$ using the side condition $x\ne y$ (see \refToSection{is}).

Also in this case we can prove a sort of necessary condition\EZ{.}

\begin{theorem} \label{theo:continuity-deterministic}
Let $\is$ be an inference system such that $\Op{\is}$ preserves the greatest lower bound, then there exists a deterministic inference system $\is' \subseteq \is$ such that $\Op{\is'} = \Op{\is}$\EZ{.}
\end{theorem}
\begin{proof}
For all $\cons\in\universe$, set $P_\cons = \{ \prem \in \wp(\universe) \mid \Rule{\prem}{\cons} \in \is\}$ and  $\prem_\cons = \bigcap P_\cons$.
We first \EZ{show} that there is a set $\prem'_\cons \subseteq \prem_\cons$ such that $\Rule{\prem'_\cons}{\cons} \in \is$. 
\EZ{Indeed, b}y hypothesis we get that $\Op{\is}(\prem_\cons) = \Op{\is}(\bigcap P_\cons) = \bigcap \Op{\is}(P_\cons)$.
Since by construction $\cons \in \bigcap \Op{\is}(P_\cons)$ we get that $\cons \in \Op{\is}(\prem_\cons)$, and so there is a rule $\Rule{\prem'_\cons}{\cons} \in \is$ such that $\prem'_\cons \subseteq \prem_\cons$.

Set $\is' $ the set of all rules $\Rule{\prem'_\cons}{\cons}$ in $\is$ \FD{for every consequence $\cons$ of a rule in $\is$}, clearly $\is' \subseteq \is$.
Now  for every $X \subseteq \universe$ we trivially have that $\Op{\is'}(X) \subseteq \Op{\is}(X)$, hence we have only to prove the other inclusion.
If $\cons \in \Op{\is}(X)$, there is a rule $\myrule \in \is$ such that $\prem \subseteq X$.
But this implies that $\prem'_\cons \subseteq \prem \subseteq X$ and so $\cons \in \Op{\is'}(X)$ as needed.
\end{proof}

These characterizations \FD{provide us with a sufficient condition to establish when} an inference system induces a downward continuous inference operator, hence if we can compute the greatest fixed point, i.e., the coinductive interpretation, iteratively applying this operator.

This description of the greatest fixed point, $\CoInd{\is} = \bigcap \Iterate{\Op{\is}}{\universe}$,  can be intuitively explain\EZ{ed} as follows: 
we start assuming that every judgement is valid, then, applying the inference operator, we loose\EZComm{remove?} all judgement\EZ{s} that cannot be proved even assuming that everything is valid, ad so on.
Hence in $\IterOp{\is}{n}$ we find all judgements that we have not been lost\EZComm{removed?} in $n$ steps.\\

We conclude the section noting that continuity is a quite strong condition on the function.
Indeed there are cases in which we can compute  iteratively the least or the greatest fixed point even if the function is not continuous.
Consider for instance the following inference system 
\[
\Rule{}{n} n>0
\BigSpace
\Rule{\{n \in \N \mid n> 0\}}{0}
\]
The induced inference operator is not continuous since it does not preserves the least upper bound of the chain $C = (I_n)_{n \in \N}$ where $I_n$ is defined by $I_0 = \emptyset$ and $I_{n+1} = I_n \cup \{n\}$.
However we have that 
\begin{align*}
\IterOp{\is}{0}(\emptyset) &= \emptyset \\
\IterOp{\is}{1}(\emptyset) &= \{n \in \N \mid n > 0 \} \\
\IterOp{\is}{2}(\emptyset) &= \N \\
\IterOp{\is}{3}(\emptyset) &= \N
\end{align*}
hence in three steps we reach the least fixed point even if $\Op{\is}$ is not continuous.

\cleardoublepage \chapter{Inference systems with coaxioms} \label{chapter:coaxioms}

When we deal with structured data types a typical issue is defining predicate{s} and/or functions  {on} them, and, since these types have a structure, we would like to exploit it  in the definition.
This definition technique is called \emph{structural recursion} and essentially provides a recursive definition of a predicate or of a function that mirrors the structure of a data type.
This recursive definition then has to be interpreted and, as we have seen, this can be typically done either inductively or coinductively. 

As briefly mentioned in the introduction, we can say there are two main classes of structured data type{s}: inductive (well-founded) data types and coinductive ({non-}well-founded) data types.
In order to deal with inductive data types, the inductive interpretation of structurally recursive definitions is enough, thanks to the well-foundedness of the data type: we can reach a base case in finitely many steps.
For coinductive data types things are more complicated: depending on what we are defining, we will need either {the} inductive or coinductive interpretation, and there are also cases where {neither} is suitable.

Recently several approaches have been proposed to provide a semantics to structurally recursive definitions on coinductive data types. In an operational style we can find proposals in all most popular paradigms:  logic \cite{SimonMBG06, SimonBMG07}, {object-oriented} \cite{AnconaZucca12, AnconaZucca13}, functional \cite{JeanninKS13, JeanninKS17} and type theory \cite{Coquand93, AbelPientka13, AbelPTS13, Mogelberg14}.
But there are also more abstract approaches, such as \cite{AdamekMV06, CaprettaUV06, CaprettaUV09}.
The majority of these proposals is characterized by a strong dichotomy between induction and coinduction, that in some cases makes semantics too rigid. 

As shown in \refToChapter{is} also inference systems suffer from this dichotomy, that makes impossible to assign a precise semantics to definitions that look very natural.
Hence {the need emerges} for a more flexible interpretation, that overcomes this dichotomy.

In this chapter we will propose an extension to inference systems, both in syntax and semantics, that will allow more flexible interpretations of them.
This extension is inspired by  some {of the} operational models mentioned above \cite{AnconaZucca12, AnconaZucca13, Ancona13} and, in our intention, will serve as an abstract framework for a better understanding of these operational models, allowing formal reasoning on them.

The key concept of this extension are \emph{coaxioms}, that are special rules that need to be specified together with the usual definition in order to control its semantics.
In other words{,} coaxioms allow {one} to choose {as interpretation} a fixed point that is not necessarily  either the least or the greatest one.
In this way we can assign a more natural semantics to definition{s} that otherwise would have a very strange meaning.
In addition we will also show that inductive and coinductive interpretations are particular cases of our extension, proving that it is actually an extension.
Another important feature is that in this framework we can interpret also inference systems where judgements that should be defined inductively and coinductively are mixed together in the same definition. 

{The rest of the chapter is organized as follows.
In \refToSection{coaxioms} we will introduce  \emph{inference systems with coaxioms}, informally explaining their semantics with a bunch of examples.
The fixed point semantics for inference systems with coaxioms is formally defined in \refToSection{coaxioms-model}.
Here we present closure and kernel systems, {which are well-know notions on the power-set}, in the more general setting of complete lattices, getting the definition of the \emph{bounded fixed point}, that represents the semantics induced by coaxioms.
In \refToSection{coaxioms-trees} we introduce several equivalent proof-theoretic semantics based on the notion of proof tree.
Particularly interesting are the two characterizations exploiting the new concept of \emph{approximated proof tree}, that will allow us to provide the semantics in terms of sequences of well-founded trees, without considering non-well-founded derivations.
Proof techniques for coaxioms to prove both completeness and soundness of definitions are discussed in \refToSection{coaxioms-reasoning}.
In particular we will introduce the \emph{bounded coinduction principle} that is a generalization of the standard coinduction principle, aimed to show the completeness of a definition expressed in terms of an inference system with coaxioms.
Finally, in \refToSection{coaxioms-examples}, we try to illustrate weaknesses and strengths of our framework, using various, more involved, examples. }

\FD{This chapter presents in more detail the work we have done in \cite{AnconaDZ17esop}.
Notably, here we discuss closures and kernels from a more general point of view (see \refToSection{ck}), in order to better frame the bounded fixed point in lattice theory. 
Furthermore, thanks to a more formal treatment of proof trees, we introduce an additional proof-theoretic characterization, using approximated proof trees (see \refToTheorem{approx-sequence}). 
We also present \EZ{another} example of application of coaxioms to graphs (see \refToSection{graph-example}). }

\section{Introduction to coaxioms} \label{sect:coaxioms}

In this section we will introduce coaxioms and try to illustrate their behaviour be means of a bunch of examples.
Recall from \refToSection{is} that $\universe$ is a universe of judgements.

\begin{definition} \label{def:coaxioms}
An \emph{inference system with coaxioms} is a pair $\Pair{\is}{\coaxioms}$ where $\is$ is an inference system and $\coaxioms \subseteq \universe$ is a set of \emph{coaxioms}{.}
\end{definition}

A coaxiom $\cons \in \coaxioms$ will be written as $\CoAxiom{\cons}$, very much like an axiom, and{,} analogously to an axiom, it can be {used} as initial assumption to derive other judgements.
However{,} coaxioms will be used in a special way, {that is,} intuitively they can be used only {``}at infinite depth'' in a derivation.
This will allow us to impose an initial assumption also to infinite proof trees, that otherwise  are not required to have such starting point. 
We will make precise this notion in next sections, now we will show some examples to illustrate  how to use coaxioms  to govern the semantics of an inference system.

\EZ{Analogously to sets of rules, sets of coaxioms can be expressed by a \emph{meta-coaxiom} with side conditions.}

Let us start with an introductory example concerning graphs, that are a widely used {non-}well-founded data type.
Consider a graph $\Pair{\Nodes}{\adj}$ where $\Nodes$ is the set of nodes and $\fun{\adj}{\Nodes}{\wp(\Nodes)}$ is the adjacency function.
We {want} to define the judgement $\Visit{\node}{\nodeset}$ stating that {nodes in the set $\nodeset$  are those} reachable from $\node$.

We define this judgement with the \EZ{following (meta-)rule and (meta-)coaxiom:}
\[
\Rule{
	\Visit{\node_1}{\nodeset_1} \Space \ldots \Space \Visit{\node_k}{\nodeset_k}
}{ \Visit{\node}{\{\node\} \cup \nodeset_1 \cup \ldots \cup \nodeset_k} } 
\adj(\node) = \{\node_1, \ldots, \node_k\}
\BigSpace
\CoAxiom{\Visit{\node}{\emptyset}} \node \in \Nodes
\]
For instance, in the case of a graph with nodes $a,b,c$, with an arc from $a$ into $b$ and conversely, and $c$ isolated, we would get the following \EZ{(meta-)}rules and coaxioms:
\begin{small}
\[
\begin{array}{c}
\Rule{\Visit{b}{\nodeset}}{\Visit{a}{\{a\}\cup\nodeset}}\BigSpace\Rule{\Visit{a}{\nodeset}}{\Visit{b}{\{b\}\cup\nodeset}}\BigSpace\Rule{}{\Visit{c}{\{c\}}}
\BigSpace\CoAxiom{\Visit{a}{\emptyset}}\BigSpace\CoAxiom{\Visit{b}{\emptyset}}\BigSpace\CoAxiom{\Visit{c}{\emptyset}}
\end{array}
\]
\end{small}

Let us ignore for a moment coaxioms and reason about the standard interpretations.
It is clear that, if we interpret the system inductively, we will only prove the judgement $\Visit{c}{\{c\}}$, because it is the only axiom and other rules do not depend on it.
In other words, the judgement $\Visit{\node}{\nodeset}$, like other judgements on graphs,  cannot be defined inductively by structural recursion, since the structure is not well-founded.
In particular the problem are cycles, where the proof may be trapped, continuously unfolding the structure of the graph without ever reaching a base case.
\FD{Usual implementations of visits on graphs rely on imperative features and correct this issue by marking already visited nodes.
In this way, they avoid visiting twice the same node, actually breaking cycles. }

On the other hand, if we interpret the \EZ{meta-rule}s coinductively (excluding again the coaxioms), then we get the correct judgements $\Visit{a}{\{a,b\}}$ and $\Visit{b}{\{a,b\}}$, but we also get the wrong judgements $\Visit{a}{\{a,b,c\}}$ and $\Visit{b}{\{a,b,c\}}$, as shown by the following derivations
\begin{small}
\[
\Rule{
	\Rule{
		\Rule{
			\vdots
		}{ \Visit{a}{\{a, b\}} }
	}{ \Visit{b}{\{a, b\}} }
}{ \Visit{a}{\{a, b\}} }
\BigSpace
\Rule{
	\Rule{
		\Rule{
			\vdots
		}{ \Visit{b}{\{a, b\}} }
	}{ \Visit{a}{\{a, b\}} }
}{ \Visit{b}{\{a, b\}} }
\BigSpace
\Rule{
	\Rule{
		\Rule{
			\vdots
		}{ \Visit{a}{\{a, b, c\}} }
	}{ \Visit{b}{\{a, b, c\}} }
}{ \Visit{a}{\{a, b, c\}} }
\BigSpace
\Rule{
	\Rule{
		\Rule{
			\vdots
		}{ \Visit{b}{\{a, b, c\}} }
	}{ \Visit{a}{\{a, b, c\}} }
}{ \Visit{b}{\{a, b, c\}} }
\]
\end{small}

We define a different interpretation, called \emph{interpretation generated by coaxioms} and denoted $\Generated{\is}{\coaxioms}$, which takes into account coaxioms in the following way.

\begin{enumerate}
\item First, we take the smallest closed superset of the set of coaxioms. In other words, we consider the inference system ${\Extended{\is}{\coaxioms}}$ obtained enriching $\is$ by judgements in $\coaxioms$ considered as axioms, and we take its inductive interpretation $\Ind{{\Extended{\is}{\coaxioms}}}$. 
\item Then, we take the largest consistent subset of $\Ind{{\Extended{\is}{\coaxioms}}}$. In other words, we take the coinductive interpretation of the inference system obtained from $\is$ by {keeping only rules with consequence in} $\Ind{{\Extended{\is}{\coaxioms}}}$, that is, we define
\[
\Generated{\is}{\coaxioms}=\CoInd{\Restricted{\is}{\Ind{{\Extended{\is}{\coaxioms}}}}}
\]
\end{enumerate}
where $\Restricted{\is}{S}$, with $\is$ inference system and $S\subseteq\universe$, denotes the inference system obtained from $\is$ by keeping only rules with 
consequence in $S$.

In the example, since the {power}-set is finite, every monotone function is continuous, hence we can compute fixed points iteratively.
Therefore, in the first phase, we obtain the following judgements (each line corresponds to an iteration of the inference operator): 
\begin{small}
\begin{quote}
$\Visit{a}{\emptyset}$, $\Visit{b}{\emptyset}$, $\Visit{c}{\emptyset}$, $\Visit{c}{\{c\}}$\\
$\Visit{a}{\emptyset}$, $\Visit{b}{\emptyset}$, $\Visit{c}{\emptyset}$, $\Visit{c}{\{c\}}$, $\Visit{a}{\{a\}}$, $\Visit{b}{\{b\}}$\\
$\Visit{a}{\emptyset}$, $\Visit{b}{\emptyset}$, $\Visit{c}{\emptyset}$, $\Visit{c}{\{c\}}$, $\Visit{a}{\{a\}}$, $\Visit{b}{\{b\}}$, $\Visit{a}{\{a,b\}},\Visit{b}{\{a,b\}}$
\end{quote}
\end{small}
The last set is closed, hence it is $\Ind{{\Extended{\is}{\coaxioms}}}$.

In the second phase, each iteration of the inference operator removes judgements which cannot be inferred from the previous step, that is, we get:
\begin{small}
\begin{quote}
$\Visit{c}{\{c\}}$, $\Visit{a}{\{a\}}$, $\Visit{b}{\{b\}}$, $\Visit{a}{\{a,b\}},\Visit{b}{\{a,b\}}$\\
$\Visit{c}{\{c\}}$, $\Visit{a}{\{a,b\}}$, $\Visit{b}{\{a,b\}}$
\end{quote}
\end{small}
This last set is consistent, hence it is $\Generated{\is}{\coaxioms}$, and it is indeed the expected result.

In terms of proof trees, judgements in $\Generated{\is}{\coaxioms}$ are those which have an arbitrary (well-founded or not) proof tree $t$ in the inference system $\is$, whose nodes all have a well-founded proof tree in $\Extended{\is}{\coaxioms}$. Note that for nodes in $t$ which are roots of a well-founded subtree this always holds (a well-founded proof tree in $\is$ is a well-founded proof tree in $\Extended{\is}{\coaxioms}$ as well), hence the condition is only significant for nodes which are roots of an infinite path in the proof tree.

For instance, in the example,  the judgement $\Visit{a}{\{a,b\}}$ has an infinite proof tree in $\is$ where each node has a finite proof tree in $\Extended{\is}{\coaxioms}$, as shown below.
\begin{small}
\[
\Rule{
	\Rule{
		\Rule{
			\vdots
		}{ \Visit{a}{\{a, b\}} }
	}{ \Visit{b}{\{a, b\}} }
}{ \Visit{a}{\{a, b\}} }
\BigSpace
\Rule{
	\Rule{
		\Rule{ }{ \Visit{a}{\emptyset} }
	}{ \Visit{b}{\{b\}} }
}{ \Visit{a}{\{a, b\}} }
\BigSpace
\Rule{
	\Rule{
		\Rule{ }{ \Visit{b}{\emptyset} }
	}{ \Visit{a}{\{a\}} }
}{ \Visit{b}{\{a, b\}} }
\]
\end{small}

Moreover, there is another important property
which will be proved in \refToSection{coaxioms-trees}: if a judgement belongs to $\Generated{\is}{\coaxioms}$, then,
for all $n\geq 0$, it has a well-founded proof tree in the inference system ${\Extended{\is}{\coaxioms}}$ where coaxioms can only be used at depth greater 
than $n$.

For instance, in the example, it is easy to see that, for any $n$, we can obtain a finite proof tree for the judgement $\Visit{a}{\{a,b\}}$ in $\Extended{\is}{\coaxioms}$ where coaxioms are used at depth greater than $n$, as shown below.
\begin{small}
\[
\Rule{
	\Rule{
		\Rule{ }{ \Visit{a}{\emptyset} }
	}{ \Visit{b}{\{b\}} }
}{ \Visit{a}{\{a, b\}} }
\BigSpace
\Rule{
	\Rule{
		\Rule{
			\Rule{ }{ \Visit{b}{\emptyset} }
		}{ \Visit{a}{\{a\}} }
	}{ \Visit{b}{\{a, b\}} }
}{ \Visit{a}{\{a, b\}} }
\BigSpace
\Rule{
	\Rule{
		\Rule{
			\Rule{
				\Rule{ }{ \Visit{a}{\emptyset} }
			}{ \Visit{b}{\{b\}} }
		}{ \Visit{a}{\{a, b\}} }
	}{ \Visit{b}{\{a, b\}} }
}{ \Visit{a}{\{a, b\}} }
\BigSpace
\ldots
\]
\end{small}

This last property motivates the name ``coaxioms''. Indeed, dually to axioms, which can be used in the proof tree at every depth, including $0$, coaxioms can only be used ``at an infinite depth'' in the proof tree. {Therefore, coaxioms filter out undesired infinite proof trees;
in other words, they bound from above the greatest fixed point corresponding to the semantics of the generalized inference system.

As {a} second example, we consider the definition of the \textit{first} sets in a grammar. Let us represent a context-free grammar by its set of terminals $T$, its set of non-terminals $N$, and all the productions $\produzioneinline{\nonterminal}{\beta_1\mid\ldots\mid\beta_n}$ with left-hand side $\nonterminal$, for each non-terminal $\nonterminal$.
Recall that, for each $\alpha\in(T\cup N)^{+}$, we can define the set
$\first{\alpha} = \{ \simb \mid \simb\in T, \alpha{\rightarrow^\star}\simb\beta\}$.
Informally, $\first{\alpha}$ is the set of the initial terminal symbols of the strings which can be derived from a string $\alpha$ in {$0$} or more steps.

We defines the judgement $\First{\alpha}{\firstset}$ \FD{\EZ{by} the following inference system with coaxioms, where} $\firstset\subseteq T$. 
\begin{small}
\[
\begin{array}{c}
\Rule{}{ \First{\simb\alpha}{\{\sigma\}} }\sigma\in T 
\BigSpace
\Rule{
	\First{\nonterminal}{\firstset}
}{ \First{\nonterminal\alpha}{\firstset} } 
\begin{array}{l}
\nonterminal\in N\\
\nonterminal{\not\rightarrow^\star}\epsilon
\end{array}
\BigSpace
\Rule{
	\First{\nonterminal}{\firstset} 
	\Space
	\First{\alpha}{\firstset'}
}{ \First{\nonterminal\alpha}{\firstset\cup\firstset'} }
\begin{array}{l}
\nonterminal\in N\\
\nonterminal{\rightarrow^\star}\epsilon
\end{array}
\\[4ex]
\Rule{}{ \First{\epsilon}{\emptyset} }
\BigSpace
\Rule{
	\First{\beta_1}{\firstset_1}
	\Space \ldots \Space
	\First{\beta_n}{\firstset_n}
}{ \First{\nonterminal}{\firstset_1\cup\ldots\cup\firstset_n} }
\produzioneinline{\nonterminal}{\beta_1\mid\ldots\mid\beta_n}
\BigSpace
\CoAxiom{ \First{\nonterminal}{\emptyset} }\nonterminal\in N
\end{array}
\]
\end{small}

The rules of the inference system correspond to the natural recursive definition of \textit{first}. 
Note, in particular, that in a string of shape $\nonterminal\alpha$, if the non-terminal $\nonterminal$ is \emph{nullable}, that is, we can derive from it the empty string, then the \textit{first} set for $\nonterminal\alpha$ should also include the \textit{first} set for $\alpha$.  

As in the previous example on graphs, the problem with this recursive definition is that, since the non-terminals in a grammar can mutually refer to each other, the function defined by the inductive interpretation can be undefined, since it may never reach a base case.
That is, a naive top-down implementation might not terminate. 
For this reason, \textit{first} sets are typically computed by an imperative bottom-up algorithm, or the top-down implementation is corrected by marking already encountered non-terminals, analogously to what is done for visiting graphs.
Again as in the previous example, the coinductive interpretation may fail to be a function, whereas, with the coaxioms, we get the expected result.

Let us now consider some examples of judgements concerning lists. 
We consider arbitrary (finite or infinite) lists of integers {and} denote {by} $\LInfSet$ the set of such lists.
We first consider the judgement $\maxElem{l}{x}$, with $l \in \LInfSet$ and $x \in \Z$,  stating that $x$ is the maximum element that {occurs in}  $l$.
This judgement has a natural definition by structural recursion we have discussed in \refToSection{is} where we have shown that neither inductive nor coinductive interpretations are able to capture the expected semantics.
Therefore in the following definition we have added coaxioms to the inference system from \refToSection{is} in order to restrict the coinductive interpretation{.}
\[
\Rule{}{ \maxElem{\Cons{x}{\EList}}{x} }
\BigSpace
\Rule{ \maxElem{l}{y} }{ \maxElem{\Cons{x}{l}}{z} } z = \max\{x, y\}
\BigSpace
\CoAxiom{ \maxElem{\Cons{x}{l}}{x} }
\]
Recall that the problem with the coinductive interpretation is that it accepts all judgements $\maxElem{l}{x}$ where $x$ is an upper bound of $l$, even if it does not {occur} in $l$.
The coaxiom, thanks to the way it is used, imposes that $\maxElem{l}{x}$ may hold only if $x$ appears somewhere in the list, hence undesired proofs are filtered out.

A similar example is given by the judgement $\elems{l}{\xs}$ where $l \in \LInfSet$ and $\xs \subseteq \Z$, stating that $\xs$ is the carrier of the list $l$, that is, the set of all elements appearing in $l$.
This judgement can be defined by structural recursion using coaxioms as follows{:}
\[
\Rule{}{ \elems{\EList}{\emptyset} }
\BigSpace
\Rule{ \elems{l}{\xs} }{ \elems{\Cons{x}{l}}{\{x\} \cup \xs} }
\BigSpace
\CoAxiom{ \elems{l}{\emptyset} }
\]
If we ignore the coaxiom {and} interpret the system coinductively{, then} we can prove $\elems{l}{\xs}$ for any {superset} $\xs$ of the carrier of $l$ if $l$ is infinite.
The coaxioms again {allow} us to filter out undesired derivations.
For instance, for $l$ the infinite list of {1s}, any judgement $\elems{l}{\xs}$ with $1 \in\xs$ can be derived. 
Indeed, for any such judgement we can construct an infinite proof tree which is a chain of applications of the last \EZ{meta-rule}.
With the coaxioms, we only consider the infinite trees where the node $\elems{l}{\xs}$ has a finite proof tree in the inference system enriched by the coaxioms. 
This is only true for $\xs=\{1\}$. 

We consider now a slight variation {of the} examples on lists from \refToSection{is} $\allPosPred{l}$ and $\memberPred{x}{l}$.
Set $\Bool = \{\True, \False\}$, we would like to define through an inference system the characteristic functions of those {two} predicates, that is, judgements $\member{x}{l}{b}$ and $\allPos{l}{b}$ with $b \in \Bool$ such that 
\begin{itemize}
\item $\member{x}{l}{\True}$ holds iff $\memberPred{x}{l}$ holds, and otherwise $\member{x}{l}{\False}$ holds 
\item $\allPos{l}{\True}$ holds iff $\allPosPred{l}$ holds, and otherwise $\allPos{l}{\False}$ holds
\end{itemize}
We can define these judgements by means of the following inference systems with coaxioms
\begin{small}
\[
\begin{array}{l}
\Rule{}{ \member{x}{\Cons{x}{l}}{\True} }
\BigSpace
\Rule{ \member{x}{l}{b} }{ \member{x}{\Cons{y}{l}}{b} } x \ne y 
\BigSpace
\CoAxiom{ \member{x}{l}{\False} }
\\[4ex]
\Rule{}{ \allPos{\EList}{\True} }
\BigSpace
\Rule{}{ \allPos{\Cons{x}{l}}{\False} } x\le 0
\BigSpace
\Rule{ \allPos{l}{b} }{ \allPos{\Cons{x}{l}}{b} }x>0
\BigSpace
\CoAxiom{ \allPos{l}{\True} }
\end{array}
\]
\end{small}
In these definitions coaxioms are essential, indeed without coaxioms for an infinite list $l$ we can derive the judgements for any $b \in \Bool$.
For instance, if $l$ is the infinite list of 1s, hence $l = \Cons{1}{l}$, the following are valid infinite proofs, obtained repeatedly applying the only rule with {non-}empty premises
\begin{small}
\[
\Rule{
	\Rule{
		\vdots
	}{ \member{2}{l}{\True} }
}{\member{2}{l}{\True} }
\BigSpace
\Rule{
	\Rule{
		\vdots
	}{ \member{2}{l}{\False} }
}{\member{2}{l}{\False} }
\BigSpace
\Rule{
	\Rule{
		\vdots
	}{ \allPos{l}{\True} }
}{ \allPos{l}{\True} }
\BigSpace
\Rule{
	\Rule{
		\vdots
	}{ \allPos{l}{\False} }
}{ \allPos{l}{\False} }
\]
\end{small}
In the interpretation generated by coaxioms, only the second and the third proofs are valid, since their nodes  are derivable starting from coaxioms, while this fact is not true for the others derivations.

\section{Fixed point semantics for coaxioms} \label{sect:coaxioms-model}

In this section we will show that the semantics of an inference system with coaxioms is indeed a fixed point of the inference operator.
Together with this characterization as fixed point we will also get a proof principle which will be a generalization of the standard coinduction principle.

\subsection{Closures and kernels} \label{sect:ck}
In this part of the section we will develop yet a little bit of lattice theory, in order to provide the theoretical background for defining the fixed point semantics of an inference system with coaxioms.
Hence in this part of the section $\Pair{\lattice}{\order}$ will be a complete lattice (see \refToDefinition{complete-lattice}) and $\fun{\function}{\lattice}{\lattice}$ a monotone function (see \refToDefinition{monotone-fun}) defined on it.
We start introducing some notions which are slight generalizations of notion that can be found in \cite{AbramskiJung94, Nation98}.

\begin{definition} \label{def:ck-system}
Let $\Pair{\lattice}{\order}$  be a complete lattice. Then 
\begin{enumerate}
\item a subset $\CSys \subseteq \lattice$ is a \emph{closure system} if for any subset $X \subseteq \CSys$, $\glb X \in \CSys$
\item a subset $\KSys\subseteq \lattice$ is a \emph{kernel system} if for any subset $X \subseteq \KSys$, $\lub X \in \KSys$
\end{enumerate}
\end{definition}

Note that with the usual convention that $\lub \emptyset = \bot$ and $\glb \emptyset = \top$, we have that for all closure systems $\CSys \subseteq \lattice$, $\top \in \CSys$, and for {all} kernel system{s} $\KSys \subseteq \lattice$, $\bot \in \KSys$. 

This definition provides a general order-theoretic account of a kind of structures that are very frequent in mathematics, in particular considering the complete lattice induced by the power-set functor.
For instance, as we have already shown in \refToSection{lattice}, given a group $G$, the set $\Sub{G}$ of all subgroups of $G$ is closed under arbitrary intersections, that is, it is closed under the meet operation.
Hence from the above definition $\Sub{G}$ is a closure system in the complete lattice $\Pair{\wp(G)}{\subseteq}$. 
This fact again holds for any algebraic structure. 
Another example comes from topology, indeed, given a topological space $\Pair{X}{\tau}$, by definition $\tau \subseteq \wp(X)$ and is closed under arbitrary unions, hence $\tau$ is a kernel system with respect to the complete lattice $\Pair{\wp(X)}{\subseteq}$. 
Moreover the set of closed {sets} in the topological space $\Pair{X}{\tau}$, that is, the set $\{X\setminus A \mid A \in \tau\}$, is closed under arbitrary intersections, hence it is a closure system.
Actually this is a general fact: if $\KSys \subseteq \wp(X)$ is a kernel system, then $\{X\setminus A \mid A \in \KSys\}$ is a closure system.
Also the converse is true.

Rephrasing \refToProposition{glb-lub-pre-post}  using this terminology we get that  for any monotone function $\fun{\function}{\lattice}{\lattice}$
\begin{itemize}
\item $\Pre{\function}$ is a closure system
\item $\Post{\function}$ is a kernel system
\end{itemize}
This observation provides us with a canonical way to associate a closure and a kernel system to a monotone function.
Let us introduce another notion{.} 

\begin{definition} \label{def:ck-op}
Let $\Pair{\lattice}{\order}$ be a complete lattice. Then 
\begin{enumerate}
\item A monotone function $\fun{\closure}{\lattice}{\lattice}$ is a \emph{closure operator}  if it satisfies the following conditions{:}
\begin{itemize}
\item for all $x \in \lattice$, $x \order \closure(x)$
\item for all $x \in \lattice$, $\closure(\closure(x)) = x$
\end{itemize}
\item A monotone function $\fun{\ker}{\lattice}{\lattice}$ is a \emph{kernel operator}  if it satisfies the following conditions{:}
\begin{itemize}
\item for all $x \in \lattice$, $\ker(x) \order x$
\item for all $x \in \lattice$, $\ker(\ker(x)) = x$
\end{itemize}
\end{enumerate}
\end{definition}

Note that since a closure operator $\fun{\closure}{\lattice}{\lattice}$  is a monotone function, we can associate with it both a closure and a kernel system, $\Pre{\closure}$ and $\Post{\closure}$. 
However, by the first condition of the definition of closure operator, we get that $\Post{\closure} = \lattice$, hence it is  not interesting, and $\Pre{\closure}=\Fix{\closure}$. 
Dually for a kernel operator $\fun{\ker}{\lattice}{\lattice}$, only $\Post{\ker} = \Fix{\ker}$ is interesting, because $\Pre{\ker} = \lattice$. 
Therefore we can say that every closure operator naturally induces a closure system   and every kernel operator naturally induces a kernel system. 

The next result shows how we can build, in a canonical way, from a closure/kernel system a closure/kernel operator.

\begin{theorem} \label{theo:ck-sys-op} 
Let $\Pair{\lattice}{\order}$ be a complete lattice.  Then 
\begin{enumerate}
\item given a closure system $\CSys \subseteq \lattice$ the function 
\[
\closure_\CSys (x) = \glb \{y \in \CSys \mid x \order y\}
\]
is a closure operator such that $\Fix{\closure_\CSys} = \CSys$
\item given a kernel system $\KSys \subseteq \lattice$ the function 
\[
\ker_\KSys (x) = \lub \{y \in \KSys \mid y \order x\}
\]
is a kernel operator such that $\Fix{\ker_\KSys} = \KSys$
\end{enumerate}
\end{theorem}
\begin{proof}
We prove only point 1, {the} proof for the other point is symmetric. \\
We first prove that $\closure_\CSys$ is monotone. 
Consider $x, y \in \lattice$ such that $x\order y$, hence $\{z \in {\CSys} \mid y \order z\}\subseteq \{z \in \CSys \mid x \order z\}$, thus, thanks to a property of $\glb$ observed in \refToSection{poset}, $\closure_\CSys (x) \order \closure_\CSys(y)$. \\
The fact that $x \order \closure_\CSys(x)$ for all $x \in \lattice$ follows from the fact that $x$ is a lower bound of the set $\{y \in \CSys \mid x \order y \}$.\\
Finally note that by definition for all $x \in \lattice$, $\closure_\CSys(x) \in \CSys$, hence in order to show that $\closure_\CSys(\closure_\CSys(x)) = \closure_\CSys(x)$ {it} is enough to show that{,} for all $z \in \CSys$, $\closure_\CSys(z) = z$, namely, $\CSys \subseteq \Fix{\closure_\CSys}$.
So consider $z \in \CSys$, we have already shown that $z \order \closure_\CSys(z)$, thus we have only to show  the other inequality.
Since $z\in\CSys$,  $z \in \{y \in \CSys \mid z \order y\}$, and this implies that $\closure_\CSys(z) \order z$.

This shows that $\closure_\CSys$ is a closure operator.
Actually we have also proved that $\CSys \subseteq \Fix{\closure_\CSys}$.
Therefore to conclude the proof it remains to show that $\Fix{\closure_\CSys} \subseteq  \CSys$, but this is trivial, since if $z = \closure_\CSys(z)$, then $z \in \CSys$ by definition. 
\end{proof}

The above theorem, considered for instance for closure systems, states that each closure system induces a closure operator having as (pre-)fixed points exactly the elements in the closure system. 
Actually we can say even more: each closure system induces a unique closure operator, that is, each closure operator is uniquely determined by its (pre-)fixed points.

\begin{theorem} \label{theo:ck-op-unique} 
Let $\Pair{\lattice}{\order}$ be a complete lattice. Then 
\begin{itemize}
\item if $\fun{\closure}{\lattice}{\lattice}$ is a closure operator then $\closure_{\Fix{\closure}} = \closure$
\item if $\fun{\ker}{\lattice}{\lattice}$ is a kernel operator, then $\ker_{\Fix{\ker}} = \ker$.
\end{itemize}
\end{theorem}
\begin{proof}
We prove only point 1, the proof for point 2 is symmetric.\\
We have to show that $\closure(x) = \closure_{\Fix{\closure}}(x)$ for all $x \in \lattice$.
By definition $\closure_{\Fix{\closure}} = \glb A$ with $A=\{y \in \Fix{\closure} \mid x\order y\}$, hence, since $x \order \closure(x)$, $\closure(x) \in A$.
In order to conclude the proof we have to show that $\closure(x)$ is the least element of $A$.
To this aim, consider $y = \closure(y) \in A$ and prove that it is above $\closure(x)$.
Note that $x\order y$, hence, by monotonicity of $\closure$, $\closure(x) \order \closure(y) = y$, as needed.
\end{proof}

In other words the above theorem tells us that to define a closure or kernel operator {it} is enough to specify a closure or a kernel system.
Therefore{,} for instance, the closure system $\Sub{G}$, where $G$ is a group, induces the closure operator $\fun{\GroupGen{-}}{\wp(G)}{\wp(G)}$, that computes for any set $X \subseteq G$  the subgroup generated by $X$.
For a topological space $\Pair{X}{\tau}$ we have that the topology $\tau$ induces a kernel operator that, for any set $A\subseteq X$, computes {its} interior, and the set of closed sets $\{X\setminus A \mid A \in \tau\}$ induces {the} topological closure operator.

\subsection{The bounded fixed point} \label{sect:bfp}
Let us now consider a monotone function $\fun{\function}{\lattice}{\lattice}$.
As we have seen, we can associate with $\function$ both a closure and a kernel system, $\Pre{\function}$ and $\Post{\function}$ respectively.
Thanks to \refToTheorem{ck-sys-op} and \refToTheorem{ck-op-unique} we know that these systems induce a unique closure and kernel operator respectively, defined below
\begin{align*}
\closure_\function(x) = \closure_{\Pre{\function}} &= \glb \{y \in \Pre{\function} \mid x \order y\}\\
\ker_\function(x) = \ker_{\Post{\function}} 	         &= \lub \{y \in \Post{\function} \mid y \order x \}
\end{align*}
We call $\closure_\function$ the \emph{closure} of $\function$ and $\ker_\function$ the \emph{kernel} of $\function$.
Intuitively, $\closure_\function(x)$ is the best pre-fixed approximation of $x$ (the least pre-fixed point above $x$), while $\ker_\function(x)$ is the best post-fixed approximation of $x$ (the greatest post-fixed point below $x$).
In this part of the section we will study some properties of these operators related to fixed points constructions.

First of all we note that from the definitions of the closure and the kernel of $\function$ we can immediately derive a generalization of both the induction and the coinduction principles described in \refToSection{KT}.
Given $\coaxioms, \bound \in \lattice$, for all $x \in \lattice$ we have 
\begin{description}
\item[\IndPrinciple] if $\function(x) \order x$ ($x$ pre-fixed) and $\coaxioms \order x$, then $\closure_\function(\coaxioms) \order x$
\item[\CoIndPrinciple] if $x \order \function(x)$ ($x$ post-fixed) and $x \order \bound$, then $x \order \ker_\function(\bound)$
\end{description}
These two principles are a generalization of standard induction and coinduction principles, because we can retrieve them through particular choices for $\coaxioms$ and $\bound$.
Indeed, if $\coaxioms = \bot$, the condition  $\coaxioms \order x$ is trivially always true, and we have $\closure_\function(\bot) = \glb \Pre{\function} = \lfp \function$ by Knaster-Tarski (\refToTheorem{KT}), hence \IndPrinciple allows us to conclude $\lfp\function \order x$ like standard induction requiring the same hypothesis.
Dually if $\bound = \top$, again the condition $x\order \bound$ is trivially always true, and $\ker_\function(\top) \lub \Post{\function} = \gfp \function$ by Knaster-Tarski, hence \CoIndPrinciple allows us to conclude $x \order \gfp \function$ like standard coinduction requiring the same hypothesis.

We now prove a result ensuring us that under suitable hypotheses we can use the closure and the kernel of a monotone function to build {a} fixed point of that function.

\begin{proposition}\label{prop:ck-fp}
Let $\coaxioms, \bound \in \lattice$.  Then 
\begin{enumerate}
\item if $\bound$ is a pre-fixed point, then $\ker_\function(\bound)$ is a fixed point
\item if $\coaxioms$ is  post-fixed point, then $\closure_\function(\coaxioms)$ is a fixed point
\end{enumerate}
\end{proposition}
\begin{proof}
We will prove only point 1, the proof for point 2 is symmetric.\\
As we noticed in \refToSection{lattice} $\LowSet{\bound}$ is a complete lattice; moreover by \refToProposition{monotone-restrict} the function $\fun{\function}{\LowSet{\bound}}{\LowSet{\bound}}$ is well-defined and monotone, since $\bound$ is a pre-fixed point.
Therefore $\ker_\function(\bound)$ is the join of all post-fixed point{s} of $\function$ in the complete lattice $\LowSet{\bound}$, hence by \refToTheorem{KT} it is a fixed point.
\end{proof}

Therefore we now know that if $\bound$ is pre-fixed $\ker_\function(\bound)$ is the greatest fixed point below $\bound$, and, if $\coaxioms$ is post-fixed, then $\closure_\function(\coaxioms)$ is the least fixed point above $\coaxioms$.

We are now able to define the \emph{bounded fixed point}.

\begin{definition}[Bounded fixed point] \label{def:bfp}
Let $\coaxioms \in \lattice$.
The \emph{bounded fixed point of $\function$ generated by $\coaxioms$}, denoted by $\Generated{\function}{\coaxioms}$,  is the greatest fixed point of $\function$ below the closure of $\coaxioms$, that is, $\Generated{\function}{\coaxioms} = \ker_\function(\closure_\function(\coaxioms))$.
\end{definition}

The bounded fixed point is well-defined since, thanks to \refToProposition{ck-fp}, there exists the greatest fixed point below $\bound$, provided that the bound $\bound$  is {a pre-fixed point.  
Since in general $\coaxioms$ might not be pre-fixed, we need to construct a pre-fixed point from $\coaxioms$. 
Note that the first step of this construction \emph{cannot} be expressed as the least fixed point of $\function$ on the complete lattice $\UpSet{\coaxioms}$, since in general $\function$ may fail to be well-defined (e.g., if $\function$ is the function which maps any element to $\bot \order  \coaxioms$ with $\coaxioms \ne \bot$). 
Indeed, $\closure_\function(\coaxioms)$ is \emph{not} a fixed point in general, but only a pre-fixed point: we need the two steps to obtain a fixed point.

Note also that the definition of bounded fixed point is asymmetric, that is, we take the greatest fixed point bounded from above by a least pre-fixed point, rather than the other way round. 
This is motivated by the intuition, explained in \refToSection{coaxioms}, that we essentially need a greatest fixed point, since we want to deal with {non-}well-founded structures, but we want to ``constrain'' in some way such greatest fixed point. 
Investigating the symmetric construction ($\closure_\function(\ker_\function(\coaxioms))$) is a matter of further work.\EZComm{ricordarsi di metterlo in future work}

The following proposition states some immediate properties of the bounded fixed point{.}
\begin{proposition} \label{prop:bfp-fun} \hspace*{\fill}
\begin{enumerate}
\item If $z \in \lattice$ is a fixed point of $\function$, then $\Generated{\function}{z}=z${.}
\item For all $\coaxioms_1, \coaxioms_2 \in \lattice$, if $\coaxioms_1\order \coaxioms_2$, then $\Generated{\function}{\coaxioms_1} \order \Generated{\function}{\coaxioms_2}${.}
\end{enumerate}
\end{proposition}
\begin{proof}
\hspace*{\fill}
\begin{enumerate}
\item If $z$ is a fixed point, {then} it is both pre-fixed and post-fixed, hence $\closure_\function(z)=z$ and $\ker_\function(z)=z$.
Thus we get that $\Generated{\function}{z} = \ker_\function(\closure_\function(z)) = \ker_\function(z) = z$.
\item Since both closure and kernel {operators} are monotone we get that 
$
\coaxioms_1\order \coaxioms_2 \Rightarrow 
\closure_\function(\coaxioms_1) \order \closure_\function(\coaxioms_2) \Rightarrow 
\ker_\function(\closure_\function(\coaxioms_1)) \order \ker_\function(\closure_\function(\coaxioms_2)) \Rightarrow 
\Generated{\function}{(\coaxioms_1} \order \Generated{\function}{\coaxioms_2}{.}
$
\end{enumerate}
\end{proof}
Therefore, by \refToProposition{ck-fp} we already know that $\Generated{\function}{\coaxioms}$ is a fixed point for any $\coaxioms \in \lattice$; the first point  of the above proposition says that all fixed point{s} of $\function$ can be generated as bounded fixed points.
In other words, considering $\Generated{\function}{-}$ as a function from $\lattice$ into itself, the first point implies that the range of this function is exactly $\Fix{\function}$.
Moreover the second point states that $\Generated{\function}{-}$ is a monotone function on $\lattice$.

An important fact is that bounded fixed points are a generalization of both least and  greatest fixed points, since they can be obtained by taking particular generators, as stated in the following proposition.

\begin{proposition} \label{prop:bfp-lfp-gfp} \hspace*{\fill}
\begin{enumerate}
\item $\Generated{\function}{\top}$ is the greatest fixed point of $\function$
\item $\Generated{\function}{\bot}$ is the least fixed point of $\function$
\end{enumerate}
\end{proposition}
\begin{proof}  \hspace*{\fill}
\begin{enumerate}
\item Note that $\closure_\function(\top)=\top$, since the only pre-fixed point above $\top$ is $\top$ itself, hence we get $\Generated{\function}{\top} = \ker_\function(\top) = \lub \Post{\function} = \gfp \function$, 
\item As already noted  $\closure_\function(\bot) = \lfp \function$, in particular $\closure_\function(\bot)$ is post-fixed, therefore we get $\Generated{\function}{\bot} = \ker_\function(\closure_\function(\bot)) = \closure_\function(\bot)$, namely it is the least fixed point of $\function${.}
\end{enumerate}
\end{proof} 

\EZ{An alternative proof for the above proposition is possible by exploiting \refToProposition{bfp-fun}.
We preferred to give the above} proof, since  this follows the asymmetry of the definition of the bounded fixed point. 
\FDComm{Dovrei dettagliare di più la prova alternativa?}\EZComm{no}

We now present a result that will be particularly useful to develop proof techniques {for} the bounded fixed point (see \refToSection{coaxioms-reasoning}).
Recall from \refToSection{continuity} that $\Iterate{\function}{x}$ denotes the set $\{ \function^n(x) \mid n \in \N\}$ where $\function^0 = \Id{\lattice}$ and $\function^{n+1} = \function \circ \function^n$. 
Moreover from \refToLemma{iterate-chain} we {know} that if $x$ is either pre-fixed or post-fixed, $\Iterate{\function}{x}$ is a chain (see \refToDefinition{chain}) and in particular {a descending chain if $x$ is pre-fixed.}

\begin{proposition} \label{prop:ker-iterate-bounds}
Let $\bound \in \lattice$ be a pre-fixed point of $\function$. Then 
\begin{enumerate}
\item for all $n \in \N$, $\ker_\function(\bound) = \ker_\function(\function^n(\bound))$
\item $\ker_\function(\bound) = \ker_\function(\glb \Iterate{\function}{\bound})$
\end{enumerate}
\end{proposition}
\begin{proof}
Note that since $\bound$ is pre-fixed, $\Iterate{\function}{\bound}$ is a descending chain, hence for all $n \in \N$ we have $\function^{n+1} (\bound) \order \function^n(\bound)$, that is, $\function^n(\bound)$ is a pre-fixed point for all $n \in \N$. 
\begin{enumerate}
\item We prove the statement by induction on $n$. If $n=0$ there is nothing to prove. 
Now, assume the thesis for $n$.
By definition $\ker_\function(\function^n(\bound))$ is a post-fixed point, hence $\ker_\function(\function^n(\bound)) \order \function(\ker_\function(\function^n(\bound)))$.
Since $\ker_\function$ is a kernel operator, by \refToDefinition{ck-op}, we have $\ker_\function(\function^n(\bound)) \order \function^n(\bound)$, hence by the monotonicity of $\function$, we get $\function(\ker_\function(\function^n(\bound))) \order \function^{n+1}(\bound)$.
Now by transitivity of $\order$ we get $\ker_\function(\function^n(\bound)) \order \function^{n+1}(\bound{)}$.
Therefore  by \CoIndPrinciple we conclude $\ker_\function(\function^n(\bound)) \order \ker_\function(\function^{n+1}(\bound))$.\\
On the other hand, since $\function^n(\bound)$ is pre-fixed, we have $\function^{n+1}(\bound) \order \function^n(\bound)$.
Thus by the monotonicity of $\ker_\function$ we get the other inequality, and this implies  $\ker_\function(\function^n(\bound)) = \ker_\function(\function^{n+1}(\bound))$.
Finally thanks to the inductive hypothesis we get the thesis.
\item By point 1 we have $\ker_\function(\bound) \order \function^n(\bound)$ for all $n \in \N$, hence $\ker_\function(\bound{)} \order \glb \Iterate{\function}{\bound}$.
Therefore by \CoIndPrinciple we get $\ker_\function(\bound) \order \ker_\function(\glb \Iterate{\function}{\bound})$.
On the other hand, we have $\glb \Iterate{\function}{\bound} \order \bound$, hence, by monotonicity of $\ker_\function$, we get the other inequality, and this implies the thesis.
\end{enumerate}
\end{proof}

Another way to read the above proposition is that, given a bound $\bound$ which is pre-fixed, we obtain the same greatest fixed point below $\bound$ if we take as bound
any element $\function^n(\bound)$ of the descending chain $\Iterate{\function}{\bound}$.
Moreover, \refToProposition{ker-iterate-bounds} says also that we obtain the same greatest fixed point  induced by $\bound$ if we take as bound  the greatest lower bound of that chain, namely, $\glb \Iterate{\function}{\bound}$.

We conclude this part of the section with a result that characterizes the closure and the kernel of respectively a post-fixed and a pre-fixed point using chains in analogy with the Kleene theorem (\refToTheorem{Kleene}).

\begin{proposition} \label{prop:ck-chain}
Let $\bound, \coaxioms \in \lattice$ be a pre-fixed and a post-fixed point respectively. Then 
\begin{enumerate}
\item if $\function$ is downward continuous, then $\ker_\function(\bound)  = \glb \Iterate{\function}{\bound}$
\item if $\function$ is upward continuous, then $\closure_\function(\coaxioms) = \lub \Iterate{\function}{\coaxioms}$
\end{enumerate}
\end{proposition}
\begin{proof}
We prove only point 1, the proof for point 2 is symmetric.\\
As we noticed in \refToSection{lattice} $\LowSet{\bound}$ is a complete lattice with top element $\bound$; 
moreover by \refToProposition{monotone-restrict} the function $\fun{\function}{\LowSet{\bound}}{\LowSet{\bound}}$ is well-defined and monotone, since $\bound$ is a pre-fixed point.
In this case it is also downward continuous, because  so is $\function$.
Therefore by \refToProposition{ck-fp}, $\ker_\function(\bound)$ is the greatest fixed point of $\function$ in the complete lattice $\LowSet{\bound}$, hence, since $\function$ is downward continuous, we get the thesis by \refToTheorem{Kleene}.
\end{proof}

Note that the above proposition requires an additional hypothesis on $\function$, that is required to be continuous, as happens for the Kleene theorem (\refToTheorem{Kleene}).
Under this assumption the above result immediately applies to the bounded fixed point, providing us with an iterative characterization of it, as the following corollary shows.

\begin{corollary} \label{cor:bfp-chain}
Let $\coaxioms \in \lattice$ and set $\bound = \closure_\function(\coaxioms)$.
If $\function$ is downward continuous, then $\Generated{\function}{\coaxioms} = \glb \Iterate{\function}{\bound}$.
\end{corollary}
\begin{proof}
By \refToDefinition{bfp} we have $\Generated{\function}{\coaxioms} = \ker_\function(\bound)$.
Since $\function$ is downward continuous, by \refToProposition{ck-chain}  we get the thesis.
\end{proof}

\subsection{Coaxioms as generators}
In this part of the section we come back to inference systems and we show that the interpretation generated by coaxioms of an inference system is indeed a fixed point of the inference operator.
In \refToSection{coaxioms} we have described two steps to construct $\Generated{\is}{\coaxioms}$, the interpretation generated by coaxioms $\coaxioms$ of an inference system $\is$.
\begin{enumerate}
\item First, we consider the inference system $\Extended{\is}{\coaxioms}$ obtained enriching $\is$ by judgements in $\coaxioms$ considered as axioms, and we take its inductive interpretation $\Ind{\Extended{\is}{\coaxioms}}$. 
\item Then, we take the coinductive interpretation of the inference system obtained from $\is$ by keeping only rules with consequence in $\Ind{\Extended{\is}{\coaxioms}}$, that is, we define
\[
\Generated{\is}{\coaxioms}=\CoInd{\Restricted{\is}{\Ind{\Extended{\is}{\coaxioms}}}}
\]
\end{enumerate}
The definition of bounded fixed point is the formulation of these two steps in the general setting of complete lattices. 
Indeed, the inference operator $\Op{\is}$ is a monotone function on the complete lattice $\Pair{\wp(\universe)}{\subseteq}$ obtained by taking set inclusion as order, 
and specifying the coaxioms $\coaxioms$ corresponds to fixing an arbitrary element of $\lattice$ as generator.
Then:
\begin{enumerate}
\item First, we construct the closure of $\coaxioms$, that is, the best closed approximation of $\coaxioms$. 
This closure plays the role of bound for the next step.
\item Then we construct the greatest fixed point below such bound.
\end{enumerate}
To show the correspondence in a precise way, we give an alternative and equivalent characterization of both the closure and the kernel of an element in $\lattice$.

\begin{proposition} \label{prop:ck-alt}
Let $\coaxioms, \bound \in \lattice$.
\begin{enumerate}
\item Consider the function $\fun{\Extended{\function}{\coaxioms}}{\lattice}{\lattice}$ defined by $\Extended{\function}{\coaxioms}(x) = \function(x) \join \coaxioms$.
Then $\closure_\function(\coaxioms) = \lfp \Extended{\function}{\coaxioms}${.}
\item Consider the function $\fun{\Restricted{\function}{\bound}}{\lattice}{\lattice}$ defined by $\Restricted{\function}{\bound}(x) = \function(x) \meet \bound$.
Then $\ker_\function(\bound) = \gfp \Restricted{\function}{\bound}$.
\end{enumerate}
\end{proposition}
\begin{proof}  
Note that both $\Extended{\function}{\coaxioms}$ and $\Restricted{\function}{\bound}$ are clearly monotone, hence the statements make sense, because \refToTheorem{KT} ensures the existence of both the least and the greatest fixed point of a monotone function.
\begin{enumerate}
\item By definition of fixed point we have that $\lfp \Extended{\function}{\coaxioms} = \Extended{\function}{\coaxioms}(\lfp \Extended{\function}{\coaxioms}) = \function (\lfp \Extended{\function}{\coaxioms} ) \join \coaxioms$, hence, by definition of $\join$, we get $\coaxioms \order \lfp \Extended{\function}{\coaxioms}$ and $\function(\lfp \Extended{\function}{\coaxioms}) \order \lfp \Extended{\function}{\coaxioms}$.
Therefore by \IndPrinciple we get $\closure_\function(\coaxioms) \order \lfp \Extended{\function}{\coaxioms}$.\\
On the other hand, by definition of $\closure_\function$, we have that $\function(\closure_\function(\coaxioms)) \order \closure_\function(\coaxioms)$ and $\coaxioms \order \closure_\function(\coaxioms)$. 
Therefore, by definition of $\join$, we get that $\Extended{\function}{\coaxioms}(\closure_\function(\coaxioms)) = \function(\closure_\function(\coaxioms)) \join \coaxioms \order \closure_\function(\coaxioms)$, hence by the induction principle we get $\lfp \Extended{\function}{\coaxioms} \order \closure_\function(\coaxioms)$.\\
Finally by antisymmetry we get the thesis.
\item \EZComm{questa prova non si pu\`o dire semplicemente che \`e simmetrica giusto?} \FDComm{quando l'ho scritta non pensavo fosse simmetrica, anche se c'è un dettaglio che è leggermente diverso, ma è di poco conto. Non so cosa sia meglio. }
By definition of fixed point we have that $\gfp \Restricted{\function}{\bound} = \Restricted{\function}{\bound} (\gfp \Restricted{\function}{\bound}) = \function(\gfp \Restricted{\function}{\bound}) \meet \bound$, hence, by definition of $\meet$, we get that $\gfp \Restricted{\function}{\bound} \order \bound$ and $\gfp \Restricted{\function}{\bound} \order \function(\gfp \Restricted{\function}{\bound})$.
Therefore by \CoIndPrinciple we get $\gfp \Restricted{\function}{\bound} \order \ker_\function(\bound)$. \\
On the other hand, by definition of $\ker_\function$, we have that $\ker_\function(\bound) \order \function (\ker_\function(\bound))$ and $\ker_\function(\bound) \order \bound$.
Therefore, by definition of $\meet$, we get that $\ker_\function(\bound) \order \function(\ker_\function(\bound)) \meet \bound = \Restricted{\function}{\bound}(\ker_\function(\bound))$, hence by the coinduction principle we get that $\ker_\function(\bound) \order \gfp \Restricted{\function}{\bound}$.\\
Finally by antisymmetry we get the thesis.
\end{enumerate}
\end{proof} 

By this alternative characterization we can formally state the correspondence with the two steps for defining $\Generated{\is}{\coaxioms}$. 

\begin{theorem} \label{theo:correspondence}
Let $\is$ be an inference system and $\coaxioms, \bound \in \wp(\universe)$,  then the following facts hold:
\begin{enumerate}
\item $\Extended{(\Op{\is})}{\coaxioms} = \Op{(\Extended{\is}{\coaxioms})}$ (so we can safely omit brackets)
\item $\Restricted{(\Op{\is})}{\bound} = \Op{(\Restricted{\is}{\bound})}$ (so we can safely omit brackets)
\item $\closure_\Op{\is}(\coaxioms) = \Ind{\Extended{\is}{\coaxioms}}$
\item $\ker_\Op{\is}(\bound) = \CoInd{\Restricted{\is}{\bound}}$
\end{enumerate}
\end{theorem}
\begin{proof}
\begin{enumerate}

\item We have to show that, for $S \subseteq \universe$, $\Extended{(\Op{\is})}{\coaxioms}(S) = \Op{(\Extended{\is}{\coaxioms})}(S)$. 
If $\cons \in \Extended{(\Op{\is})}{\coaxioms}(S)$, then either $\cons \in \coaxioms$ or $\cons \in \Op{\is}(S)$; 
in the former case there exists $\Rule{}{\cons} \in \Extended{\is}{\coaxioms}$ by definition of $\Extended{\is}{\coaxioms}$, 
in the latter there exists $\Rule{\prem}{\cons} \in \is$ such that $\prem \subseteq S$, and this  implies $\Rule{\prem}{\cons} \in \Extended{\is}{\coaxioms}$. 
Therefore in both cases $\cons \in \Op{(\Extended{\is}{\coaxioms})}(S)$. \\
Conversely, if $\cons \in \Op{(\Extended{\is}{\coaxioms})}(S)$, then there exists $\Rule{\prem}{\cons} \in \Extended{\is}{\coaxioms}$ such that $\prem \subseteq S$. 
By definition of $\Extended{\is}{\coaxioms}$, either $\Rule{\prem}{\cons} \in \is$ or $\cons \in \coaxioms$ and $\prem = \emptyset$, 
therefore in the former case $\cons \in \Op{\is}(S)$ and in the latter $\cons \in \coaxioms$, thus in both cases $\cons \in \Extended{(\Op{\is})}{\coaxioms}(S)$.  

\item We have to show that, for $S \subseteq \universe$, $\Restricted{(\Op{\is})}{\bound}(S) = \Op{(\Restricted{\is}{\bound})}(S)$. 
If $\cons \in \Restricted{(\Op{\is})}{\bound}(S)$, then we have $\cons \in \bound$ and $\cons \in \Op{\is}(S)$, hence there is $\myrule \in \is$ such that $\prem \subseteq S$;
therefore, by definition of $\Restricted{\is}{\bound}$, we get $\myrule \in \Restricted{\is}{\bound}$, and this implies that $\cons \in \Op{(\Restricted{\is}{\bound})}(S)$.\\
Conversely, if $\cons \in \Op{(\Restricted{\is}{\bound})}(S)$, then there exists $\myrule  \in \Restricted{\is}{\bound}$ such that $\prem \subseteq S$. 
By definition of $\Restricted{\is}{\bound}$, we have that $\myrule \in \is$ and $\cons \in \bound$,
therefore $\cons \in \Op{\is}(S)$ and  $\cons \in \bound$, thus $\cons \in \Restricted{(\Op{\is})}{\bound}(S)$.  

\item By \refToProposition{ck-alt} we get that $\closure_\Op{\is}(\coaxioms) = \lfp \Extended{\Op{\is}}{\coaxioms}$, that corresponds to the \emph{inductive interpretation} of $\Extended{\is}{\coaxioms}$, $\Ind{\Extended{\is}{\coaxioms}}$, by \refToTheorem{ind-eq} and point 1 of this theorem.

\item By \refToProposition{ck-alt} we get that $\ker_\Op{\is}(\bound) = \gfp \Restricted{\Op{\is}}{\bound}$, that corresponds to the \emph{coinductive interpretation} of $\Restricted{\is}{\bound}$, $\CoInd{\Restricted{\is}{\bound}}$, by \refToTheorem{coind-eq} and point 2 of this theorem{.}

\end{enumerate}
\end{proof}

Thanks to \refToTheorem{correspondence}, we can conclude that, given an inference system with coaxioms $\Pair{\is}{\coaxioms}$:
\[
\Generated{\is}{\coaxioms}=
\CoInd{\Restricted{\is}{\Ind{\Extended{\is}{\coaxioms}}}} = 
\ker_\Op{\is} (\closure_\Op{\is}(\coaxioms)) = 
\Generated{\Op{\is}}{\coaxioms}
\]
that is, the  interpretation generated by coaxioms $\coaxioms$ of the inference system $\is$ is exactly the bounded fixed point of $\Op{\is}$ generated by $\coaxioms$.

Finally applying \refToProposition{bfp-lfp-gfp} we get that the inductive and the coinductive interpretations of $\is$ are particular cases of the interpretation generated by coaxioms.
Indeed we get the inductive interpretation when $\coaxioms = \emptyset$ and we get the coinductive interpretation when $\coaxioms = \universe$, as shown below.
\begin{align*}
\Generated{\is}{\emptyset} &= \Generated{\Op{\is}}{\emptyset} = \lfp \Op{\is} = \Ind{\is} \\
\Generated{\is}{\universe} &= \Generated{\Op{\is}}{\universe} = \gfp \Op{\is} = \CoInd{\is}
\end{align*}

\section{Proof trees for coaxioms} \label{sect:coaxioms-trees}

In this section we formalize several proof-theoretic characterizations   for the semantics of inference systems with coaxioms, proving their equivalence with the fixed point semantics presented in \refToSection{coaxioms-model}.
All presented proof-theoretic characterizations are based on the notion of proof tree introduced in \refToDefinition{proof-tree}.

The first characterization is based on the following theorem which slightly generalizes  the standard result about the correspondence between the fixed point and the proof-theoretic semantics of inference systems in the coinductive case (see \refToTheorem{coind-eq}).

\begin{theorem} \label{theo:ker-trees}
Let $\is$ be an inference system and $\bound \subseteq \universe$ a closed set of judgements.
Then for all $\judg \in \universe$ the following are equivalent\EZ{:}
\begin{enumerate}
\item $\judg \in \ker_\Op{\is}(\bound)$
\item there exists a proof tree $t$ for $\judg$ in $\is$ such that each node of $t$ is (labelled) in $\bound$
\end{enumerate}
\end{theorem}
\begin{proof}
By \refToTheorem{correspondence} and \refToTheorem{coind-eq} we have that $\ker_\Op{\is}(\bound) = \gfp \Op{\Restricted{\is}{\bound}} = \CoInd{\Restricted{\is}{\bound}}$.
Hence $\judg \in \ker_\Op{\is}(\bound)$ iff there exists a proof tree $t$ for $\judg$ in $\Restricted{\is}{\bound}$.
By \refToDefinition{proof-tree} every node in $t$ is (labelled by) a consequence $\cons$ of a rule in $\Restricted{\is}{\bound}$, hence $\cons \in \bound$, by definition of $\Restricted{\is}{\bound}$. 
\end{proof}

As a particular case we get our first proof-theoretic characterization of $\Generated{\is}{\coaxioms}$.

\begin{corollary} \label{cor:proof-trees-1}
Let $\Pair{\is}{\coaxioms}$ be an inference system with coaxioms.
Then the following are equivalent
\begin{enumerate}
\item $\judg \in \Generated{\is}{\coaxioms}$
\item there exists a proof tree $t$ for $\judg$ in $\is$ such that each node of $t$ has a well-founded proof tree in $\Extended{\is}{\coaxioms}$
\end{enumerate}
\end{corollary}
\begin{proof}
We have that $\Generated{\is}{\coaxioms} = \ker_\Op{\is}(\Ind{\Extended{\is}{\coaxioms}})$, hence by \refToTheorem{ker-trees} we get that $\judg \in \Generated{\is}{\coaxioms}$ iff there is a proof tree $t$ for $\judg$ in $\is$ whose nodes are all in $\Ind{\Extended{\is}{\coaxioms}}$.
Therefore all nodes of $t$ have a well-founded proof tree in $\Extended{\is}{\coaxioms}$ by  \refToDefinition{ind-coind}.
\end{proof}

For the second proof-theoretic characterization, we need to define \emph{approximated proof trees} in an inference system with coaxioms.
In the definition below we write $t_\judg$ to denote an arbitrary tree rooted in $\judg$.

\begin{definition} \label{def:approx-trees}
Let $\Pair{\is}{\coaxioms}$ be an inference system with coaxioms, the sets $\T_n$ of \emph{approximated proof trees of level $n$ in $\Pair{\is}{\coaxioms}$}, for $n\in \N$, are inductively defined as follows:
\begin{center}
\begin{tabular}{ll}
$t\in\T_0$ & if $t$ well-founded proof tree in $\Extended{\is}{\coaxioms}$\\[2ex]
$\Rule{\T}{{\cons}} \in \T_{n+1}$ & if $\Rule{\prem}{\cons}\in\is$ and $\T{=}\{t_\judg \mid \judg \in \T_n \in \prem\}{\subseteq} \T_n$
\end{tabular}
\end{center}
\end{definition}
\FD{Note that an approximated proof tree is actually a proof tree, since the set $\T$ and the set $\prem$ in the inductive step are in bijection: elements $t_\judg$ in $\T$ are indexed by judgements in $\prem$, hence there is a surjective map from $\prem$ to $\T$; moreover, if $t_\judg = t_{\judg'}$, then $\judg = t_\judg(\EString) = t_{\judg'}(\EString) = t_{\judg'}$, hence this map is also injective. }

In other words,  an approximated proof tree of level $n$ in $\Pair{\is}{\coaxioms}$ is a well-founded proof tree in $\Extended{\is}{\coaxioms}$ where coaxioms can only be used at depth  $\geq n$.
Therefore if $t \in \T_n$ is an approximated proof tree of level $n$, \EZ{then,} for all $\alpha \in \dom(t)$ with $\Len{\alpha} < n$, $t(\alpha)$ is the consequence of a rule in $\is$, more precisely $\Rule{ \{t(\beta) \mid \beta \in \chl(\alpha) \} }{t(\alpha)} \in \is$.

Another simple property of approximated proof trees is state\EZ{d} in the following proposition\EZ{.}
\begin{proposition} \label{prop:approx-subtree}
If $t \in \T_n$, $\alpha \in \dom(t)$ and $\Len{\alpha} = k \le n$, then $t_\alpha \in \T_{n-k}$.
\end{proposition}
\begin{proof}
We proceed by induction on $\Len{\alpha}$.
If $\Len{\alpha} = 0$, then $\alpha= \EString$, hence $t_\EString = t \in \T_n$.
Assume $\Len{\alpha} = k+1$, hence $\alpha = \beta a$, hence $\beta \in \dom(t)$ and $\Len{\beta}  = k$.
Therefore by inductive hypothesis $t_\beta \in \T_{n-k}$, hence  $t_\alpha = {t_\beta}_a \in \dsub(t_\beta)$, and this implies, by \refToDefinition{approx-trees}, that $t_\alpha \in \T_{n-k-1}$.
\end{proof}

The following theorem states that approximated proof trees  of level $n$ correspond to the $n$-th element of the descending chain $\Iterate{\Op{\is}}{\bound}=\{ \IterOp{\is}{n}(\bound) \mid n \in \N\}$, with  $\bound=\closure_\Op{\is}(\coaxioms)=\Ind{\Extended{\is}{\coaxioms}}$.

\begin{theorem} \label{theo:approx-trees}
Let $\Pair{\is}{\coaxioms}$ be an inference system with coaxioms, and $\judg\in\universe$ a judgement. 
We have that, for all $n \in \N$\EZ{,} the following are equivalent\EZ{:}
\begin{enumerate}
\item $\judg\in \IterOp{\is}{n}(\closure_\Op{\is}(\coaxioms))$
\item $\judg$ has an approximated proof tree of level $n$ in $\Pair{\is}{\coaxioms}$
\end{enumerate}  
\end{theorem}
\begin{proof}
Let $\bound$ be $\closure_\Op{\is}(\coaxioms)$. 
We prove the thesis by induction on $n$.
\begin{description}

\item[Base] If $n=0$, then, by \refToTheorem{correspondence}, $\bound = \closure_\Op{\is}(\coaxioms)$ corresponds to the inductive interpretation of $\Extended{\is}{\coaxioms}$, hence the equivalence hold\EZ{s by} \refToDefinition{ind-coind}.

\item[Induction] We assume the equivalence for $n$ and prove it for $n+1$. 
We prove separately the two implications.
\begin{description}

\item[$1\Rightarrow 2$] If $\cons \in \IterOp{\is}{n+1}(\bound)$, then there exists $\myrule \in \is$ such that $\prem \subseteq \IterOp{\is}{n}(\bound)$. 
Hence, by inductive hypothesis, each judgement in $\prem$ has an approximated proof tree of level $n$, that is, for all $\judg \in \prem$ there is an approximated proof tree $t_\judg \in \T_n$ rooted in $\judg$. 
Set $\T = \{t_\judg \in \T_n \mid \judg \in \prem\}$.
Hence, $t = \Rule{\T}{\cons}$ is a proof tree for $\cons$, and by \refToDefinition{approx-trees}, $t\in \T_{n+1}$.

\item[$2\Rightarrow 1$] If $t \in \T_{n+1}$ is an approximated proof tree for $\cons \in \universe$, then, by definition, there exists $\myrule \in \is$ such that $t = \Rule{\T}{\cons}$, $\T=\{t_\judg \in \T_n \mid \judg \in \prem\}$, and $\T \subseteq \T_n$, where $t_\judg$ is a tree rooted in $\judg$.
By inductive hypothesis we have $\prem\subseteq \IterOp{\is}{n} (\bound)$, and, by definition  of $\Op{\is}$, this implies $\cons \in \IterOp{\is}{n+1}(\bound)$ as needed. 
\end{description}
\end{description}
\end{proof}

The second proof-theoretic characterization of the interpretation generated by coaxioms  is an immediate consequence of the above theorem.

\begin{corollary} \label{cor:tree2}
Let $\Pair{\is}{\coaxioms}$ be an inference system with coaxioms, and $\judg \in \universe$ a judgement. 
Then the following are equivalent:
\begin{enumerate}
\item $\judg \in \Generated{\is}{\coaxioms}$
\item there exists a proof tree $t$ for $\judg$ in $\is$ such that each node of $t$ has an approximated proof tree of level $n$ in $\Pair{\is}{\coaxioms}$, for all $n \in \N$.
\end{enumerate}
\end{corollary}
\begin{proof}
By \refToTheorem{correspondence}, \refToProposition{ker-iterate-bounds}, and \refToTheorem{ker-trees}, we get that,  for all $\judg \in \universe$, $\judg \in \Generated{\is}{\coaxioms}$ iff there exists a proof tree $t$ for $\judg$ in $\is$ such that each node $\judg'$ of $t$ is in $\bigcap \Iterate{\Op{\is}}{\bound}$ with $\bound = \closure_\Op{\is}(\coaxioms)$.  
By \refToTheorem{approx-trees}, $\judg' \in \bigcap \Iterate{\Op{\is}}{\bound}$ iff has an approximated proof tree of level $n$, for all $n \in \N$.
\end{proof}

If the hypotheses of \refToCorollary{bfp-chain} are satisfied, then we get a simpler equivalent proof-theoretic characterization.

\begin{corollary}\label{cor:tree3}
Let $\Pair{\is}{\coaxioms}$ be an inference system with coaxioms, and $j \in \universe$ a judgement.
\EZ{I}f $\Op{\is}$ is downward continuous, then the following are equivalent:
\begin{enumerate}
\item $j \in \Generated{\is}{\coaxioms}$
\item $\judg$ has an approximated proof tree of level $n$ in $\Pair{\is}{\coaxioms}$, for all $n \in \N$.
\end{enumerate}
\end{corollary}
\begin{proof}
Let $\bound$ be the set $\closure_\Op{\is}(\coaxioms)$. 
By \refToTheorem{correspondence} and \refToCorollary{bfp-chain}, we get that $\Generated{\is}{\coaxioms} = \bigcap \Iterate{\Op{\is}}{\bound}$, therefore the thesis follows immediately from \refToTheorem{approx-trees}.
\end{proof}

\EZComm{caratterizzazione nuova, capire come si possa usare come tecnica e se migliora le cose}
\FDComm{è il caso di fare un sottosezione qui?}
In order to define the last proof\EZ{-}theoretic characterization (\refToTheorem{approx-sequence}), we need to introduce  a richer structure on trees.
In particular we will define a partial order on $\Tree{\Labels}$ which is a slight \FD{relaxation} of the canonical order introduced in \cite{Courcelle83}.

Consider trees $t, t' \in \Tree{\Labels}$ we define 
\[
t \TOrder t' \Longleftrightarrow \dom(t) \subseteq \dom(t') \mbox{ and } \forall \alpha \in \dom(t).\ t(\alpha) = t'(\alpha)
\]
It is easy to see that $\TOrder$ is a partial order, \FD{actually it is function inclusion. }
Indeed, reflexivity and transitivity follow from the same properties of $\subseteq$ and $=$\EZ{, and a}antisymmetry can be proved noting that if $t\TOrder t'$ and $t' \TOrder t$ we have that $\dom(t) = \do(t')$ (by antisymmetry of $\subseteq$) and $t(\alpha) = t'(\alpha)$ for all $\alpha \in \dom(t)$, hence $t=t'$.

\FD{Intuitively, $t \TOrder t'$ means that $t$ can be obtained from $t'$ by pruning some branches.
Alternatively, considering trees as graphs, $t\TOrder t'$ means that $t$ is a subgraph of $t'$.
In any case $\TOrder$ \EZ{expresses} a very strong relation among trees, actually too strong for our aims, hence we need to relax it a little bit. }

\FD{We relax it by considering what we call its \emph{$n$-th approximation},} defined below.
Given a tree $t$, we denote by $\dom_n(t)$ the set $\{\alpha \in \dom(t) \mid \Len{\alpha} \le n\}$.
The $n$-th approximation of $\TOrder$, denoted by $\ApproxTOrder{n}$, is defined as follows:
\[
t \ApproxTOrder{n} t' \Longleftrightarrow 
\dom_n(t) \subseteq \dom_n(t') \mbox{ and } 
\forall \alpha \in \dom_n(t).\ t(\alpha) = t'(\alpha)
\]
Intuitively $\ApproxTOrder{n}$ is identical to $\TOrder$, but limited to nodes at level $\le n$.
We call it the $n$-th approximation of $\TOrder$ since $\ApproxTOrder{n}$ is coarser than $\TOrder$,namely, if $t\TOrder t'$ then $t \ApproxTOrder{n} t'$ for all $n \in \N$.
Actually we can say even more: $t \TOrder t'$ if and only if $t \ApproxTOrder{n} t'$ for all $n \in \N$.
Moreover if $t \ApproxTOrder{n} t'$ then for all $k \le n$ we have $t \ApproxTOrder{k} t'$, that is, $\ApproxTOrder{n}$ is a finer approximation than $\ApproxTOrder{k}$.
In the end note that $\ApproxTOrder{n}$ is reflexive and transitive, but it fails to be antisymmetric, because we compare only node until level $n$, hence we cannot conclude an equality between the whole trees.

We now state a result that is crucial for our proof\EZ{-}theoretic characterization.

\begin{theorem} \label{theo:limit-tree-sequence}
Let $(t_n)_{n \in \N}$ be a sequence of trees,
such that\EZ{,} for all $n \in \N$, $t_n \ApproxTOrder{n} t_{n+1}$. 
Then, there exists a tree $t$ such that $\forall n \in \N.\ t_n \ApproxTOrder{n} t$, and\EZ{,} for any other tree $t'$ such that $\forall n \in \N.\ t_n \ApproxTOrder{n} t'$\EZ{,} we have $t \TOrder t'$.
\end{theorem}
\begin{proof}
We define the function $\fun{t}{L}{\Labels}$ where $L = \bigcup_{n \in \N} \dom_n(t_n)$  and for all $\alpha \in L$, $t(\alpha) = t_{k(\alpha)}(\alpha)$, where $ k(\alpha) = \min D_\alpha$ with $D_\alpha = \{ n \in \N \mid \alpha \in \dom_n(t_n) \}$.
Note that $k(\alpha)$ is well-defined, because $D_\alpha \ne \emptyset$, since $\alpha \in L$ and, by construction of $L$, there is at least an index $n \in \N$ such that $\alpha \in \dom_n(t_n)$.
Moreover $L$ is a tree language, since if $\alpha a \in L$, then $\alpha a \in \dom_n(t_n)$ for some $n \in \N$, that is a tree language, hence $\alpha \in \dom_n(t_n) \subseteq L$.
Therefore $t$ is a tree.\\
Fix now $n \in \N$, we have to show that $t_n \ApproxTOrder{n} t$. 
By construction we have $\dom_n(t_n) \subseteq \dom_n(t)$, and if $\alpha \in \dom_n(t_n)$, then by construction $k(\alpha) \le n$.
Therefore we have that $t_{k(\alpha)} \ApproxTOrder{k(\alpha)} t_n$, hence $t_{k(\alpha)}(\alpha) = t_n(\alpha)$, thus $t(\alpha) = t_n(\alpha)$ and this implies $t_n \ApproxTOrder{n} t$. 

Consider now a tree $t'$ such that $\forall n \in \N.\ t_n \ApproxTOrder{n} t'$.
Therefore we have that for all $n \in \N$, $\dom_n(t_n) \subseteq \dom_n(t') \subseteq \dom(t')$, hence $\dom(t) \subseteq \dom(t')$.
Then, if $\alpha \in \dom(t)$, there is $n \in \N$ such that $\alpha \in \dom_n(t_n)$ and $t(\alpha) = t_n(\alpha)$.
Since $t_n \ApproxTOrder{n} t'$, we have that $t_n(\alpha) = t'(\alpha)$, hence $t(\alpha) = t'(\alpha)$ and this implies $t \TOrder t'$.
\end{proof}

\FD{It is easy to see that a tree $t$ having the property expressed in  the above theorem is unique.
Indeed, if $t$ and $t'$ have that property for a sequence $(t_n)_{n \in \N}$, then  we have  both $t \TOrder t'$ and $t' \TOrder t$, hence $t = t'$.
Therefore we denote such a tree by $\TLub_{n \in \N} t_n$. }

The above theorem ensures the existence of a sort of least upper bound of an ascending chain of trees: \FD{$\TLub_{n \in \N} t_n$  behaves like a least upper bound, but for approximations of a partial order. 
However, since $\ApproxTOrder{n}$ is an approximation of $\TOrder$, it can by shown that if $(t_n)_{n \in \N}$ is a chain with respect to $\TOrder$, then $\TLub_{n \in \N} t_n$ is indeed the least upper bound of the chain, as \EZ{stated} in the following corollary. } 

\begin{corollary}
Let $(t_n)_{n \in \N}$ be a sequence of trees,
such that for all $n \in \N$, $t_n \TOrder t_{n+1}$. 
Then, \FD{$\TLub_{n \in \N} t_n$ is the least upper bound of the sequence $(t_n)_{n \in \N}$. }
\end{corollary} 
\begin{proof}
\FD{Since $\ApproxTOrder{n}$ is an approximation of $\TOrder$ we have that $t_n \ApproxTOrder{n} t_{n+1}$ for all $n \in \N$. Setting $t = \TLub_{n \in \N} t_n$, by \refToTheorem{limit-tree-sequence}, we get $t_n \ApproxTOrder{n} t$ for all $n \in \N$. 
We have to show that $t$ is an upper bound of $(t_n)_{n \in \N}$, hence consider $\alpha \in \dom(t_n)$ and suppose $\Len{\alpha} = k$. 
We have two cases\EZ{:}
\begin{itemize}
\item if $k \le n$, then $\alpha \in \dom_n(t_n)$, hence $\alpha \in \dom_n(t) \subseteq \dom(t)$ and $t_n(\alpha) = t(\alpha)$
\item if $k > n$, then, since $t_n \TOrder t_k$, $\alpha \in \dom_k(t_k) \subseteq \dom_k(t) \subseteq \dom(t)$, and $t_n(\alpha) = t_k(\alpha) = t(\alpha)$
\end{itemize}
Therefore we get $t_n \TOrder t$. }

\FD{To show that $t$ is the least upper bound, consider  an upper bound $t'$, hence $t_n \TOrder t'$ for all $n  \in \N$, and this implies that $t_n \ApproxTOrder{n} t'$ for all $n \in \N$.
Therefore, by \refToTheorem{limit-tree-sequence}, we get $t \TOrder t'$. }
\end{proof}

We now consider the equivalence relations induced by each $\ApproxTOrder{n}$, defined as follows:
\[
t \ApproxEq{n} t' \Longleftrightarrow t \ApproxTOrder{n} t' \mbox{ and } t' \ApproxTOrder{n} t
\]
\EZ{or, more explicitly:}
\[
t \ApproxEq{n} t' \Longleftrightarrow \dom_n(t) = \dom_n(t') \mbox{ and } \forall \alpha \in \dom_n(t).\ t(\alpha) = t'(\alpha)
\]
These equivalence relations are an approximation of the equality relation, indeed $t = t'$ if and only if $t \ApproxEq{n} t'$ for all $n \in \N$. 

\FD{The relations $\ApproxTOrder{n}$ and $\ApproxEq{n}$ look very similar: they are both an approximation of another relation, they are both reflexive and transitive and they both do not care about levels higher than $n$.
However, the fact that $\ApproxEq{n}$ is an equivalence relation \EZ{makes} it different.
Indeed, if $t \ApproxEq{n} t'$, then the first $n$ levels of $t$ and $t'$ are forced to be equal, while\EZ{,} if $t \ApproxTOrder{n} t'$, then the first $n$ levels of $t'$ must contain also those of $t$, but can have also  additional branches.
In other words, with $\ApproxEq{n}$ we can change only the depth of the trees, while with $\ApproxTOrder{n}$ we can change both the depth and the breadth. }
\FDComm{tentativo di spiegazione delle differenze}\EZComm{ok}

As a consequence of \refToTheorem{limit-tree-sequence} we get the following theorem.

\begin{theorem} \label{theo:eq-tree-sequence}
Let $(t_n)_{n \in \N}$ be a sequence of trees,
such that, for all $n \in \N$, $t_n \ApproxEq{n} t_{n+1}$. 
Then, there exists a unique tree $t$ such that $\forall n \in \N.\ t_n \ApproxEq{n} t$.
\end{theorem}
\begin{proof}
By definition of $\ApproxEq{n}$ we have that $t_n \ApproxTOrder{n} t_{n+1}$ and $t_{n+1} \ApproxTOrder{n} t_n$ for all $n \in \N$. 
Therefore, by \refToTheorem{limit-tree-sequence}, we get that there is a tree $t = \TLub_{n \in \N} t_n$ such that, for all $n \in \N$, $t_n \ApproxTOrder{n} t$, hence we have only to prove that $t \ApproxTOrder{n} t_n$ \FD{and that $t$ is unique. }
Since we know that $\dom_n(t_n) \subseteq \dom_n(t)$ and for all $\alpha \in \dom_n(t_n)$, $t_n(\alpha) = t(\alpha)$, it is enough to show that $\dom_n(t) \subseteq \dom_n(t_n)$.
Thus, consider $\alpha \in \dom_n(t)$.
By construction of $t$ (see the proof of \refToTheorem{limit-tree-sequence}), there is an  index $k\in \N$ such that $\alpha \in \dom_{k}(t_k)$.
We have two cases:
\begin{itemize}
\item If $k\le n$, then by hypothesis $\dom_k(t_k) \subseteq\dom_k(t_n) \subseteq \dom_n(t_n)$, hence $\alpha \in \dom_n(t_n)$.
\item Otherwise, \EZ{that is,} if $n<k$, since $\alpha\in \dom_n(t)$, we have $\Len{\alpha} \le n<k$, hence $\alpha \in \dom_n(t_k) \subseteq \dom_n(t_n)$, because, by hypothesis $t_n \ApproxEq{n} t_k$.
\end{itemize}
Therefore we have $\dom_n(t) \subseteq \dom_n(t_n)$ as needed.

\EZ{To prove that $t$ is unique,} consider a tree $t'$ such that $t_n \ApproxEq{n} t'$ for all $n\in \N$, by transitivity we get $t \ApproxEq{n} t'$ for all $n \in \N$, and this implies $t=t'$.
\end{proof}

\FD{It is well known that trees carry a complete metric space structure \cite{ArnoldNivat80, Courcelle83} and, even if our notion of tree is more general than that adopted in these works, we can recover the same metric on our trees, using the equivalence relations introduced earlier.
The metric is defined as follows:
\[
d(t, t') = 2^{-h} \BigSpace h = \min\{n \in \N \mid t \not\ApproxEq{n} t'\}
\]
with assumptions $\min \emptyset = \infty$ and $w^{-\infty} = 0$. 
It is easy to see that a sequence $(t_n)_{n \in \N}$ such that $t_n \ApproxEq{n} t_{n+1}$, like that considered in \refToTheorem{eq-tree-sequence}, is a Cauchy sequence in the metric space; indeed $d(t_n, t_{n+1}) \le 2^{-n}$. 
Therefore such sequences converge also in the metric space, and the limit is the same.
However our notion of convergence seems to be more general: sequences like those considered in \refToTheorem{limit-tree-sequence} are not necessarily Cauchy sequences, but they admit a limit in our framework.
For instance consider the sequence $(t_n)_{n \in \N}$ of children injective trees labelled on $\N$ and rooted in $0$ is, defined\footnote{It is enough to provide a definition for the domain since trees are children injective.} by 
\[
\dom(t_0) = \{\EString\} \BigSpace \dom(t_{n+1}) = \dom(t_n) \cup  \{n\}
\]
It is easy to check that $t_n \ApproxTOrder{n} t_{n+1}$, hence, by \refToTheorem{limit-tree-sequence}, it converges to $\TLub_{n\in \N} t_n$.
However, it is not a Cauchy sequence, since $d(t_n, t_{n+1}) = 2^{-1}$ for all $n \in \N$, and $\TLub_{n \in \N} t_n$ is not a limit of the sequence in the metric space.
A deeper comparison between these relation\EZ{s} and the standard metric structure on trees will be matter of further work. }

We can now introduce the concept that will allow the last proof-theoretic characterization.

\begin{definition} \label{def:approx-sequence}
Let $\Pair{\is}{\coaxioms}$ be an inference system with coaxioms and $\judg \in \universe$ a judgement. Then\EZ{:}
\begin{enumerate}
\item An \emph{approximating proof sequence} for $\judg$ is a sequence of proof trees $(t_n)_{n \in \N}$ for $\judg$ such that $t_n \in \T_n$ and  $t_n \ApproxTOrder{n} t_{n+1}$ for all $n \in \N$.
\item A \emph{strong approximating proof sequence} for $\judg$ is a sequence of proof trees $(t_n)_{n \in \N}$ for $\judg$ such that $t_n \in \T_n$ and $t_n \ApproxEq{n} t_{n+1}$ for all $n \in \N$.
\end{enumerate}
\end{definition}

Obviously every strong approximating proof  sequence is also an approximating proof sequence.
Note also that all trees in these sequences are well-founded proof trees in $\Extended{\is}{\coaxioms}$.
Intuitively, both notions represent the growth of a proof for $\judg$ in $\is$ approximated using coaxioms.
The difference is that trees in an approximating proof sequence can grow both in depth \EZ{and} in breadth, while in a strong approximating proof sequence they can grow only in depth.
We now prove our last theorem.

\begin{theorem} \label{theo:approx-sequence}
Let $\Pair{\is}{\coaxioms}$ be an inference system with coaxioms and $\judg\in \universe$ a judgement. 
Then the following are equivalent
\begin{enumerate}
\item $\judg \in \Generated{\is}{\coaxioms}$
\item $\judg$ has a strong approximating proof sequence
\item $\judg$ has an approximating proof sequence
\end{enumerate}
\end{theorem}
\begin{proof} \hspace*{\fill}
\begin{description}

\item[$1\Rightarrow 2$] 
\FD{We define trees $t_{\judg, n}$ for $\judg \in \Generated{\is}{\coaxioms}$ and $n \in \N$ such that $t_{\judg, n}(\EString) = \judg$ by induction on $n$.
By \refToCorollary{proof-trees-1} we know that every judgement $\judg \in \Generated{\is}{\coaxioms}$ has a well-founded proof tree in $\Extended{\is}{\coaxioms}$,  that is, a proof tree in $\T_0$ rooted in $\judg$: we select one of these trees and call it $t_{\judg, 0}$.
Furthermore, since $\Generated{\is}{\coaxioms}$ is a post-fixed point, for any $\judg \in \Generated{\is}{\coaxioms}$ we can select a rule $\Rule{\prem_\judg}{\judg} \in \is$ with $\prem_\judg \subseteq \Generated{\is}{\coaxioms}$; hence $t_{\judg, n+1}$ \EZ{can be} defined as follows\EZ{:} }
\[
t_{\judg, n+1} = \Rule {
	\{ t_{\judg', n} \mid \judg' \in \prem_\judg \}
}{\judg}
\]
\FDComm{Ho provato a spiegare meglio la costruzione}\EZComm{ok}
Clearly by construction for all $\judg \in \Generated{\is}{\coaxioms}$ and for all $n \in \N$, $t_{\judg, n} \in \T_n$.
We show by induction on $n$ that for all $n \in \N$ and for all $\judg \in \Generated{\is}{\coaxioms}$,  $t_{\judg, n} \ApproxEq{n} t_{\judg, n+1}$.
\EZComm{controllare: usare questo stile ovunque nelle prove per induzione}
\begin{description}
\item[Base] If $n=0$, then $\dom_0(t_{\judg, 0}) = \dom_0(t_{\judg, 1}) = \{\EString\}$ and by construction $ t_{\judg, 0}(\EString) = t_{\judg, 1}(\EString) = \judg$, hence $t_{\judg, 0} \ApproxEq{0} t_{\judg, 1}$. 
\item[Induction] We assume the thesis for $n-1$ and prove it for $n$, hence we have to show that $t_{\judg, n} \ApproxEq{n} t_{\judg, n+1}$.
By construction we have $t_{\judg, n} = \Rule{ \{ t_{\judg', n-1} \mid \judg' \in \prem_\judg \} }{\judg}$ and $t_{\judg, n+1} = \Rule{ \{ t_{\judg', n} \mid \judg' \in \prem_\judg \} }{\judg}$.
By inductive hypothesis we get $t_{\judg', n-1} \ApproxEq{n-1} t_{\judg', n}$ for all $\judg' \in \prem_\judg$.
Therefore we have  
\[\begin{split}
\dom_n(t_{\judg, n}) &= \{\EString\} \cup \bigcup_{\judg' \in \prem_\judg} \judg' \dom_{n-1}(t_{\judg', n-1})  \\
					&= \{\EString\} \cup \bigcup_{\judg' \in \prem_\judg} \judg' \dom_{n-1}(t_{\judg', n}) \\
					&= \dom_n(t_{\judg, n+1}) 
\end{split}\] 
Consider now $\alpha \in \dom_n(t_{\judg, n})$, we have two cases\EZ{:}
\begin{align*}
\alpha &= \EString 								& & t_{\judg, n}(\alpha) = \judg = t_{\judg, n+1}(\alpha) \\
\alpha &= \judg' \beta,\ \judg' \in \prem_\judg	& & t_{\judg, n}(\alpha) = t_{\judg', n-1}(\beta) = t_{\judg', n}(\beta) = t_{\judg, n+1}(\alpha)
\end{align*}
and this shows $t_{\judg, n} \ApproxEq{n} t_{\judg, n+1}$ as needed. 
\end{description}

\item[$2\Rightarrow 3$] Trivial, by \refToDefinition{approx-sequence}.

\item[$3\Rightarrow 1$] By \refToTheorem{limit-tree-sequence} we know that there is a tree $t$ such that $t_n \ApproxTOrder{n} t$ for all $n \in \N$.
We show that $t$ is a proof tree in $\is$ for $\judg$.
Obviously $\judg = t_0(\EString) = t(\EString)$.
Consider $\alpha \in \dom(t)$, then, by construction of $t$,  there are natural number $m, n \in \N$ such that $\alpha \in \dom_m(t_m\EZ{)}$ and $\chl(\alpha) \subseteq \dom_n(t_n)$ with $\Len{\alpha} < m \le n$.
Therefore, since $t_m \ApproxTOrder{m} t_n$, we get $\alpha \in \dom_m(t_n) \subseteq \dom_n(t_n)$.
Since $t_n \in \T_n$ and $\Len{\alpha} < n$ the rule $\Rule{ \{t(\beta) \mid \beta \in \chl(\alpha)\} }{t(\alpha)}$ is a rule in $\is$ by \refToDefinition{approx-trees}, thus $t$ is a proof tree in $\is$.\\
Now consider a node $\alpha \in \dom(t)$, then there is $k \in \N$ such that $\alpha \in \dom_k(t_k\EZ{)}$, and so $\Len{\alpha} \le k$ and $\alpha \in \dom_m(t_m)$ for all $m \ge k$.
We define the sequence $(t_n^\alpha)_{n \in \N}$ such that $t_n^\alpha = {t_{n+k}}_\alpha$.
By \refToProposition{approx-subtree} we get $t_n^\alpha \in \T_{n+k-\Len{\alpha}} \subseteq \T_n$.
This observation shows that every node in $t$ has an approximated proof tree of level $n$ for all $n \in \N$, hence by \refToTheorem{approx-trees} we get $\judg \in \Generated{\is}{\coaxioms}$.

\end{description}
\end{proof}

\section{Reasoning with coaxioms} \label{sect:coaxioms-reasoning}

In this section we discuss proof techniques for inference systems with coaxioms.

Assume that $\DefSet=\Generated{\is}{\coaxioms}$ \FD{(for "defined")}  is the interpretation generated by coaxioms for some $\Pair{\is}{\coaxioms}$, and that $\Spec$ (for ``specification'') is the intended set of judgements, called \emph{valid} in the following.

Typically, we are interested in proving $\Spec\subseteq\DefSet$ (\emph{completeness}, that is, each valid judgement can be derived) and/or  $\DefSet\subseteq\Spec$ (\emph{soundness}, that is, each derivable judgement is valid). Proving both properties amounts to say that the inference system with coaxioms actually defines the intended set of judgements.

In the following, set $\bound=\closure_\Op{\is}(\coaxioms)=\Ind{\Extended{\is}{\coaxioms}}$.

\paragraph{Completeness proofs}
To show completeness, we can use \CoIndPrinciple. 
Indeed, since $\DefSet= \ker_\Op{\is} (\beta)$, if $\Spec \subseteq  \beta$ and $\Spec$ is a post-fixed point of $\Op{\is}$, by \CoIndPrinciple we get that $\Spec \subseteq \DefSet$. 
That is, using the notations of inference systems, to prove completeness it is enough to show that:
\begin{itemize}
\item $\Spec \subseteq \Ind{\Extended{\is}{\coaxioms}}$
\item $\Spec \subseteq \Op{\is}(\Spec)$
\end{itemize}}
We call this principle the \emph{bounded coinduction principle}.

We illustrate the technique on the inference system with coaxioms $\Pair{\is}{\coaxioms}$ which defines the judgement $\allPos{l}{b}$ (see \refToSection{coaxioms}). 
Let $\SpecAllPos$ be the set of judgements $\allPos{l}{b}$ where $b$ is $\True$ if all the elements in $l$ are positive, $\False$ otherwise.
Completeness means that the judgement  $\allPos{l}{b}$ can be proved, for all $\allPos{l}{b}\in\SpecAllPos$. By the {bounded coinduction} principle, it is enough to show that
\begin{itemize}
\item $\SpecAllPos \subseteq \Ind{\Extended{\is}{\coaxioms}}$
\item $\SpecAllPos \subseteq \Op{\is}(\SpecAllPos)$
\end{itemize}
To prove the first condition, we have to show that, for each $\allPos{l}{b}\in\SpecAllPos$, $\allPos{l}{b}$ has a \emph{finite} proof tree in $\Extended{\is}{\coaxioms}$.
This can be easily shown, indeed:
\begin{itemize}
\item If $l$ contains a (first) non-positive element, hence\\ $l=\Cons{x_1}{ \Cons{\ldots}{ \Cons{x_n}{ \Cons{x}{l'} } } }$ with $x_i>0$ for $i\in [1..n]$, $x \leq 0$, and $b=F$\\
then we can reason by arithmetic induction on $n$. Indeed, for $n=0$, $\allPos{l}{b}$ is the consequence of the second rule with no premises, and for $n>0$ it is the consequence of the third rule where we can apply the inductive hypothesis to the premise.
\item If $l$ contains only positive elements, hence $b=T$, then $\allPos{l}{b}$ is a coaxiom, hence it is the consequence of a rule with no premises in $\Extended{\is}{\coaxioms}$.
\end{itemize}
To prove the second condition, we have to show that, for each $\allPos{l}{b}\in\SpecAllPos$, $\allPos{l}{b}$ is the consequence of a rule with premises in $\SpecAllPos$. This can be easily shown, indeed:
\begin{itemize}
\item If $l=\Lambda$, hence $b=T$, then $\allPos{\Lambda}{\True}$ is the consequence of the first rule with no premises.
\item If $l=\Cons{x}{l'}$ with $x\leq 0$, hence $b=F$, then $\allPos{l}{\False}$ is the consequence of the  second rule with no premises.
\item If $l=\Cons{x}{l'}$ with $x > 0$, and $b=T$, hence $\allPos{l'}{\True}\in\SpecAllPos$, then $\allPos{l}{\True}$ is the consequence of the  third rule with premise $\allPos{l'}{\True}$, and analogously if $b=F$.
\end{itemize}

\paragraph{Soundness proofs} 
To show soundness, it is convenient to use the alternative characterization in terms of approximated proof trees given in \refToSection{coaxioms-trees}, as detailed below.
First of all, from \refToProposition{ker-iterate-bounds}, $\DefSet\subseteq\bigcap \{\IterOp{\is}{n}(\beta)\mid n \geq 0\}$.
Hence, to prove $\DefSet\subseteq\Spec$, it is enough to show that $\bigcap \{\IterOp{\is}{n}(\beta)\mid n \geq 0\}\subseteq\Spec$. 
Moreover, by \refToTheorem{approx-trees}, for all $n\in \N$, judgements in $\IterOp{\is}{n}(\beta)$ are those which have an approximated proof tree of level  $n$.
Hence, to {prove} set inclusion, we can show that all judgements which have an approximated proof tree of level $n$ for each $n$ are in $\Spec$ or equivalently, by \EZ{contraposition},  that 
judgements which are not in $\Spec$, that is, non-valid judgements, fail to have an approximated proof tree of level $n$ for some $n$.

We illustrate the technique again on the example of \allPosName. 
We have to show that, for each $\allPos{l}{b}\not\in\SpecAllPos$, there exists $n\geq 0$ such that $\allPos{l}{b}$ cannot be proved by using coaxioms at level greater than $n$. By cases:

\begin{itemize}
\item If $l$ contains a (first) non-positive element, hence\\ $l=\Cons{x_1}{ \Cons{\ldots}{ \Cons{x_n}{ \Cons{x}{l'} } } }$ with $x_i>0$ for $i\in [1..n]$, $x \leq 0$, then, assuming that coaxioms can only be used at a level greater  than $n+1$, $\allPos{l}{b}$ can only be derived by instantiating $n$ times the third rule, and once the second rule, hence $b$ cannot be $\True$.
\item If $l$ contains only positive elements, then it is immediate to see that there is no finite proof tree for $\allPos{l}{\False}$.
\end{itemize}
\section{Taming coaxioms: advanced examples}\label{sect:coaxioms-examples}

In this section we will present some \EZ{more} examples of situations where coaxioms can help to \EZ{define} judgements on non well-founded structures.
These more complex examples will serve as evidence for explaining how to use coaxioms and which kind of problems they can cope with.

\subsection{Mutual recursion}
Circular definitions involving inductive and coinductive judgements have no semantics in standard inference systems, 
because all judgements have to be interpreted either inductively, or coinductively.
The same problem arises in the context of coinductive logic programming \cite{SimonBMG07},
where a logic program has a well-defined semantics only if inductive and coinductive predicates can be stratified:
each stratum defines only inductive or coinductive predicates (possibly defined in a mutually recursive way), 
and cannot depend on predicates defined in upper strata. 
Hence, it is possible to define the semantics of a logic program only if there are no mutually recursive definitions involving both inductive and coinductive predicates.

We have already seen that an inductive inference system corresponds to an inference system with coaxioms where there are no coaxioms,
while a coinductive one corresponds to the case where coaxioms consist of all judgements in $\universe$;
however, between these two extremes, coaxioms offer many other possibilities thus allowing a finer control on the
semantics of the defined judgements.
    
There exist cases where two or more related judgements need to be defined recursively,
but for some of them the correct interpretation is inductive, while for others is coinductive
\cite{SimonMBG06,SimonBMG07,Ancona13}. 
In such cases, coaxioms may be employed to
provide the correct definition in terms of a single inference system with no stratification, although special care is required
to get from the inference system the intended meaning of judgements.     
In order to see this, let us consider the judgement $\pathZero(t)$, where $t$ is an infinite tree\footnote{For the purpose of this example, 
we only consider trees with infinite depth and branching.} over $\{0,1\}$, which
holds iff there exists a path starting from the root of $t$  and containing just $0$s. 
Trees are represented as infinite terms of shape $\tree(n,l)$, where $n\in\{0,1\}$ is the
root of the tree, and $l$ is the infinite list of its direct subtrees. 
For instance, if $t_1$ and $t_2$ are the trees defined by the
syntactic equations 
\begin{quote}
$t_1=\tree(0,l_1)$ \BigSpace
$l_1=\Cons{t_2}{ \Cons{t_1}{l_1} }$\BigSpace
$t_2=\tree(0,l_2)$\BigSpace
$l_2=\Cons{\tree(1,l_1)}{l_2}$
\end{quote}
 then we expect
$\pathZero(t_1)$ to hold, but not $\pathZero(t_2)$.

To define $\pathZero$, we introduce an auxiliary judgement $\isinZero(l)$ testing whether an infinite list $l$ of trees contains a tree
$t$ such that $\pathZero(t)$ holds.
Intuitively, we expect $\pathZero$ and $\isinZero$ to be interpreted coinductively and inductively, respectively;
this reflects the fact that $\pathZero$ checks a property universally quantified over an infinite sequence 
(a \emph{safety} property in the terminology of concurrent systems): all the elements of the path must \EZ{be equal to} $0$; 
on the contrary, $\isinZero$ checks a property existentially 
quantified over an infinite sequence (a \emph{liveness} property in the terminology of concurrent systems): the list must contain a 
tree $t$ with a specific property (that is, $\pathZero(t)$ must hold).
Driven by this intuition, one could be tempted to define the following inference system with coaxioms for all judgements
of shape $\pathZero(t)$, and no coaxioms for judgements of shape $\isinZero(l)$:
\[
\Rule{\isinZero(l)}{\pathZero(\tree(0,l))}{}\BigSpace 
\CoAxiom{\pathZero(t)}\BigSpace
\Rule{\pathZero(t)}{\isinZero(\Cons{t}{l})} \BigSpace
\Rule{\isinZero(l)}{\isinZero(\Cons{t}{l})}
\]
Unfortunately, because of the mutual recursion between $\isinZero$ and $\pathZero$, the inference system above does not capture the intended behaviour:
$\isinZero(l)$ is derivable for every infinite list of trees $l$, even when $l$ does not contain a tree $t$ with an infinite path starting from its root
and containing just $0$s. 

To overcome this problem, we can break the mutual dependency between judgements,
replacing the judgement $\isinZero$ with the more general one $\isin$, such that $\isin(t,l)$ holds iff the infinite list $l$ contains the tree $t$.
Consequently, we can define the following inference system with coaxioms: 
\[
\Rule{\isin(t,l) \Space \pathZero(t)}{\pathZero(\tree(0,l))}{} \BigSpace
\CoAxiom{\pathZero(t)}\BigSpace 
\Rule{}{\isin(t,\Cons{t}{l})} \BigSpace
\Rule{\isin(t,l)}{\isin(t,\Cons{t'}{l})}
\]
 
Now the semantics of the system corresponds to the intended one, and we do not need to stratify the definitions into two separate inference systems.

Following the characterization in terms of proof trees and the proof techniques provided in \refToSection{coaxioms-trees} and 
\refToSection{coaxioms-reasoning}, we can sketch a proof of correctness. 
Let $\Spec$ be the set  where
elements have either shape $\pathZero(t)$, where $t$ represents a tree with an infinite path of just $0$s starting from its root, or $\isin(t,l)$, where $l$ represents an infinite list containing the tree $t$; then a judgement belongs to $\Spec$ iff it can be derived in the 
inference system with coaxioms defined above.

\paragraph{Completeness:} We first show that the set $\Spec$ is a post-fixed point, that is, it is consistent w.r.t.
the inference rules, coaxioms excluded. 
Indeed, if $t$ has an infinite path of $0$s, then it has necessarily  shape
$\tree(0,l)$, where $l$ must contain a tree $t'$ with an infinite path of $0$s. 
Hence, the inference rule for $\pathZero$ can be applied with premises
$\isin(t',l)\in \Spec$, and $\pathZero(t')\in \Spec$.
If an infinite list contains a tree $t$, then  
it has necessarily shape $\Cons{t'}{l}$ where, either $t=t'$, and hence the axiom for $\isin$ can be applied,
or $t\neq t'$ and $t$ is contained in $l$, and hence the inference rule for $\isin$ can be applied
with premise $\isin(t,l)\in \Spec$.

We then show that $\Spec$ is bounded by the closure of the coaxioms. 
For the elements of shape  $\pathZero(t)$ it suffices to directly apply the corresponding coaxiom; for the elements
of shape $\isin(t,l)$ we can show that there exists a finite proof tree built possibly also with the coaxioms by induction on the
first position (where the head of the list corresponds to \EZ{$0$}) in the list where $t$ occurs.
If the position is \EZ{$0$} (base case), then $l=\Cons{t}{l'}$, and the axiom can be applied;
if the position is $n>0$ (inductive step), then $l=\Cons{t'}{l'}$ and $t$ occurs in $l'$ at position $n-1$,
therefore, by inductive hypothesis, there exists a finite proof tree for $\isin(t,l')$, therefore
we can build a finite proof tree for $\isin(t,l)$ by applying the inference rule for $\isin$.

\paragraph{Soundness:} We first observe that the 
only finite proof trees that can be derived for $\isin(t,l)$ are obtained by 
application of the axiom for $\isin$, hence $\isin(t,l)$ holds iff there
exists a finite proof tree for $\isin(t, l)$ built with the inference rules
for $\isin$. 
Then, we can prove that, if $\isin(t,l)$ holds, then $t$ is contained in
$l$ by induction on the inference rules for $\isin$. 
For the axiom (base case) the claim trivially holds, while for the other inference rule we
have that if $t$ belongs to $l$, then trivially $t$ belongs to $\Cons{t'}{l}$.  

For the elements of shape $\pathZero(t)$ we first observe that by the coaxioms they all trivially belong to the closure
of the coaxioms. 
Then, any proof tree for $\pathZero(t)$ must be infinite, because there are no axioms but only one inference rule
for $\pathZero$ where $\pathZero$ is referred in the premises; furthermore, such a rule is applicable only if the
root of the tree is \EZ{$0$}. We have already proved that if $\isin(t,l)$ is derivable, then $t$ belongs to $l$,
therefore we can conclude that if $\pathZero(t)$ is derivable, then $t$ contains an infinite path starting from its root,
and containing just $0$s.\\

We conclude this example by providing an alternative (even though a bit redundant)
inference system, that allows us to recover judgements of shape $\isinZero$, without breaking the correctness of the definition.\EZComm{in che senso? vuoi dire semplicemente che posso definirlo direttamente?} 
\FDComm{volevo dire che posso reintrodurre il judgement  del sistema con ricorsione mutua}
We can introduce the auxiliary predicate $\isinZero$ defined in terms of
$\isin$ and $\pathZero$, and still get the intended semantics, since we have removed the dangerous cyclic rule for $\isinZero$:

\[
\begin{array}{c}
\Rule{\isinZero(l)}{\pathZero(\tree(0,l))}\BigSpace
\CoAxiom{\pathZero(t)}\\[4ex]
\Rule{\isin(t,l) \Space \pathZero(t)}{\isinZero(l)}\BigSpace 
\Rule{}{\isin(t,\Cons{t}{l})} \BigSpace
\Rule{\isin(t,l)}{\isin(t,\Cons{t'}{l})}
\end{array}
\]

\subsection{A numerical example}
It is well\EZ{-}known that real numbers in the closed interval $[0,1]$ can be represented
by infinite sequences $(d_i)_{i\in\N^+}$ of decimal\footnote{Of course the example can be generalized to any base $B\geq 2$.} digits,
where $\N^+$ denotes the set of all positive natural numbers. 
Indeed, $(d_i)_{i\in\N^+}$ represents the real number which is the limit of the series $\sum_{i=1}^{\infty}10^{-i}d_i$ in the standard
complete metric space of real numbers (such a limit always exists by completeness, 
because the associated sequence of partial sums is always a Cauchy sequence). 
Such a representation is not unique for all rational numbers in $[0,1]$ (except for the bounds $0$ and $1$) that can be represented by a finite sequence of digits followed by an infinite sequence of $0$s; 
for instance, $0.42$ can be represented either by the sequence $42\bar{0}$, or by the sequence $41\bar{9}$, where
$\bar{d}$ denotes the infinite sequence containing just the digit $d$.

For brevity, for $r=(d_i)_{i\in\N^+}$, $\sem{r}$ denotes $\sum_{i=1}^{\infty}10^{-i}d_i$ (that is, the real number represented by $r$).
We want to define the judgement $\add(r_1,r_2,r,c)$ which holds iff $\sem{r_1}+\sem{r_2}=\sem{r}+c$ with $c$ an integer number; that is,
$\add(r_1,r_2,r,c)$ holds iff the addition of the two real numbers represented by the sequences $r_1$ and $r_2$ yields the real number
represented by the sequence $r$ with carry $c$. 
We will soon discover that, to get a complete definition for $\add$,
$c$ is required to range over a proper superset of the set $\{0,1\}$, differently from what one could initially expect.

We can define the judgement $\add$ with the following inference system with coaxioms, where $\div$ and $\bmod$ denote
the integer division, and the remainder operator, respectively.
 
\[
\Rule{\add(r_1,r_2,r,c)}{\add(\Cons{d_1}{r_1},\Cons{d_2}{r_2},\Cons{(s \bmod 10)}{r},s \div 10)}{s=d_1+d_2+c} \BigSpace
\CoAxiom{\add(r_1,r_2,r,c)}
\]

A real number in $[0,1]$ is represented by an infinite list of decimal digits, which, therefore,
can always be decomposed as $\Cons{d}{r}$, where $d$ is the first digit (corresponding to the exponent $-1$),
and $r$ is the rest of the list of digits. 
Here, $r_1$, $r_2$, and $r$ range over the set
of infinite lists of decimal digits, while the carry must range over $\{-1,0,1,2\}$ to support a complete definition. 
As clearly emerges from the proof of completeness provided
below, besides the obvious values $0$ and $1$, the values $-1$ and $2$
have to be considered for the carry to ensure a complete definition of $\add$
because both $\add(\bar{0},\bar{0},\bar{9},-1)$  and $\add(\bar{9},\bar{9},\bar{0},2)$ 
hold, and, hence, should be derivable; these two judgements allow the derivation 
of an infinite number of other valid judgements, as, for instance, $\add(1\bar{0},1\bar{0},1\bar{9},0)$ and
$\add(1\bar{9},1\bar{9},4\bar{0},0)$, respectively.

Also in this case we can sketch a proof of correctness: for all infinite sequences of decimal digits $r_1$, $r_2$ and
$r$, and all $c\in\{-1,0,1,2\}$, $\add(r_1,r_2,r,c)$ is derivable iff
$\sem{r_1}+\sem{r_2}=\sem{r}+c$.

\paragraph{Completeness:}
By the coaxioms we trivially have that each element of shape
$\add(r_1,r_2,r,c)$ such that $\sem{r_1}+\sem{r_2}=\sem{r}+c$ with $c\in\{-1,0,1,2\}$ belongs
to the closure of the coaxioms. 

To show that the unique inference rule of the system is consistent with the set of all valid  judgements,
let us assume that $\sem{r'_1}+\sem{r'_2}=\sem{r'}+c'$ with $r'_1=\Cons{d_1}{r_1}$, $r'_2=\Cons{d_2}{r_2}$, $r'=\Cons{d}{r}$ and $c'\in\{-1,0,1,2\}$.
Let us set $s=10c'+d$, and $c=s-d_1-d_2$, then $s\bmod 10=d$ and $s\div 10=c'$, and we get the desired conclusion of the
inference rule, and the side condition holds;
it remains to show that $\sem{r_1}+\sem{r_2}=\sem{r}+c$ with $c\in\{-1,0,1,2\}$.

We first observe that by the properties of limits w.r.t. the usual arithmetic operations, and
by definition of $\sem{-}$, for all infinite sequence $r$ of decimal digits, if $r=\Cons{d}{r'}$, then
$\sem{r}=10^{-1}(d+\sem{r'})$; then, from the hypotheses we get the equality $d_1+\sem{r_1}+d_2+\sem{r_2}=d+\sem{r}+10c'$,
hence $d_1+\sem{r_1}+d_2+\sem{r_2}=\sem{r}+s$, 
and, therefore, $\sem{r_1}+\sem{r_2}=\sem{r}+c$; finally,
$c$ is an integer because it is an algebraic sum of integers, and, since $c=\sem{r_1}+\sem{r_2}-\sem{r}$, and $0\leq\sem{r_1},\sem{r_2},\sem{r}\leq1$, we get $c\in\{-1,0,1,2\}$. 

\paragraph{Soundness:}
Let $r'_1=\Cons{d_1}{r_1}$, $r'_2=\Cons{d_2}{r_2}$, and $r'=\Cons{d}{r}$ be infinite sequences of decimal digits, and $c'\in\{-1,0,1,2\}$;
we note that the judgement $\add(r'_1,r'_2,r',c')$ can only be derived from the unique inference rule where the premise is the judgement
$\add(r_1,r_2,r,c)$ with  $c$  equal to  $10c'+d-d_1-d_2$ and must range over $\{-1,0,1,2\}$.

To prove soundness we show that if $\sem{r'_1}+\sem{r'_2}\neq\sem{r'}+c'$, then  the judgement 
$\add(r'_1,r'_2,r',c')$ cannot be derived in the inference system. 
Let us set $\delta'=|\sem{r'}+c'-\sem{r'_1}-\sem{r'_2}|$; obviously, under the hypothesis
$\sem{r'_1}+\sem{r'_2}\neq\sem{r'}+c'$, we get $\delta' >0$.
In particular, the following fact holds:  if $\delta'\geq 4\cdot10^{-1}$, then
$10c'+d-d_1-d_2\not\in\{-1,0,1,2\}$. 
Indeed, by the identity $\sem{r}=10^{-1}(d+\sem{r'})$ already
used for the proof of completeness, we have that 
$\delta'=10^{-1}\delta$ with $\delta = |\sem{r}+c-\sem{r_1}-\sem{r_2}|$, with $c=10c'+d-d_1-d_2$;  
$10^{-1}(\sem{r}+c-\sem{r_1}-\sem{r_2}) \geq 4\cdot10^{-1}$
implies $c\geq 3$
($\sem{r_1},\sem{r_2},\sem{r}\in[0,1]$), and, hence, $c=10c'+d-d_1-d_2\not\in\{-1,0,1,2\}$. On the other hand, 
$10^{-1}(\sem{r}+c-\sem{r_1}-\sem{r_2})\leq -4\cdot10^{-1}$ implies  
$c\leq -2$, hence $c=10c'+d-d_1-d_2\not\in\{-1,0,1,2\}$.

By virtue of this fact, and thanks to the hypotheses, we can prove by arithmetic induction over $n$ that for all $n\geq 1$, if
 $\delta'\geq 4\cdot10^{-n}$, then there exist only finite proof trees for $\add(r'_1,r'_2,r',c')$ where the coaxioms are applied at most at depth $n-1$.
The base case is directly derived from the previously proven fact. Indeed, for $n=1$, we can only derive $\add(r'_1,r'_2,r',c')$ by directly applying the coaxiom.
For the inductive step we observe that all derivation of depth greater than $1$ are built applying the inference rule as first step.
If such rule is applicable for deriving the conclusion $\add(r'_1,r'_2,r',c')$, then  we can apply the inductive
hypothesis for the premise $\add(r_1,r_2,r,c)$ since we have already shown that 
$\delta'=10^{-1}\delta$, therefore  
$\delta\geq 4\cdot10^{-(n-1)}$.

We can now conclude by observing that if $\sem{r'_1}+\sem{r'_2}\neq\sem{r'}+c'$, then there exists
$n$ such that $\delta'\geq 4\cdot10^{-n}$, therefore, by the previous result, we deduce that it is not possible
to build a finite tree for $\add(r'_1,r'_2,r',c')$ where the coaxioms are applied at arbitrary depth $k$ 
(in particular, $k$ is bounded by $n-1$); therefore $\add(r'_1,r'_2,r',c')$ cannot be derived in the inference system.\\

From the proof of soundness we can also deduce that if we let $c$ range over $\Z$, then
 the inference system becomes unsound; for instance, 
 it would be possible to build the following infinite proof for $\add(\bar{0},\bar{0},\bar{0},1)$ 
where all nodes clearly belong to the closure of the coaxioms,
and, hence, 
$\add(\bar{0},\bar{0},\bar{0},1)$ would be derivable, but $\sem{\bar{0}}+\sem{\bar{0}}\neq\sem{\bar{0}}+1$:
\[
\Rule
    {
      \Rule
          {
            \vdots
          }
          {\add(\bar{0},\bar{0},\bar{0},10^1)}
    }
    {\add(\bar{0},\bar{0},\bar{0},10^0)}
\]

\subsection{Distances and shortest paths on weighted graphs} \label{sect:graph-example} 
In \refToSection{is} we have presented a definition for the predicate $\dist{\node}{\anode}{\delta}$ stating that the distance between node $\node$ and $\anode$ in a \EZ{graph} is $\delta$.
In that case the standard coinductive interpretation was enough to capture the intended semantics.
Here we consider a more \EZ{complex problem}: to compute distances between nodes in a \emph{weighted graph}.

Let us introduce the notion of weights for graphs.
In a graph $\Pair{\Nodes}{\adj}$ the set of edges is the set $\Edges \subseteq \Nodes \times \Nodes$  defined \EZ{by} $\Edges = \{ \Pair{\node}{\anode} \in \Nodes \times \Nodes \mid \anode \in \adj(\node) \}$. 
We will often write $\node\anode$ for an edge $\Pair{\node}{\anode} \in \Edges$.
A weight function is a function $\fun{w}{\Edges}{\N}$.
Here we consider natural numbers as codomain, however we could have considered any other set of non-negative numbers.
Hence, a weighted graph is a graph $\Pair{\Nodes}{\adj}$ together with a weight function $w$.

In a weighted graph $G$, the weight of a path $\alpha$ is the sum of the weights of the edges \FD{(counting repetitions)} \EZ{determined by} $\alpha$, we denote this by $w(\alpha)$.
Note that in general the weight of a path $\alpha$ is different from its length, defined as  the number of edges \FD{(counting repetitions)}  \EZ{determined by} the path and denoted by $\|\alpha\|$.
The distance between nodes $\node$ and $\anode$ is defined as the minimum weight of a path connecting $\node$ to $\anode$, it is infinite if no such path exists.
Below we \EZ{show} the inference system with coaxioms defining the judgement $\dist{\node}{\anode}{\delta}$ on a weighted graph, where $\delta \in \N \cup \{\infty\}$.
\begin{small}
\[
\begin{array}{c}
\Rule{}{ \dist{\node}{\node}{0} }
\BigSpace
\Rule{}{ \dist{\node}{\anode}{\infty} }
{\begin{array}{l}
\node \ne \anode \\
\adj(\node) = \emptyset
\end{array} }
\BigSpace
\CoAxiom{ \dist{\node}{\anode}{\infty} }
\\[4ex]
\Rule{
	\dist{\node_1}{\anode}{\delta_1} 
	\Space \ldots \Space
	\dist{\node_k}{\anode}{\delta_k}
}{ \dist{\node}{\anode}{\delta} }
{\begin{array}{l}
\node \ne \anode \\
\adj(\node) = \{\node_1, \ldots, \node_k\} \ne \emptyset \\
\delta = \min\{w(\node\node_1) + \delta_1, \ldots,  w(\node\node_k) + \delta_k\}
\end{array} }
\end{array}
\]
\end{small}
In order to show that we cannot simply consider the coinductive interpretation of the above inference system, hence we need coaxioms, let us consider a weighted \EZ{version} of the simple graph \EZ{shown} in \refToSection{is}.
\begin{center}
\begin{tikzcd}[column sep=large, row sep=large]
e             & b \ar[d, bend right, swap, "0"]         &   \\
d \ar[r, "2"] & a \ar[u, bend right, swap, "0"] \ar[ul, "5"] & c \ar[l, swap, "1"]
\end{tikzcd}
\end{center}
If we interpret the inference system coinductively we can derive judgements like $\dist{c}{e}{\delta}$ for any $\delta \in [1..5]$ or $\dist{a}{d}{\delta}$ for any $\delta \in \N \cup \{\infty\}$, as shown in \refToFigure{dist-wrong-trees}.
\begin{figure}
\caption{Proof trees for $\dist{c}{e}{1+\delta_1}$ and $\dist{a}{d}{\delta_2}$} 
\label{fig:dist-wrong-trees}
\[
\Rule{
	\Rule{
		\Rule{}{ \dist{e}{e}{0} }
		\Space
		\Rule{
			\vdots
		}{ \dist{b}{e}{\delta_1} }
	}{ \dist{a}{e}{\delta_1} } { \delta_1\le 5}
}{ \dist{c}{e}{1+\delta_1} }
\BigSpace
\Rule{
	\Rule{}{ \dist{e}{d}{\infty} }
	\Space
	\Rule{
		\vdots
	}{ \dist{b}{e}{\delta_2} }
}{ \dist{a}{d}{\delta_2}  }
\]
\end{figure}
The issue here is the cycle that, having total weight equal to $0$, allows us to build cyclic proofs without increasing the value of $\delta$.
Therefore the coaxiom is needed to filter out such proofs.
Indeed, it is easy to see that it is not possible to build a finite proof tree  for judgements proved in \refToFigure{dist-wrong-trees}  starting from the coaxiom.

Now we will sketch a proof of correctness. 
First of all we \EZ{prove} two useful facts.
\begin{quote}
\StarOne For all proof trees $t$ for a judgement $\dist{\node}{\anode}{\delta}$, there exists a path $\alpha$ from $\node$ to $\anode$ with $\|\alpha\| = n$ iff there exists a node in $t$ at level $n$ labelled by $\dist{\anode}{\anode}{0}$.
\end{quote}
Let $t$ be a proof tree for $\dist{\node}{\anode}{\delta}$. 
We prove separately the two implications.
\begin{description}
\item[$\Rightarrow$] Let $\alpha$ be a path from $\node$ to $\anode$.
We proceed by induction \EZ{on} the length of $\alpha$.
If $\|\alpha\|=0$ (base case), then $\node=\anode$, hence $\dist{\node}{\anode}{\delta}$ has been derived by applying the first axiom, and this implies $\delta = 0$. 
Therefore the root of $t$ (at level $0$) is labelled by $\dist{\anode}{\anode}{0}$.
If $\|\alpha\| = n+1$ (inductive step), then $\alpha = \node \beta$ where \EZ{$\beta$} is a path from a node $\node'$ \EZ{to} $\anode$ of length $n$.
Therefore $\dist{\node}{\anode}{\delta}$ has been derived by applying the inference rule, hence there is a direct subtree of $t$ rooted in $\dist{\node'}{\anode}{\delta'}$, where, by inductive hypothesis, $\dist{\anode}{\anode}{0}$ occurs at level $n$.
Thus in $t$ that judgement occurs at level $n+1$ as needed.
\item[$\Leftarrow$] We proceed by induction on the level $n$.
If $\dist{\anode}{\anode}{0}$ occurs at level $0$ (base case), then it is the root of $t$, hence $\node = \anode$  and the searched path is the singleton path $\anode$.
If it occurs at level $n+1$ (inductive step), then the depth of $t$ is greater than $0$, hence $\dist{\node}{\anode}{\delta}$ has been derived by applying the inference rule.
Therefore $\dist{\anode}{\anode}{0}$ belongs to a direct subtree $t'$ of $t$ rooted in $\dist{\node'}{\anode}{\delta'}$ and it occurs in $t'$ at level $n$.
Thus by inductive hypothesis there is a path $\beta$ from $\node'$ to $\anode$ of length $n$, hence the path $\node \beta$ of length $n+1$ connects $\node$ to $\anode$.
\end{description}
\begin{quote}
\StarTwo For all proof tree $t$, $t$ is rooted in $\dist{\node}{\anode}{\infty}$ iff all nodes in $t$ are of shape $\dist{\node'}{\anode}{\infty}$\EZ{.}
\end{quote}
Consider a proof tree $t$.
The implication $\Leftarrow$ is trivial. 
Let us prove the other one.
Consider a node in $t$, we proceed by induction on the level of such node.
If the level is $0$ (base case), then the thesis follows immediately by hypothesis.
If the level is $n+1$ (inductive step), then the selected node is the child of a node at level $n$, that, by inductive hypothesis, is of shape $\dist{\node'}{\anode}{\infty}$.
Therefore the inference rule has been applied, and, since the conclusion is $\dist{\node'}{\anode}{\infty}$, all premises, including the selected node, must be of shape $\dist{\node''}{\anode}{\infty}$, because we take the minimum. \\

We can formulate the correctness statement as follows: $\dist{\node}{\anode}{\delta}$ is derivable iff $\delta$ is the minimum of $w(\alpha)$ for all paths $\alpha$ from $\node$ to $\anode$.

\paragraph{Completeness: }
Let us consider a judgement $\dist{\node}{\anode}{\delta}$ where $\delta$ is the minimum of $w(\alpha)$ for $\alpha$ a path from $\node$ to $\anode$.
If $\delta=0$ or $\delta = \infty$ or $\adj(\node) = \emptyset$, then $\dist{\node}{\anode}{\delta}$ is the consequence of an axiom.
Otherwise, note that $\alpha = \node\beta$ where $\beta$ is a path from a node $\node' \in \adj(\node)$ \EZ{to} $\anode$, hence $w(\alpha) = w(\node\node') + w(\beta)$.
Furthermore, if there were another path $\beta'$ from a node $\node'$ to $\anode$ with $w(\beta') < w(\beta)$, then the path $\node\beta'$ would be such that $w(\node\beta') < w(\alpha) = \delta$\EZ{,} that is absurd, hence $\dist{\node'}{\anode}{w(\beta)}$ is a valid judgement.
Moreover, note that for any other $\node_i \in \adj(\node)$, with $\dist{\node_i}{\anode}{\delta_i}$ a valid judgement, we have $\delta \le w(\node\node_i) + \delta_i$, since, otherwise, we could build a path from $\node$ to $\anode$ with weight less than $\delta$\EZ{,} that is absurd.
Therefore $\dist{\node}{\anode}{\delta}$ is the consequence of \EZ{the} inference rule and \EZ{its} premises are valid judgements, and this shows that the specification  \EZComm{volevi dire un'altra cosa?} is a consistent set.
\FDComm{in che senso un'altra cosa? Devo provare che è post-fisso}

In order to show the boundedness condition, we have to build a finite proof tree for $\dist{\node}{\anode}{\delta}$ (chosen as before) using coaxioms as axioms. 
If $\delta = 0$ or $\delta = \infty$  we simply apply either  the axiom or the coaxiom, and the same holds when $\adj(\node) = \emptyset$.
Otherwise, we know that there is a path $\alpha$ from $\node$ to $\anode$ with $w(\alpha) = \delta$.
Let us assume $\alpha = \node_0\cdots \node_n$ with $\node = \node_0$ and $\anode = \node_n$, hence $\node_{i+1} \in \adj(\node_i)$ for all $i \in [0..n-1]$.
At each level $i$ we apply the inference rule with consequence $\dist{\node_i}{\anode}{\delta_i}$  and premises $\dist{\node'}{\anode}{\infty}$, for all $\node' \in \adj(\node_i)$ with $\node'\ne \node_{i+1}$, and $\dist{\node_{i+1}}{\anode}{\delta_{i+1}}$.
The only exception is level $n$ where, instead of the inference rule, since $\node_n = \anode$, we apply the axiom $\dist{\anode}{\anode}{0}$. 
In this way we have that for all $i$, $\delta_i = w(\node_i\cdots \node_n)$, hence $\delta_0 = w(\alpha) = \delta$ as needed.

\paragraph{Soundness: }
Here we have to show that each derivable judgement is valid.
For judgements of shape $\dist{\node}{\anode}{\infty}$ the thesis follows immediately from \StarOne and \StarTwo.
Hence, let us assume $\delta \in \N$.

We assume that $\delta$ is not the minimum weight of a path $\alpha$ from $\node$ to $\anode$, and we prove that the judgement $\dist{\node}{\anode}{\delta}$ cannot be derived with an approximated proof tree of level $n$ for some $n$.

We first note that if $\dist{\node}{\anode}{\delta}$ has an approximated proof tree of  depth\footnote{\EZ{H}ere we consider just the depth of the proof tree, not its level as in \refToDefinition{approx-trees}\EZ{.}} $n$, then $\delta$ is the minimum weight of a path $\alpha$ from $\node$ to $\anode$ with $\|\alpha\|  \le n$. 
We proceed by induction on $n$.
If $n=0$ (base case), then $\node = \anode$ and $\delta = 0$, since the only applicable rule is the first axiom, and this is enough since the only path of length \EZ{$0$} is the trivial one, that has weight $0$.

If the depth is $n+1$ (inductive step), then, assuming $\adj(\node) = \{\node_1, \ldots, \node_k\}$, $\dist{\node_i}{\anode}{\delta_i}$ has an approximated proof tree of depth $n$.
Therefore, by inductive hypothesis, we get \EZ{that} $\delta_i$ is the minimum weight of a path $\alpha$ from $\node_i$ to $\anode$ with $\|\alpha\| \le n$.
Hence $\delta$ is the minimum of the $w(\node\node_i) + \delta_i$.

Let now $\alpha$ be a path from $\node$ to $\anode$ with $\|\alpha\| \le n+1$, hence $\alpha = \node \beta$ where $\beta$ is a path from $\node_i$ to $\anode$ for some $i$ with $\|\beta\| \le n$.
Therefore we have $\delta \le w(\node \node_i) + \delta _i \le w(\node\node_i\EZ{)} + w(\beta) = w(\alpha)$, by inductive hypothesis.

So, if $\dist{\node}{\anode}{\delta}$ is not valid, \EZ{then} there is a path $\alpha$ from $\node$ to $\anode$ such that $w(\alpha) < \delta$.
\EZ{Assuming} that $\alpha$ has length equal to $n$, we have just shown that $\dist{\node}{\anode}{\delta}$ cannot have an approximated proof tree of depth greater than $n$.
Now an approximated proof tree of level $n$ either \EZ{does not} use coaxioms or uses coaxioms and \EZ{has} depth greater than $n$.
From what we have just proved, the latter case is not possible.
In the former case, by \StarOne, since $\alpha$ is a path in the graph, there is a node in the proof tree at level $n$ labelled by $\dist{\anode}{\anode}{0}$, \EZ{and, as} the proof tree has at least depth equal to $n$\EZ{,} that is absurd. 
Therefore $\dist{\node}{\anode}{\delta}$ cannot have an approximated proof tree of level $n$. \\

The notion of distance is tightly related to paths in a graph $G$.
Actually, from the above proofs, it \EZ{is} easy to see that a proof tree for a judgement $\dist{\node}{\anode}{\delta}$ explores all possible paths from $\node$ to $\anode$ in the graph in order to compute the distance.
Therefore, in some sense, it also \EZ{finds} the shortest path from $\node$ to $\anode$.
Hence, with a slight variation of the inference system  for the distance, we can get an inference system for the judgement $\minPath{\node}{\anode}{\alpha}{\delta}$ stating that $\alpha$ is the shortest path from $\node$ to $\anode$ with weight $\delta$. 
We add to paths the special one $\bot$ that represents the absence of paths between two nodes, with the assumption that $\node \bot = \bot$.
\begin{small}
\[
\begin{array}{c}
\Rule{}{ \minPath{\node}{\node}{\node}{0} }
\BigSpace
\Rule{}{ \minPath{\node}{\anode}{\bot}{\infty} }
{\begin{array}{l}
\node \ne \anode \\
\adj(\node) = \emptyset
\end{array} }
\BigSpace
\CoAxiom{ \dist{\node}{\anode}{\bot}{\infty} }
\\[4ex]
\Rule{
	\minPath{\node_1}{\anode}{\alpha_1}{\delta_1} 
	\Space \ldots \Space
	\minPath{\node_k}{\anode}{\alpha_k}{\delta_k}
}{ \minPath{\node}{\anode}{\node \alpha_i}{w(\node\node_i) + \delta_i} }
{ \begin{array}{l}
\scriptstyle
\node \ne \anode \\
\scriptstyle
\adj(\node) = \{\node_1, \ldots, \node_k\} \ne \emptyset \\
\scriptstyle
i = \argmin\{w(\node\node_1) + \delta_1, \ldots,  w(\node\node_k) + \delta_k\}
\end{array} }
\end{array}
\]
\end{small}

\subsection{Big-step operational semantics with divergence} \label{sect:op-sem}
It is well-known that divergence cannot be captured by the big-step operational semantics of a programming language when
semantic rules are interpreted inductively (that is, in the standard way) \cite{LeroyGrall09,Ancona12,Ancona14}.
When rules are interpreted coinductively some partial result can be obtained under suitable hypotheses,
but  
a practical way to capture divergence  with a big-step operational semantics
is to introduce two different forms of judgement \cite{CousotCousot92,LeroyGrall09}: one corresponds to the standard big-step
evaluation relation, and is defined inductively, while the other one captures divergence, and is defined coinductively
in terms of the inductive judgement, thus requiring stratification. 

With coaxioms a unique judgement can be defined in a more direct and compact way.
We show\footnote{This example was inspired by Bart Jacobs.} how this is possible for the standard call-by-value operational semantics of the \LambdaCalculus in 
big-step style.
\begin{figure}
\begin{center}
Syntax of terms and values\\[2ex]
\begin{math}
\begin{array}{l}
e ::= v \mid x \mid e\ e \qquad
v ::= \lambda x.e \qquad
\infv ::= v \mid \infty 
\end{array}
\end{math}\\[2ex]
Semantic rules \\[2ex]
\begin{math}
\begin{array}{c}
\CoAxiomName
{coax}
{\eval{e}{\infty}}
\qquad
\RuleName
{val}
{}
{\eval{v}{v}}
\qquad
\RuleName
{app}
{\eval{e_1}{\lambda x.e}\quad\eval{e_2}{v}\quad\eval{\subs{e}{x}{v}}{\infv}}
{\eval{e_1\ e_2}{\infv}}
\\[4ex]
\RuleName
{l-inf}
{\eval{e_1}{\infty}}
{\eval{e_1\ e_2}{\infty}}
\qquad
\RuleName
{r-inf}
{\eval{e_1}{v}\quad\eval{e_2}{\infty}}
{\eval{e_1\ e_2}{\infty}}
\end{array}
\end{math}
\end{center}
\caption{Call-by-value big-step semantics of $\lambda$-calculus with divergence}\label{fig:lambda}
\end{figure}
\refToFigure{lambda} defines syntax, values, and semantic rules. The meta-variable $v$ ranges over standard values, that is,
lambda abstractions, while $\infv$ includes also divergence, represented by $\infty$. 
The evaluation judgement has the general
shape $\eval{e}{\infv}$, meaning that either $e$ evaluates to a value $v$ (when $\infv\neq\infty$) or diverges (when $\infv=\infty$).  

For what concerns the semantic rules, only a coaxiom is needed, stating that every expression may diverge. 
This ensures that
$\infty$ can be the only allowed outcome for the evaluation of an expression which diverges; this can only happen
when the corresponding derivation tree is infinite.
Rule (val) is standard. Rule (app) deals with the evaluation of application when both expressions $e_1$ and $e_2$ 
do not diverge; the meta-variable $v$ is required for the judgement \eval{e_2}{v} to guarantee
convergence of $e_2$, while $\infv$ is used for the result of the whole application, since
the evaluation of the body of the lambda abstraction could diverge. As usual, $\subs{e}{x}{v}$ denotes capture-avoiding substitution
modulo $\alpha$-renaming. Rules (l-inf) and (r-inf) cover the cases when either $e_1$ or $e_2$ diverges when trying to evaluate application,
assuming that a left-to-right evaluation strategy has been imposed.

We show that the only judgement derivable for $e_\Delta=(\lambda x.x\ x) \lambda x.x\ x$ is \eval{e_\Delta}{\infty}.
To this aim, we first disregard the coaxiom and exhibit an infinite derivation tree 
for the judgement \eval{e_\Delta}{\infv}, derivable for all $\infv$:
\vspace{-1em}
\begin{scriptsize}
\[
\begin{array}{l}
\RuleName
{app}
{
  \RuleName
  {val}
  {}
  {\eval{\lambda x.x\ x}{\lambda x.x\ x}}
  \quad
  \RuleName
  {val}
  {}
  {\eval{\lambda x.x\ x}{\lambda x.x\ x}}
  \quad 
  \RuleName
  {app}
  {\vdots}
  {\eval{\subs{(x\ x)}{x}{\lambda x.x\ x}}{\infv}}}
{\eval{\subs{(x\ x)}{x}{\lambda x.x\ x}=e_\Delta}{\infv}}
\end{array}
\]
\end{scriptsize}

In this particular case the derivation tree is also regular, but of course there are examples of divergent computations
whose derivation tree is not regular. The vertical dots indicate that the derivation continues with the same repeated pattern.  
The derivation corresponds to the coinductive interpretation of the standard big-step semantics rules \cite{LeroyGrall09,Ancona12},
which may exhibit non-deterministic behaviour as happens for this example; however, here the coaxiom plays a crucial role
by filtering out all undesired values, and, thus, leaving only the value $\infty$ representing divergence; indeed, by employing also the coaxiom, 
finite derivation trees can be built for \eval{e_\Delta}{\infv} only when $\infv=\infty$. 
By \refToTheorem{approx-sequence}
we can get an infinite sequence of approximating sequence of arbitrarily increasing level:
\begin{scriptsize}
\[
\begin{array}{l}
\RuleName
{coax}
{}
{\eval{e_\Delta}{\infty}}
\\[4ex]
\RuleName
{app}
{
  \RuleName
  {val}
  {}
  {\eval{\lambda x.x\ x}{\lambda x.x\ x}}
  \quad
  \RuleName
  {val}
  {}
  {\eval{\lambda x.x\ x}{\lambda x.x\ x}}
  \quad 
  \RuleName
  {coax}
  {}
  {\eval{\subs{(x\ x)}{x}{\lambda x.x\ x}}{\infty}}}
{\eval{\subs{(x\ x)}{x}{\lambda x.x\ x}=e_\Delta}{\infty}}
\\[4ex]
\vdots
\end{array}
\]
\end{scriptsize}

As a consequence, in the inference system with the coaxiom a valid infinite derivation tree
can be built for  \eval{e_\Delta}{\infv} only when $\infv=\infty$.

\cleardoublepage \chapter{Related work} \label{chapter:related}

Inference systems \cite{Aczel77} are widely adopted to formally define operational semantics, language translations, type systems,
subtyping relations, deduction calculi, and many other relevant judgement\EZ{s}. 
Although inference systems have been introduced for dealing with inductive definitions,
in the last two decades several authors have focused on their coinductive interpretation.

\citet{CousotCousot92}  define divergence of programs by coinductive interpretation of an inference system that extends the big-step operational semantics.
The same approach is followed by other authors \cite{HughesMoran95, Schmidt98, LeroyGrall09}.
\citet{LeroyGrall09} analyse two kinds of coinductive big-step operational semantics for the call-by-value \LambdaCalculus, and study their relationships with the small-step and denotational semantics, and their suitability for compiler correctness proofs.
Coinductive big-step semantics is used as well to reason on cyclic objects stored
in memory \cite{MilnerTofte91,LeroyRouaix98}, and  to prove type soundness in Java-like languages \cite{Ancona12,Ancona14}.
Coinductive inference systems are also considered in the context of type analysis and subtyping for object-oriented languages
\cite{AnconaLagorio09, AnconaCorradi14}. 

More recently several approaches have been proposed to extend existing programming languages to support coinductive data types and corecursion. 
Therefore, these proposals are more focused on operational aspects, and their corresponding implementation issues; we can find contributions  in all most popular paradigms:  logic \cite{SimonMBG06, SimonBMG07}, object-oriented \cite{AnconaZucca12, AnconaZucca13}, \EZ{and} functional \cite{JeanninKS13, JeanninKS17}.
Also in  type theory \cite{Coquand93, AbelPientka13, AbelPTS13, Mogelberg14} and category theory \cite{AdamekMV06, CaprettaUV06, CaprettaUV09} we can find some research effort in this direction with a more abstract perspective.

The following discussion of related work will be divided into these two areas: language support for programming with coinductive data types and abstract models for dealing with such data types.

\section{Programming with coinductive data types}

\paragraph{Logic paradigm}
A logic program is a set of Horn clauses $\Clause{A}{B_1, \ldots, B_n}$, specifying that the atom $A$ (the head) is valid if atoms $B_1, \ldots, B_n$ (the body) are valid.
The execution of a program consists in the \emph{resolution} of a goal (a set of atoms) against the program:
the interpreter tries to build a valid derivation of the atoms in the goal using the clauses in the program.

\FD{Objects manipulated by a logic program are ground\footnote{That is, with no free variables.}  terms and ground atoms built on them, hence the declarative semantics of a logic program is a set of such atoms. 
Actually, a logic program is a particular inference system on a universe consisting of ground atoms, hence we can reuse  the model-theoretic semantics of inference systems to define the declarative semantics of a logic program. }

\FD{In standard logic programming (\LP), terms are finite, hence  atoms are finite too, thus the universe is the \emph{Herbrand base}, that is\EZ{,} the set of all finite ground atoms. 
Since everything is finite, the most reasonable choice for the semantics of a logic program is its inductive interpretation, that is,  the least fixed point of the inference operator induced by the logic program. }

\FD{Therefore, only finite derivations are valid, hence a sound resolution procedure has to try to build a finite derivation for all atoms in the goal. }
More precisely, the standard resolution of \LP, called \emph{SLD resolution},  is performed in three steps: 
\begin{enumerate}
\item First, an atom $A$ from the current goal is selected.
\item Then, the interpreter looks for a clause $\Clause{A'}{B_1, \ldots, B_n}$  in the program whose head unifies with the selected atom (we need unification since both $A$ and $A'$ may contain free variables).
\item Finally, all atoms in the body $B_1, \ldots, B_n$ are added to the goal and the substitution deriving from the unification is applied to the goal.
\end{enumerate}
These steps are iterated until we get an empty goal. 

\FD{In \emph{coinductive logic programming} (\coLP), introduced by \citet{SimonBMG07}, also infinite terms \EZ{are considered}, hence infinite atoms.
Therefore, \coLP programs are interpreted in a different universe: the \emph{complete Herbrand base}, that consists of all finite and infinite ground atoms.
This is the reason why the declarative semantics of a \coLP program is its coinductive interpretation, that is, the greatest fixed point of the induced inference operator.
In this way also definitions of coinductive predicates are supported. }

From an operational perspective, first of all we have to represent infinite terms.
\EZ{More precisely}, only \emph{regular terms} are considered, that is, terms with a finite number of subterms, that can be naturally represented through unification as a finite set of syntactic equations \cite{AdamekMV06b}.
Then, a sound resolution procedure \cite{SimonMBG06, SimonBMG07, AnconaDovier15}  can be defined, based on cycle detection (\emph{coresolution}).
That is, during the resolution the interpreter keeps track of all encountered atoms and, when it selects from  the current goal an atom that has already been encountered, it simply accepts it. 
In this way also cyclic derivations can be built  enabling coinductive logic programming.

In \coLP only standard coinduction is supported. 
In \cite{Ancona13, MantadelisRM14} more flexible operational models are provided.
In particular, the notion of \lstinline!finally! clause, introduce by \citet{Ancona13}, allows the programmer to specify a fact that \EZ{should be} resolved when a cycle is detected, instead of simply accepting the atom. 
In this way,  predicates that are neither purely inductive nor purely coinductive can be defined and used  in a logic program.

The notion of \lstinline!finally! clause has inspired coaxioms \EZ{as described} in this thesis.
However, despite the existing strong correlation with coaxioms, the semantics of \lstinline!finally! clauses does not always coincide with a fixed point of the inference operator induced by the program\EZ{. This} is a \EZ{relevant} difference with coaxioms, that, instead, always generate a fixed point.

\paragraph{Functional paradigm} 
A functional program consists in a set of function definition\EZ{s} together with an expression \EZ{to} be evaluated.
In \EZ{existing} functional languages we can find two main evaluation strategies: \FD{eager evaluation (call-by-value) and  lazy evaluation (call-by-need).
In the former approach, all arguments of a function call are immediately evaluated and their value, if present, is \EZ{returned}; instead, the latter approach delays the evaluation of arguments in a function call until their value is needed for the computation. }

Functional languages  based on lazy evaluation, such as Haskell, naturally support infinite data types and the definition of corecursive functions on them, both by means of standard function definition.
For instance\EZ{,} the stream of all natural numbers can be defined by the function $f(n) = \Cons{n}{f(n+1)}$, hence $f(0)$  represents the stream of all natural numbers.
This definition makes sense because the recursive call $f(n+1)$ is evaluated only when we access the tail of the stream, hence, without causing non termination.

However, in this way, infinite objects are never fully available; in other words, they are ``potential'' objects that can be deconstructed infinitely many times, but functions that require the whole object to be computed cannot be defined.
For instance, if we want to compute the maximal element of an infinite list, we need to inspect the whole list, \EZ{and this} cannot be done lazily.

In the ML family, instead, which adopts eager evaluation, a different approach has been considered: as in the logic paradigm, infinite terms are restricted to regular ones, and a different semantics is defined for corecursive functions. 
This idea is implemented in CoCaml \cite{JeanninKS17}.

CoCaml, as already OCaml, allows the programmer to declare regular objects through the let-rec construct. 
A corecursive call to a function is performed by building a system of equations that  will be solved by an equation solver specified in the function definition. 
The system of equations is constructed  associating with each function call  a variable and partially evaluating the body of the function, by replacing each corecursive call with the associated variable. 
Solvers can be either pre-defined or directly written by the programmer in order to enhance flexibility.
Among proposed pre-defined solvers in \cite{JeanninKS17}, the \emph{iterator solver} seems to have an expressive power very similar to \lstinline!finally! clauses.

From a more abstract point of view, corecursive functions are solutions to an appropriate hylo diagram, whose existence is guaranteed by the fact that  the codomain with the equation solver \EZ{should} form an Elgot algebra \cite{AdamekMV06}, see next section\EZ{. H}owever\EZ{,} in \cite{JeanninKS13, JeanninKS17} there is no formal proof of this.
The intuition suggests that choosing a solver corresponds to choose a specific partial
order on the codomain of the function, in such a way that the desired function is a fixed point in the corresponding CPO. 

However, he spirit of our work is very different from that \EZ{of} CoCaml, since we do not aim to extend a practical language with corecursion, 
but, rather, to provide a very general framework which smoothly extends the well-known notion of inference system, and that could be used in many useful contexts, as shown in \refToSection{coaxioms-examples}. 
The foundation of CoCaml  is based on the theory of recursion in the framework of { category theory and in particular using coalgebras. }
Our approach, instead, relies on the standard complete lattice of subsets, with set inclusion as partial order. 
In this way, a single and simple model, based on classical results, works uniformly for any possible recursive definition expressed in terms of an inference system wit coaxioms.

\paragraph{Object-oriented paradigm}
In the object-oriented paradigm cyclic objects are usually managed relying on imperative features, thus the language does not provide any native support for computing with such objects.
The programmer has to implement ad-hoc machinery to deal with cyclic objects in an appropriate way, and this is often involved and error-prone.

In order to overcome these difficulties, \citet{AnconaZucca12, AnconaZucca13} have proposed an extension of Featherweight Java (\FJ) \cite{IgarashiPW99}: corecursive Featherweight Java (\coFJ).
This is a purely functional core calculus for Java-like languages supporting cyclic objects and corecursive methods.

Cyclic objects are represented by syntactic equations. 
They cannot be directly written by the programmer, 
but only built during the execution by corecursive methods.
Analogously to the mechanism we have described for coresolution, each corecursive  call is evaluated in an environment associating to already encountered calls a unique label. 
If the call is in the environment, then the associated label is returned as result, otherwise a fresh label is associated to the current call, and the method body is evaluated in the extended environment;
finally, an equation for this new label is returned as result.

To make the mechanism more flexible, like in the logic paradigm,  the authors introduce a \lstinline!with! clause associated either to the expression or to the method definition, that is, an expression that will be evaluated when a cycle is detected instead of simply returning the label\EZ{, and} this provides support for methods that are neither purely recursive nor purely corecursive.
Again like \EZ{in} the logic paradigm, this feature has inspired coaxioms and is strongly related to them, however the semantics of \lstinline!with!  clauses may not always correspond to a fixed point, while coaxioms always generate a fixed point.

\section{Category-theoretic and type-theoretic models}

\paragraph{Algebras, coalgebras and (co)recursion}
The category-theoretic solution to interpret (co)inductive definitions is based on the notions of   algebras and coalgebras of an endofunctor $\function$ on a given category \cite{JacobsRutten97}.
An \emph{$\function$-algebra} is a pair $\Pair{A}{\alpha}$ where $A$ and $\fun{\alpha}{\function A}{A}$ are respectively an object and an arrow in the base category;
dually, an \emph{$\function$-coalgebra} is a pair $\Pair{C}{\gamma}$ where $C$ and $\fun{\gamma}{C}{\function C}$ are respectively an object and an arrow in the base category.} 

Here, the endofunctor represents the structure of the definition, for instance in the example of natural number\EZ{s} the functor is $\function X = 1 + X$, where $1$ is the terminal object in the base category and $+$ denotes the coproduct;
or for streams of type $A$ the functor is $\function X = A\times X$, where $A$ is an object and $\times$ denotes the product.

The object inductively defined by $\function$ is the \emph{initial $\function$-algebra} $\Pair{\lfp{\function}}{\InAr}$ \cite{GoguenThatcher74}, while the coinductively defined object is the \emph{final $\function$-coalgebra} $\Pair{\gfp{\function}}{\FinAr}$ \cite{Rutten00,Jacobs16}. 
They are an algebra and a coalgebra satisfying the following universal property:
for each $\function$-algebra $\Pair{A}{\alpha}$ (resp. $\function$-coalgebra $\Pair{C}{\gamma}$) {there} exists a unique arrow $\fun{\alpha^*}{\lfp{\function}}{A}$ (resp. $\fun{\gamma_*}{C}{\gfp{\function}}$) such that the following diagram commutes:
\begin{center}
\begin{tikzcd}
\lfp{\function} \ar[r, "\alpha^*"] & A \\
\function \lfp{\function} \ar[u, "\InAr"] \ar[r, "\function \alpha^*"] & \function A \ar[u, "\alpha"]
\end{tikzcd}
\hspace{2ex}
\begin{tikzcd}
C \ar[d, "\gamma"] \ar[r, "\gamma_*"] & \gfp{\function}  \ar[d, "\FinAr"] \\
\function C  \ar[r, "\function \gamma_*"] & \function \gfp{\function} 
\end{tikzcd}
\end{center}

This universal property is a form of \emph{(co)recursion principle}, that is, it allows us to assign a meaning to (co)recursive definitions of certain types of functions. 
Furthermore, thanks to the Lambek's lemma \cite{Lambek68}, $\InAr$ and $\FinAr$ are \emph{iso}s, thus we can equivalently express the commutativity of the diagrams as $\alpha^* = \InAr^{-1} \cdot \function \alpha^* \cdot \alpha$ and $\gamma_* = \gamma \cdot \function \gamma_* \cdot \FinAr^{-1}$, where $\cdot$ denotes arrow composition.
So these diagrams turn out to be particular instances of a more general recursion scheme described in the following diagram, sometimes called \emph{hylo diagram}:

\begin{center}
\begin{tikzcd}
C \ar[d, "\gamma"] \ar[r, "f"] & A  \\
\function C  \ar[r, "\function f"] & \function A  \ar[u, "\alpha"] 
\end{tikzcd}
\end{center}

This diagram describes a \emph{divide-et-impera} approach, that is, we first decompose the input by the $\function$-coalgebra $\gamma$, then we perform recursive calls, and finally we aggregate results through an $\function$-algebra $\alpha$. 

Unfortunately, in the general case, that is, when $\Pair{A}{\alpha}$ and $\Pair{C}{\gamma}$ are an arbitrary algebra and coalgebra, respectively, we are not guaranteed  that such an $f$ exists, neither that it is unique.  
For such reason, algebras and coalgebras that ensure unique solutions have been studied  under the name of \emph{recursive coalgebras} \cite{CaprettaUV06} and  \emph{corecursive algebras} \cite{CaprettaUV09}, respectively. 

A slightly different perspective is adopted by \citet{AdamekMV06}: rather than requiring the existence of unique solution\EZ{s} to recursion schemes, they focus on canonical solutions.
They represent a recursive definition as a  system of equation\EZ{s}, that, in categorical terms, is an arrow $X \rightarrow  \function X + A$ where $X$ is an object representing variables, $A$ is an $\function$-algebra representing parameters and $\function$ is a structure functor.
In other words, a system of equations is an $H$-coalgebra, where $H$ is the functor $HX = \function X +A$.
A solution to a system of equation\EZ{s} $\fun{e}{X}{HX}$ is an arrow $\fun{e^*}{X}{A}$ that make\EZ{s} the following hylo digram commute\EZ{:}

\begin{center}
\begin{tikzcd}
X\ar[d, "e"] \ar[r, "e^*"] & A  \\
\function X + A  \ar[r, "He^*"] & \function A + A \ar[u, "\Pair{\alpha}{\Id{A}}"] 
\end{tikzcd}
\end{center}

In order to study canonical solutions, \citet{AdamekMV06} define the notion of \emph{Elgot algebra}, that is a triple $\Triple{A}{\alpha}{(-)^*}$ where $\Pair{A}{\alpha}$ is an $\function$-algebra and $(-)^*$ is a function mapping each system of equation\EZ{s} to its (canonical) solution.

\paragraph{Type theory}
Type theories are formal theories  where primitive concepts are types  and terms. 
These theories are very versatile: they can be used as foundational framework for mathematics, to reason about rich type systems (e.g., polymorphic and dependent type systems), and   to implement powerful proof assistants, such as Coq and Agda.

We consider type theories here since they can serve as a foundational framework for programming languages, and actually can be regarded as special programming languages.
Indeed type theories bring together a logical system and a (functional) programming language,
however the latter is quite different from usual languages: here  all functions must be total, that is, they must terminate.

However, non-termination can be encoded relying on infinite objects: a non-terminating computation is one that builds an infinite result. 
Therefore, the system has to guarantee that defined functions either terminate or produce an infinite result. 
The second requirement is achieved by ensuring that the function definition is \emph{productive}, that is, if we inspect the result finitely many times, then we always get a result.

Several approaches to ensure productivity have been proposed, with different representations of infinite objects. 
First attempts represent objects as non-well-founded terms over a signature of constructors, and ensure productivity through  \emph{syntactic checks}, as, e.g., in \cite{Coquand93}.
Here  productivity is guaranteed by checking that each recursive reference is \emph{guarded} by  at least one constructor.
In this way, we are sure that each recursive call unfolds  some constructor, that is, produces some data.

A strong limitation of this approach is that it is not compositional, since the composition of guarded recursive functions may give rise to a non-guarded definition.
To recover compositionality, type-based  productivity checks have been proposed, as, for instance, in \cite{AtkeyMcBride13, Mogelberg14}, where productivity is achieved by a type system with guarded recursive types, represented through a type constructor inspired by modal logic.

In \cite{AbelPientka13,AbelPTS13} a different approach is considered. 
Authors  accept duality between finite and infinite objects and make it even stronger.
Indeed, the former are built using constructors, while the latter are observed using \emph{destructors}. 
Formally, finite objects are instances of recursive variant types (each variant is a constructor), while infinite objects are instances of recursive record types (each field is a destructor) and their semantics is given by an initial algebra and by a final coalgebra\EZ{,} respectively.
Moreover, functions over finite objects are defined by pattern matching, while those that produce infinite objects by \emph{copattern matching}, that is, by defining the behaviour of the function under all possible observations. 
Under this approach, productivity reduces to termination, indeed a function definition  is productive if any finite number of observations on its result terminates, and termination is ensured by the type system using sized types. \\

Despite these categorical and type-theoretic models address analogous problems as coaxioms, the relationship \EZ{with} them is not clear and it is matter of further work.

\cleardoublepage \chapter{Conclusions} \label{chapter:conclu}

Inference systems are a general and versatile framework that is well-known and widely used.
It allows to define several kinds of judgements from operational semantics to type systems, from deduction calculi to language translations.
They can also serve as theory to reason about recursive definitions, providing a rigorous semantics in a quite simple way.

We have described two well-known equivalent semantics for inference systems: one in a model-theoretic style and the other in a proof-theoretic style.
The former defines the interpretation of an inference system as a fixed point of the inference operator, the latter, instead, as the set of judgements  that has a certain witness of their acceptability.
These {witnesses} are called proof trees (or derivations) and represent \FD{the steps, each \EZ{one} justifies\EZ{d} by a rule, that we have to take in order to obtain the desired judgement\EZ{, starting from some assumptions  if necessary. }}

In literature we have not found  a {rigorous enough} (for our aims) treatment of the proof-theoretic semantics, hence we have {provided} it starting from a very precise definition of tree (see \refToSection{trees-graphs}).
Thanks to this precise notion of tree we proved \refToTheorem{graph-tree}{,} that states the existence of a canonical homomorphism of graphs, mapping each node of a given graph to a tree rooted in such node.
This result {has} allowed us to give a new, at the best of our knowledge, proof of the equivalence between the model-theoretic and proof-theoretic semantics in the coinductive case.

The core of this thesis, however, is the concept of inference system with coaxioms (\refToChapter{coaxioms}): a generalized notion of inference system, that subsumes the standard one, supporting  flexible definitions of judgements by structural recursion on non-well-founded data types.
Indeed standard inference systems suffer from a strong rigidity: their interpretation is dichotomous, either inductive (the least one) or coinductive (the greatest one), but what can we do if we need something in the middle?
One may ask if this is a real issue, but the examples we have provided shows that there are {many interesting} cases in which we need a fixed point that is neither the least nor the greatest one, and {standard} inference system{s} are not able to provide such flexibility.

Our work starts from the operational models, closely related to each other,  introduced by \citet{AnconaZucca12, AnconaZucca13} and \citet{Ancona13}.
As already discussed, these operational semantics {introduce} some flexibility for interpreting predicates and functions recursively defined on non-well-founded data types.
The initial {objective of our work} was {to provide} a more abstract semantics for such operational models, hence we developed a first model in \cite{AnconaDZ16}  focused on this aim.
However{,} the result was not satisfactory, since we managed to capture the semantics of a restricted class of definitions, with a model that was quite tricky.

Then, we decided to take a more abstract perspective, considering inference systems as reference framework.
In this context we discover{ed} the notion of coaxioms, that convinced us to be the right one.
We proposed {it} in \cite{AnconaDZ17esop} and discussed it in more detail in the present thesis.

In order to finely describe coaxioms, we have generalized the meta-theory of inference systems by
providing two equivalent semantics, one based on fixed points in a complete
lattice, and the other on the notion of proof tree. 
In the former case, the semantics of an inference system with coaxioms is the greatest fixed point of its corresponding one step
inference operator, below the least pre-fixed point containing the coaxioms; in
the latter case, the standard notion of proof tree for the coinductive case is
generalized by requiring coaxioms to be applicable "at an infinite depth".

More precisely, in order to define the model-theoretic semantics, we have considered closure and kernel systems, usually defined only for the power-set lattice, in the general context of complete lattices, proving some properties of theirs.
Then, we have studied in more detail pre-fixed and post-fixed points of a monotone function on a complete lattice, that form a closure system and a kernel system respectively.
Using these notions we have managed to define the bounded fixed point, that is the fixed point that captures the semantics of inference systems with coaxioms.

From the proof-theoretic perspective, we have provided three different and equivalent  characterizations.
All of them essentially impose a condition on coinductive proof trees\footnote{Here we mean proof trees valid for the coinductive interpretation, hence both well-founded and non-well-founded proof trees\EZ{.}}   to be accepted. 
In other words, all these conditions allows us to filter out undesired derivations. 
The first characterization requires that each judgement in the tree is derivable with a well-founded proof tree  in the extended inference system (the inference system where coaxioms are considered as axioms).
The other two characterizations are based on the notion of approximated proof tree of level $n$, that are well-founded proof trees in the extended inference system, where coaxioms can only be used at depth greater than $n$.
The second proof-theoretic characterization requires all judgements in the coinductive proof to have an approximated proof tree for each level.

The last characterization is quite different: it does not require any coinductive proof tree.
The crucial notion here is that of (strong) approximating proof sequence, that is a sequence of approximated proof trees of increasing level, having an initial portion in common that grows with the level of the approximated proof tree.
The result is that, if we provide such a sequence for a judgement, then it is acceptable.
Indeed, we have defined a notion of limit for approximated proof sequence and proved that this limit is a valid coinductive proof tree.

We have also developed proof techniques to reason with coaxioms, in particular we have defined the bounded coinduction principle, that generalizes the standard coinduction principle, and allows us to prove  completeness for a definition.
We have also provided a proof technique to prove soundness, based on approximated proof trees and a reasoning by contraposition.

\paragraph{Further work}
Starting from this thesis, in order to develop possible extensions and applications of coaxioms, we identify three main directions for further investigations:
\begin{enumerate}
\item deepening the comprehension of coaxioms, 
\item defining language constructs to support flexible (co)inductive  definitions of data types, predicates and functions,
\item and developing applications to model infinite behaviours of programs and systems.
\end{enumerate} 

For what concerns the model, a first compelling direction for further developments is exploring other proof techniques \cite{HurNDV13} for coaxioms and their mechanization in proof assistant.
To this aim, it would be useful exploring the relationship between coaxioms ad type theories, since several proof assistant are based on such theories.

We will also try to investigate the dual to the notion studied here: one could consider the
least fixed point above the greatest post-fixed point contained in the coaxioms,
instead of the greatest fixed point below the least pre-fixed point containing the
coaxioms. In particular, it would be interesting studying inference systems for
which the two different semantics coincide, since in that case we would get a generalization of the induction principle providing us with a proof technique to show soundness.

\FD{An open problem concerning the interpretation generated by coaxioms is its computability.
It is quite obvious that in general this set cannot be computed, however it could be interesting studying conditions and/or restrictions that ensure at least  that this set is semi-decidable.
To this aim it could be useful trying to provide another proof-theoretic characterization based on partial proof trees, that are proof trees with assumptions, and form a complete partial order.\footnote{Special thanks go to Eugenio Moggi for his useful comments   to highlight the importance of computability issues and for his suggestions to start the development in this direction.}  }

Another interesting development is to investigate a variant of the model able to directly capture the definition of functions, rather than representing them as functional relations. 
This would be relevant to more appropriately model language constructs to support flexible (co)recursion in  functional languages.
This variant could also imply a change of framework, moving from lattice theory to domain or category theory, where the semantics of (co)recursive definitions of functions is better supported.
Therefore a deeper comparison  between coaxioms and category-theoretic or type-theoretic models could be useful.

Considering language support for flexible (co)induction and (co)recursion, the first step would be providing a support for coaxioms to the logic paradigm.
As we have already noted, a logic program is much like an inference system, hence the translation of coaxioms in this paradigm seems not to be too complex.
Indeed, we have already done some steps in this direction in \cite{AnconaDZ17coalp}, where we have provided an extension to \LP, defining both a declarative and a sound operational semantics: the former is based on the bounded fixed point, the latter on a combination of SLD and coSLD resolutions.
We have also implemented a prototype meta-interpreter in SWI-Prolog\footnote{Available at \url{http://www.disi.unige.it/person/AnconaD/Software/co-facts.zip}}.

Extending the notion of coaxioms to the setting of object-oriented and functional paradigms is more challenging, due to the gap between the underlying theories.
Indeed, these paradigms deal with functions rather than relations, that are, instead, the objects managed by inference systems with coaxioms; since functions can be seen as particular relations, we can represent them in our model, however we have always to ensure that the generated fixed point is actually a function, and this is not always guaranteed.

For the object-oriented paradigm  a starting point could be the revision of the operational semantics of \coFJ on the basis of the abstract model provided by coaxioms; 
in particular to guarantee that the function denoted by a function definition in \coFJ is actually a fixed point of  the induced monotone operator.
The extension to the functional paradigm is even more challenging: the model does not directly support higher order functions, that are a key feature of functional programming languages.
Hence, to these aims, a deeper study of the abstract model is surely required.

Finally, starting from the example in \refToSection{op-sem}, it could be interesting to better study the capabilities of coaxioms to model  non-termination.
We have already done a first step in this direction in \cite{AnconaDZ17oopsla}, where we apply the approach sketched in \refToSection{op-sem} to an imperative \FJ-like language, studying in particular application of proof techniques for coaxioms to prove the soundness of predicates (such as typing relations) with respect to the operational semantics.

A further extension of this application would be applying coaxioms to define trace-based operational semantics \cite{NakataUustalu09}, that allows to capture finer characterizations of the behaviour of non-terminating programs.

\backmatter
\cleardoublepage
\markboth{\spacedlowsmallcaps{\bibname}}{\spacedlowsmallcaps{\bibname}} 
\phantomsection
\addtocontents{toc}{\protect \vspace{\chapSpace}}
\addcontentsline{toc}{chapter}{\tocEntry{\bibname}}
\printbibliography

\end{document}